\documentclass[graybox]{svmult}

\usepackage{type1cm}        
\usepackage{makeidx}         
\usepackage{graphicx}        
\usepackage{multicol}        
\usepackage[bottom]{footmisc}

\usepackage{newtxtext}       %
\usepackage{newtxmath}       

\usepackage{url}

\usepackage{float}
\usepackage{graphicx}
\usepackage{subfigure}
\usepackage{subcaption} 
\usepackage{caption} 
\usepackage{comment}
\usepackage{booktabs}

\usepackage{tabularx}
\usepackage{multirow}

\usepackage[]{hyperref} 

\makeatletter
\def\blfootnote{\xdef\@thefnmark{}\@footnotetext}
\makeatother

\usepackage[sorting=none]{biblatex}
\bibliography{main.bib}

\makeindex             

\begin{document}

\title*{Unboxing Diffusion Models for the Arts: Interactive Model Bending and Practice-Based Explainability}
\titlerunning{Unboxing Diffusion Models for the Arts}
\author{Ahmed M. Abuzuraiq and Philippe Pasquier}

\institute{Ahmed M. Abuzuraiq\at School of Interactive Arts and Technology, Surrey, Canada \email{aabuzura@sfu.ca}
\and Philippe Pasquier \at School of Interactive Arts and Technology, Surrey, Canada \email{pasquier@sfu.ca}}
%
%

\maketitle

\abstract*{Explainable AI (XAI) in creative practice can be less about technocentric explanation and more about enabling artists to inspect, modify, and debug models as part of making. Yet large-scale text-to-image diffusion systems are typically presented as opaque end-to-end tools, limiting this kind of material engagement. We argue that even large models can function as creative materials when their internal structure is made visible and manipulable. To support this, we propose a hands-on approach to explainability centred on experimentation and intervention. We instantiate this approach with a model bending and an interactive (inspection) interface integrated into ComfyUI’s node-based workflow, including interactive layer selection and intervention controls. Through qualitative and quantitative analysis of bending interventions in Stable Diffusion 1.5, we show how manipulating specific components of a diffusion pipeline produces relatively consistent families of visual effects, allowing artists to build practical, layer-level intuition about how different parts of the model shape generated images.}

\abstract{Explainable AI (XAI) in creative practice can be less about technocentric explanation and more about enabling artists to inspect, modify, and debug models as part of making. Yet large-scale text-to-image diffusion systems are typically presented as opaque end-to-end tools, limiting this kind of material engagement. We argue that even large models can function as creative materials when their internal structure is made visible and manipulable. To support this, we propose a hands-on approach to explainability centred on experimentation and intervention. We instantiate this approach with a model bending and an interactive (inspection) interface integrated into ComfyUI’s node-based workflow, including interactive layer selection and intervention controls. Through qualitative and quantitative analysis of bending interventions in Stable Diffusion 1.5, we show how manipulating specific components of a diffusion pipeline produces relatively consistent families of visual effects, allowing artists to build practical, layer-level intuition about how different parts of the model shape generated images.}

\section{Introduction}
Explainable AI \blfootnote{\textbf{Under review for "Explainable AI for the Arts" (N. Bryan-Kinns, Ed.), Springer.}} (XAI) traditionally focuses on demystifying machine learning systems for auditing, transparency, or safety~\cite{gunning2019xai}. However, explainability in creative contexts can serve different roles, including making models modifiable and debuggable for a meaningful artistic engagement~\cite{bryan2023exploring}, and creating the conditions for a sustained artistic practice that goes beyond mere amusement~\cite{tecks_2024_ExplainabilityPathsSustained}. The last workshop on XAIxArts~\cite{bryan-kinns_2025_XAIxArtsManifestoExplainablea} culminated in a manifesto that called artists and researchers to explore alternatives to the technocentric explanations of AI that value artistic practices and use intentional hacking, glitches, and imperfections as creative tools. This work continues along those lines.

It is argued that working with curated (small) datasets and human-scale models can enhance artists’ agency~\cite{vigliensoni_2022_smalldatamindsetgenerative}. Broadening the scope of the artistic process with AI to include both model training and inference can also reinforce artists' trust in AI models~\cite{tecks_2024_ExplainabilityPathsSustained}. Both approaches afford artists more control over AI models. However, artists' ability to craft with AI models as materials for creation~\cite{caramiaux2022explorers, broad_2024_using, rozental_2025_HowArtistsUse} is diminished when working with large-scale generative models~\cite{abuzuraiq_2024_SeizingMeansProduction}. If small-data and model training present an alternate explainable way for artists working with human-scale generative models, what options are available for artists wanting to work with large-scale models such as text-to-image diffusion models? Although this material relationship is easier to build with small models, where data and model structure/training are accessible and malleable, it becomes more challenging at scale. However, we argue that large models can also be treated as material, provided their structure is exposed and can be manipulated.

We propose that treating AI models as materials, especially within interfaces that expose the model's components like the node-based interface of ComfyUI, can foster a craft-based relationship between artists and generative systems. In particular, through long-term engagement and hands-on manipulation of a model's components, artists can develop a form of tacit understanding and familiarity with their AI materials akin to that found in traditional crafts, i.e. a form of explainability that is rooted in doing. We explore the application of these ideas for large-scale models through a plugin and interface for model bending~\cite{broad_2021_NetworkbendingExpressive} that is implemented into the node-based interface of ComfyUI~\cite{2024_ComfyAI}. 

We argue that interactive model bending can expose the material structure of large generative models to artists, enabling creative exploration and practice-based explainability through direct intervention. To support this claim, we present a model-bending plugin integrated into ComfyUI, an interactive inspection interface for navigating and manipulating diffusion model components, and a systematic qualitative and quantitative exploration of bending outputs in Stable Diffusion 1.5. Together, these contributions point toward how large-scale diffusion models can be engaged with creative materials.

\section{Background}

\subsection{Image Generation Models}
Generative models for image generation aim to learn the underlying distribution of visual data so that new, realistic images can be synthesized. Among the various approaches, diffusion-based generative models have emerged as a powerful and flexible framework, particularly for high-fidelity and text-conditioned image synthesis. 

\subsubsection{Latent Diffusion}
Diffusion models gradually add noise to real data using a fixed forward process, then train a neural network to reverse that process step by step so it can turn random noise into coherent samples at generation time. Latent diffusion does the same thing in a compact latent space, e.g., learned by a variational autoencoder (VAE) model, where the encoder maps images into low-dimensional latents for diffusion, and the decoder maps the denoised latents back to pixels, making generation more efficient. The outputs of the text encoder of a contrastive model (e.g., CLIP or T5) can be fed into the diffusion model via cross-attention, where they are used to guide the generation to adhere to a user prompt. In addition to text conditioning, each denoising step is associated with a discrete time index that indicates the current noise level. These timesteps are embedded , via the \textit{time\_embed} layers in SD1.5, into continuous representations and provided to the model alongside the noisy latents. This allows the network to condition its denoising behaviour on the current noise level, ensuring a coherent progression from high-noise initial states to low-noise, structured outputs.

\subsubsection{Transformers and Attention}
Transformers are a neural network architecture built around attention
mechanisms, originally introduced for sequence modelling and now widely
used across text, image, audio, and multimodal tasks.

Attention mechanisms form the backbone of diffusion models by allowing information to flow between different spatial regions based on content relevance. 
When mathematically described, attention is defined as: 
\begin{equation}
\text{Attention}(Q, K, V) = \text{softmax}\!\left(\frac{QK^{T}}{\sqrt{d}}\right)V,
\end{equation}
where $Q$, $K$, and $V$ denote the query, key, and value matrices, respectively, and $d$ is the dimensionality of the key vectors. When a model’s output is conditioned on an external signal, such as a text prompt, a variant known as cross-attention is employed, in which the key and value matrices are derived from text embeddings while the query matrix originates from image features. This formulation aligns linguistic information with visual representations, enabling text-conditioned image generation.

Earlier latent diffusion models, e.g., Stable Diffusion v1.5, for image generation utilized a UNet model for noise prediction at each step in the denoising process. A UNet model is composed of gradually downsampled input blocks that feed to middle blocks, which then expand again to recreate the input images through output blocks. UNet models also feature skip connections that relay information from input blocks to output blocks to preserve spatial details lost during downsampling and overall for better model training. More recent models (Stable Diffusion v3 onwards) have shifted to using a Diffusion Transformer (DiTs) architecture for noise prediction.

\subsection{Network Bending}
Network bending refers to the deliberate manipulation of an AI model’s internal activations or computational pathways to intervene in its generative process, typically for expressive or exploratory purposes~\cite{broad_2021_NetworkbendingExpressive}. 
This is achieved by injecting a \textit{bending operator} into a model to transform its intermediate outputs during inference, without the need for additional training or data. Bending operators can perform operations such as adding noise, multiplying/adding scalar values, rotation, and scaling; morphological operations like erosion and dilation, or other custom operations. Network bending is inspired by circuit bending, a practice originating in experimental electronic music, in which artists intentionally short-circuit or rewire electronic devices (often children's toys or synthesizers) to produce unpredictable or expressive behaviours~\cite{ghazala2005circuitbending}.

Previous work on network bending has applied interventions at various layers of the synthesizer network in StyleGAN models~\cite{broad_2021_NetworkbendingExpressive, kraasch_2022_AutolumeLiveTurningGANs, 2024_AutolumeNeuralnetworkbased}. More recently, Pavlov's exploration focuses on exchanging the parameter-free activation functions of StyleGAN2 and BigGAN models~\cite{pavlov_2025_Controllingimagegeneration}. In the context of diffusion models, bending operations have been implemented at user-specified timesteps within the denoising process~\cite{dzwonczyk_2024_NetworkBendingDiffusion} (as inferred from source code), and at select layers of the noise prediction model (UNet)~\cite{curtis_PromptsDevelopingExpressive}. Though not labelled as model bending, the work by Grabe et al. on Hidden Layer Interaction similarly explores the manipulation of attention layers in a Stable Diffusion 1.4 model~\cite{grabe_2025_PatchExplorerInterpreting}. In this work, we introduce a set of tools for model bending within ComfyUI that enable artists to bend various components of the latent diffusion pipeline, including and extending beyond the components covered by previous work~\cite{broad_2021_NetworkbendingExpressive, dzwonczyk_2024_NetworkBendingDiffusion, curtis_PromptsDevelopingExpressive}. Our approach does not utilize clustering of features to capture semantic directions in the model, instead we find that large-scale text-to-image diffusion models appear to exhibit visually rich latent spaces that are sufficient for productive bending~\footnote{Though certain parts  (\textit{h-space}) seem to be more potent for semantic manipulation\cite{kwon_2023_DiffusionModelsalready}}. 

In the music domain, Collins recently explored the impact of bending on the musical features of music generated by the Stable Audio 1.0 model~\cite{collins_2025_UnstableAudioCode}. Earlier work by Yee-King and McCallum explored network bending for the Differentiable Digital Signal Processing (DDSP) model~\cite{yee-king_2021_Studioreportsound} for sound synthesis. Aside from exploring bending within creative domains, multiple generic frameworks exist for tracing and then intervening on the internal states of PyTorch models, including NNSight~\cite{fiotto-kaufman_2025_NNsightNDIFDemocratizing}, Pyvene~\cite{wu-etal-2024-pyvene} and baukit~\cite{bau_2022_Baukit}. Model bending can be easily implemented using any of these frameworks.

Shared among the above approaches is the manipulation of model weights without requiring additional training, e.g., in contrast to the work by Aldegheri et al.~\cite{aldegheri_2023_Hackinggenerativemodels} insert a small trainable module to push the model towards creative outputs. However, we focus on training-free approaches as we find them more accessible to non-technical artists~\footnote{The Wekinator for predictive models and Low Rank Adaptation (LoRAs) for generative models present successful examples of accessible and effective training-based approaches}. 


\subsection{Bending Interfaces}
Model bending exposes all of the model's parameters, which can range from millions (in StyleGAN) to billions (in Stable Diffusion 1.4), that are distributed across numerous parts of the model, such as fully connected, convolutional, normalization or attention layers, to name a few. Furthermore, multiple types of operations can be applied to the bent parts. This presents a challenge in effectively applying model bending in practice. As a response, we find multiple systems for model bending (or related) that introduce interfaces to assist in specifying the parts of the model to bend. These systems include Autolume~\cite{sobhan_2025_AutolumeGANbasedNoCoding}, which allows users to visualize and transform intermediate layers in StyleGAN2 models, Pavlov's interface~\cite{pavlov_2025_Controllingimagegeneration}, where users pick neural layers and specify how they wish to change their activation functions, and Grabe's interface for hidden layer interaction~\cite{grabe_2025_HiddenLayerInteraction}, which enables users to target specific ranges within a convolutional channel of a Deep Convolutional GAN model. Lastly, the PatchExplorer interface~\cite{grabe_2025_PatchExplorerInterpreting} allows users to visualize and manipulate attention heatmaps in diffusion models. All of these interfaces present the final model inference outputs~\cite{grabe_2025_HiddenLayerInteraction}, but some also visualize the model layers' structure~\cite{pavlov_2025_Controllingimagegeneration}, while others visualize select intermediate feature maps~\cite{sobhan_2025_AutolumeGANbasedNoCoding}, or all the attention heatmaps for a given prompt~\cite{grabe_2025_PatchExplorerInterpreting} to facilitate semantic and targeted edits. Our interface focuses on presenting the structure and bending outputs of the model instead of its intermediate representations. By being implemented in ComfyUI, our system can be extended to cover a wide range of generative models or custom bending operators. 

\section{Diffusion Bending: System and Interface}
To explore the potential of model bending with diffusion models, we implement custom nodes that enable intervening on diffusion model parts in ComfyUI, and build an interactive interface that enables model bending within the context of ComfyUI. In what follows, we briefly describe the key elements of our approach.

\subsection{Stable Diffusion Models}
In this work, we explore bending the Stable Diffusion 1.5 (SD1.5) model~\cite{rombach_2022_HighResolutionImageSynthesisb}, an early large-scale text-to-image system. While newer and more powerful models such as Flux or Z-Image have since been developed, the overall components of these systems remain largely the same, including text conditioning, encoding and decoding in a compressed latent space, multi-step denoising, and the use of a neural network—whether a UNet or a Diffusion Transformer—for noise prediction.

Stable Diffusion 1.5 employs a UNet architecture for denoising, composed of multiple input blocks, a middle block, and multiple output blocks. These blocks contain residual blocks, which consist of convolutional layers with skip connections that pass feature maps from the encoder to the decoder at corresponding resolutions. These skip connections help preserve spatial detail and stabilize training by enabling the direct reuse of low-level features during reconstruction. The Spatial Transformers incorporate both self-attention and cross-attention layers, allowing image features to interact with each other and with text embeddings used for conditioning. In addition, the UNet includes downsampling and upsampling operations that progressively reduce and then restore spatial resolution, enabling the model to capture both fine-grained details and high-level semantic structure during the denoising process. 

Finally, we note that components beyond the UNet in the Stable Diffusion pipeline can also be manipulated. These include the Variational Autoencoder (VAE) used for image encoding and decoding, the text embeddings produced by the Contrastive Language–Image Pre-training (CLIP) model, intermediate latent representations, LoRA matrices appended to the UNet, and even activations at selected timesteps during sampling. Bending operations can be applied at these points to introduce controlled variations while keeping other generation parameters fixed. Although we provide custom nodes to support these manipulations, they are beyond the scope of this paper and are therefore not discussed further.

\subsection{ComfyUI}
ComfyUI is an open-source, node-based interface for designing and executing advanced diffusion pipelines~\cite{2024_ComfyAI}. Unlike other interfaces such as AUTOMATIC1111~\cite{2024_Automatic1111WebUI}, which prioritizes simplicity and abstracts away the model’s internal workings, ComfyUI adopts a low-code, modular approach that decomposes the latent diffusion process into discrete nodes. These nodes represent stages such as loading base models and applying LoRA adaptations, encoding images into latent representations, generating textual embeddings (e.g., via a CLIP model), and configuring the denoising process in diffusion models. ComfyUI’s backend also manages model loading in a way that helps circumvent GPU memory constraints.

\begin{figure}
    \centering
    \includegraphics[width=1\linewidth]{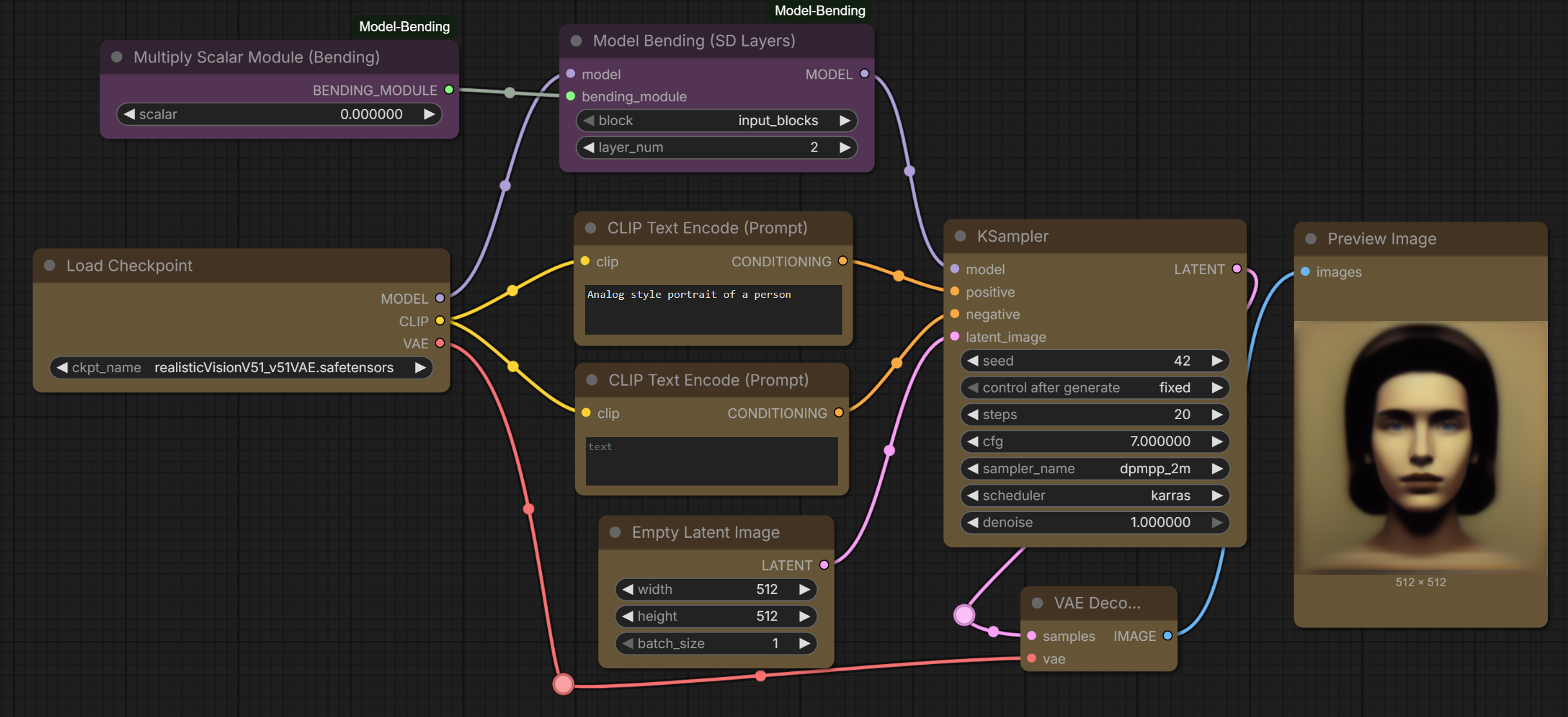}
    \caption{Basic workflow in ComfyUI (gold). The UNet bending nodes are shown in purple.}
    \label{fig:basic_workflow}
\end{figure}

We chose ComfyUI because it exposes the inner components of the diffusion pipeline, encouraging users to explore, customize, and develop an understanding of each component. Additionally, ComfyUI often supports the most up-to-date models for image and video generation, making it a reasonable choice for research-driven or artistic experimentation. A basic workflow in ComfyUI is illustrated in~\autoref{fig:basic_workflow}, and includes steps such as model loading, prompt encoding with CLIP, sampling with user-defined parameters (e.g., number of steps, CFG scale), and decoding the resulting latent image.

\subsection{Diffusion Bending Custom Nodes}
In this work, we implement model bending through custom nodes in ComfyUI. The overarching goals of this work are: (1) to provide tools for introducing visual variations and diversity into the text-to-image diffusion generative process, (2) facilitate a better understanding of the generative process and its parts, (3) lower the barrier of entry to model manipulations, and (4) integrate into an existing ecosystem of tools and access the latest generative models. We address goals by developing a diverse set of custom nodes that enable interventions at different stages of the latent diffusion process, alongside an interactive interface for visualizing and manipulating model components. By integrating the system directly into ComfyUI, the tool benefits from ComfyUI's growing community, support for the most recent models, and access to a wide range of existing plugins. The tool can be downloaded manually through a git repository~\footnote{\url{https://github.com/abuzreq/ComfyUI-Model-Bending}} or by installing it directly through ComfyUI's Nodes Manager. 

In general, bending works by manipulating the outputs of parts of the generative process before passing them downstream. Users specify the model to be bent, the parts of the model they wish to manipulate through paths that describe the hierarchy leading to these parts (e.g. \textit{diffusion\_model.middle\_block.0.in\_layers}), as well as choosing the bending operation to be applied to the tensor output at that part. The bending operations are PyTorch modules, and can be chosen out of a predefined collection (e.g. multiplication, rotation, or noise addition, among others) or by creating custom operations as new PyTorch modules. Lastly, users may optionally provide the range of denoising steps within the diffusion process at which to bend; by default, bending is applied to all steps. 

With these inputs in place, the bending node copies the model definition (so bending changes only impact downstream parts of the workflow) and assigns hooks (event listeners) with a function that calls the bending module when any of the bending paths are reached during inference. Alternatively, but functionally equivalent, users can use NNSight as a backend for implementing their specified model interventions (i.e. bending operations).

\begin{figure}
    \centering
    \includegraphics[width=0.5\linewidth]{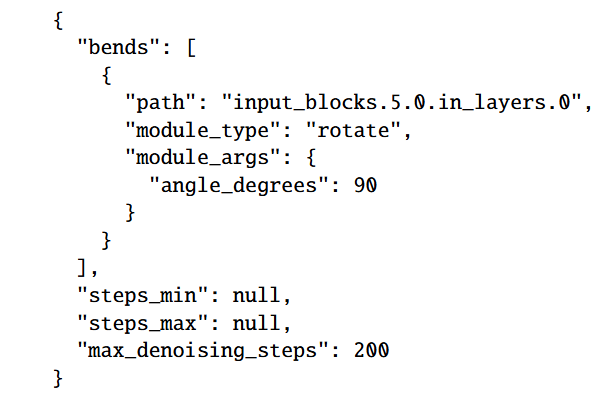}
    \caption{JSON description of bending operations.}
    \label{sec:json_format}
\end{figure}

We provide multiple ways for specifying the bending paths and operations. The simplest \textit{Model Bending (SD Layers)} node as in \autoref{fig:basic_workflow}) allows users to choose from a pre-computed list of convolutional layers in the UNet. Alternatively, users can use the \textit{Model Bending} or \textit{Model Bending (NNSight)}, which enable user to input a list of comma-separated paths at which to bend, and a bending operation to apply. Lastly, users can specify the bending parameters in a JSON format, including multiple operations at multiple paths, and the range of timesteps to bend at (\autoref{sec:json_format}). 

\subsection{Interactive Bending}
\label{sec:interactive_bending}
Considering the complexity and scale of modern diffusion models, as well as the wide range of bending parameters, we found it important to support users in interactively identifying the parts they wish bend. We describe these supports next.

\subsubsection{Model Inspection}
\begin{figure}
    \centering
    \includegraphics[width=0.5\linewidth]{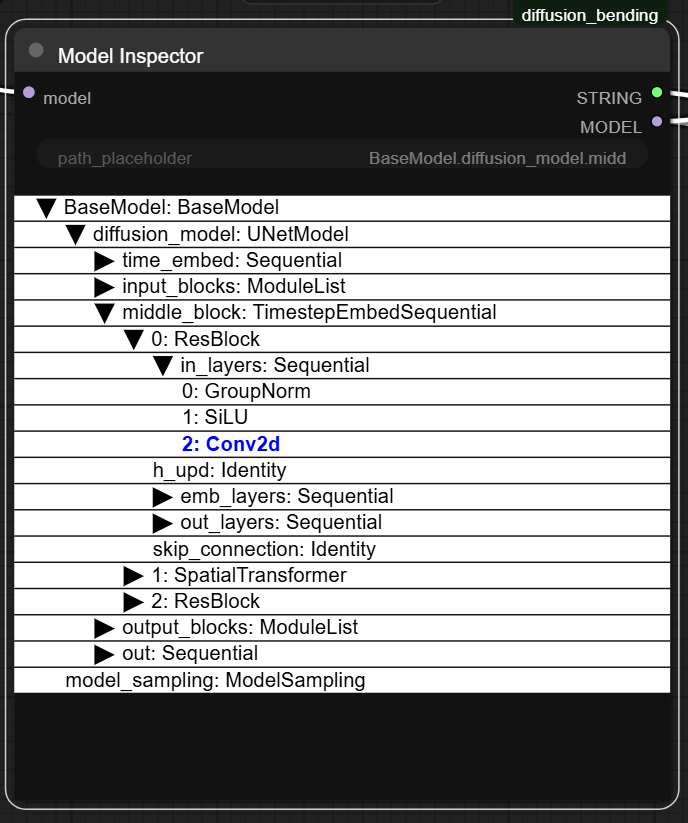}
    \caption{With the model inspector node, artists can interactively pick layers from the model hierarchy to manipulate. Picked layers are highlighted.}
    \label{fig:model_inspector}
\end{figure}

For more granular control, including the ability to select specific blocks or layers, we provide the Model Inspector node (\autoref{fig:model_inspector}). This tool displays the model’s architecture as a nested, expandable tree, allowing users to visually navigate and select any bending path to pass to the model bending node. It currently supports both UNet and VAE models. 

\subsubsection{Interactive Bending}
\begin{figure}
    \centering
    \includegraphics[width=1\linewidth]{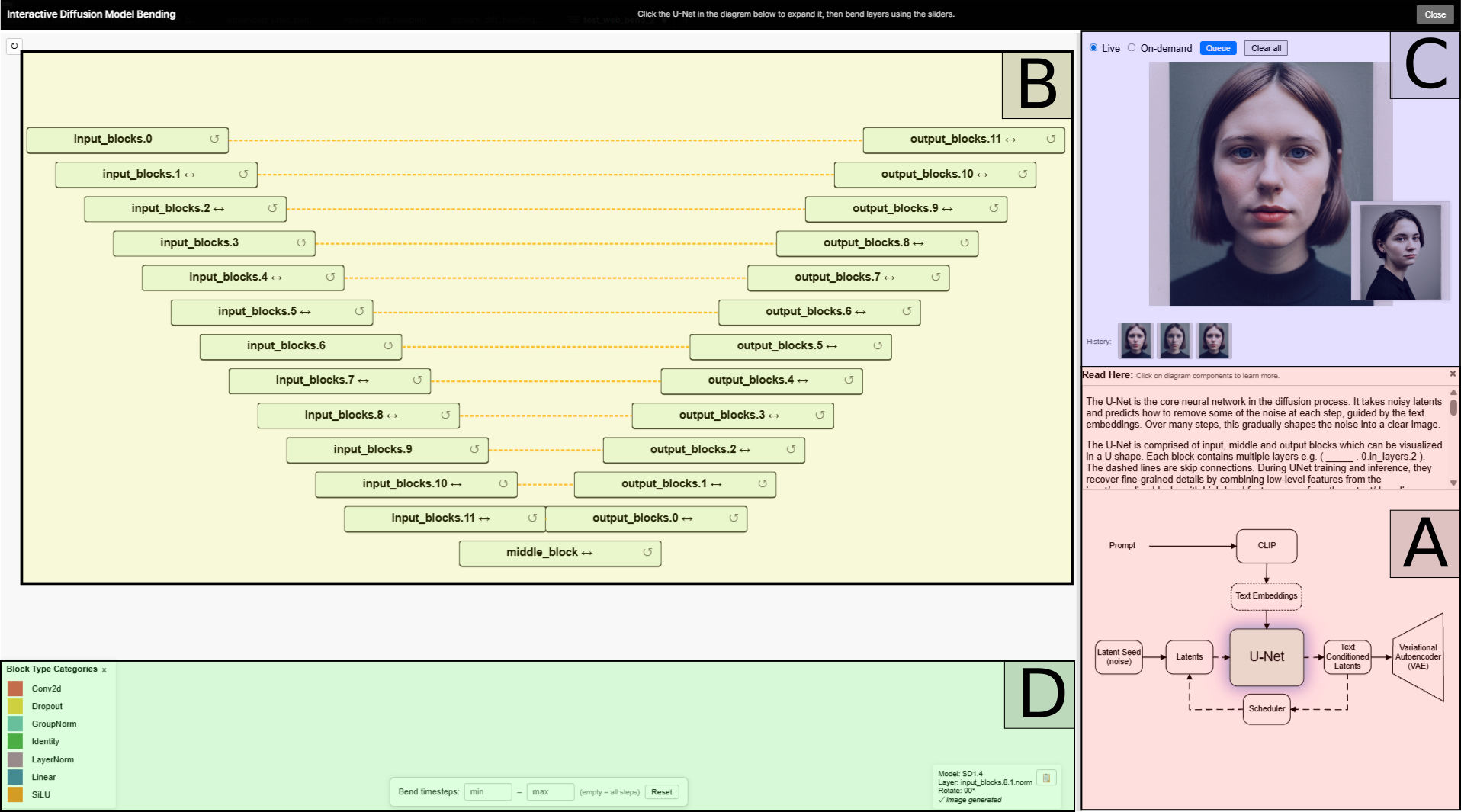}
    \caption{Interactive Interface for Bending Diffusion Models. The interface shows the structure of the UNet (B), presents the current and past bending outputs (C), and shows the high-level structure of the entire latent diffusion process (A).}
    \label{fig:interactive_bending_web}
\end{figure}

The Interactive Bending Web Interface is a web-based tool that allows users to explore and apply interventions (bending operations) to the UNet of a latent diffusion model. The interface presents a model structure diagram as an SVG visualization of the diffusion pipeline, including components such as the UNet, VAE, CLIP, and scheduler, with a brief explanation revealed when each part is clicked (\autoref{fig:interactive_bending_web} - A). A dedicated UNet layer explorer displays the network in an interactive U-shaped layout, where major sections like input blocks, the middle block, and output blocks can be expanded (\autoref{fig:interactive_bending_web} - B). Within each block, layers are organized by module type -such as ResBlock, SpatialTransformer, Upsample, and Downsample- using collapsible panels for easier navigation. An early version of the interface, with pre-computed results, can be found at ~\url{https://diffusion-bending-demo.netlify.app}. To access and run the interactive interface as well as the scripts used to generate the results in the in Section~\ref{sec:preliminary_study}, please download the project from \url{https://drive.google.com/file/d/1DJx3WVfdQLsd73VEoe-T95OAzbMpw6OD/view?usp=sharing}.\footnote{This is prepared for the review process and will be made public later.}

\begin{figure}
    \centering
    \includegraphics[width=0.6\linewidth]{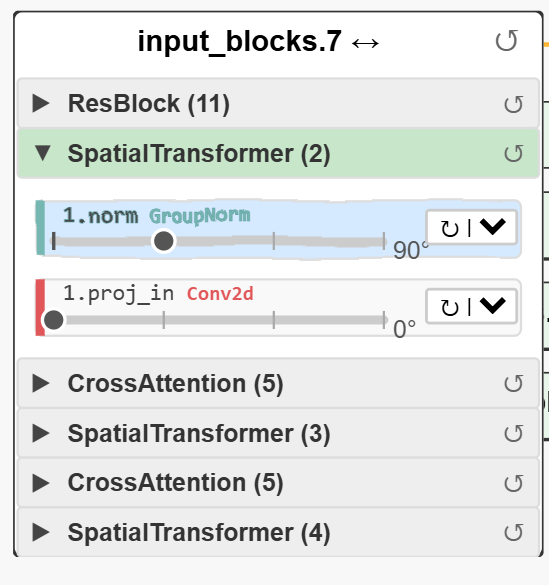}
    \caption{In each UNet block, users can drill down into one of the containers (higher-level components), pick a low-level layer (e.g. Normalization layer), specify a bending operation from a pulldown, and drag the slider to apply it. Multiple layers can be bent at once, and current bending operation can be reset on a block or container level.}
    \label{fig:interactive_bending_block}
\end{figure}

For each bendable layer, the interface provides clear selection and bending controls that display the layer name alongside a slider (\autoref{fig:interactive_bending_block}). The slider allows users to choose the type and amount of intervention, including rotation-based bending (0°, 90°, 180°, or 270°), additive Gaussian noise (with standard deviations of 0, 1, or 2), and multiplicative scaling (0 or ablation, 0.5, 1, 1.5, or 2). Layers are colour-coded by type according to a legend, making it easy to distinguish different kinds of modules at a glance. At the bottom of the UNet diagram (\autoref{fig:interactive_bending_web} - D) are the legend mapping colours to module types, controls for the range of denoising steps to impact, and a brief summary of currently applied operations with the option to copy these operations as a JSON string (c.f., \autoref{sec:json_format})

A preview panel displays both the unbent (default) output and the bent output based on the current configuration, while an output history strip shows thumbnails of past generations; selecting a thumbnail restores the bending settings that produced that result. Queue controls allow users to choose between a live mode, where prompts are automatically queued when sliders change, and an on-demand mode, where generation is triggered manually with a button (\autoref{fig:interactive_bending_web} - C).

Users can open the interface directly from a ComfyUI node (Interactive Bending WebUI) through a button or context menu. The node then reads the model structure, locates the UNet and visualizes it, and then applies the user-chosen bending parameters before passing the modified model downstream. Bending parameters are transmitted from the Web UI to the backend over HTTP, and the extension exposes REST endpoints for retrieving model structure, managing bending selections, clearing settings, and accessing image history. 

\section{Preliminary Study}
\label{sec:preliminary_study}
We present a systematic study of the outputs of bending different parts of a latent diffusion model and analyze it qualitatively and quantitatively.

\subsection{Study Setup}
We conducted a systematic study of bending interventions by enumerating combinations of bend location within the UNet, seeds, prompts, and denoising step ranges. As a bending operator, we use multiplication by a scalar, with values of 0.0 (i.e. ablation), 0.5, 1.5, and 2.0.

For each configuration, one aspect was fixed while varying the others resulting in: (1) Same Prompt/Seed, Different Layers (Section~\ref{sec:results_multiple_paths}), (2) Same Prompt, Different Seeds (Section~\ref{sec:results_multiple_seeds}), (3) Same Seed, Different Prompts (Section~\ref{sec:results_multiple_prompts}), (4) Same Seed/Prompt/Layer, Different Timesteps (Section~\ref{sec:results_multiple_timesteps}). In all, the unbent (default) run executed first to serve as a baseline. All bent outputs were recorded along with their associated parameters, generated image, and their final latent representation (i.e., ones passed to the decoder). This setup enables the interactive WebUI bending interface to be driven from cached experimental results for quick feedback and supports subsequent quantitative analysis of the outcomes.

To quantify the effects of bending, we measured deviation from the unbent baseline using the cosine distance between the latents of the results (i.e. processed latents produced through the denoising process). Metrics were computed post hoc for all experimental conditions and associated with each generated sample.


Results were analyzed by grouping bent outputs according to structural and procedural factors such as UNet region, bend type, and denoising step range, and by summarizing metric distributions within each group using standard descriptive statistics. This framework was designed to support large-scale, automated sweeps of bending interventions, enable quantitative comparison across diverse manipulation regimes, and facilitate reproducible analysis of how localized interventions propagate through the diffusion process.

\subsection{Qualitative Results}
In the following sections, we present results from bending experiments across multiple configurations. All experiments use a variant of Stable Diffusion v1.5 -specifically, RealisticVisionV51\_v51VAE—which has been fine-tuned for realistic image generation, with a particular emphasis on faces~\footnote{\url{https://huggingface.co/lllyasviel/fav_models/blob/main/fav/realisticVisionV51_v51VAE.safetensors}}. We deliberately chose a face-capable model because humans are especially sensitive to subtle variations in facial features. This makes faces a useful domain for illustrating how bending can introduce visual differences and experimental aesthetics that may be difficult to achieve through prompting alone. Furthermore, all denoising experiments were conducted using seed 42 (unless otherwise specified), with 20 denoising steps, a Classifier-Free Guidance (CFG) scale of 7, the DPMPP\_2m sampler, and the Karras scheduler. 

The prompt "Analog style portrait of a person" was deliberately constructed to separate style, composition, and subject. The phrase "analog style" specifies a broad stylistic regime, "portrait" constrains the image format and composition, and "of a person" defines the semantic subject. This decomposition allows us to examine how bending affects each facet (or does not). An analog style was selected specifically for its aesthetic flexibility and tolerance for variation.

\subsubsection{Same Prompt/Seed, Different Layers}
\label{sec:results_multiple_paths}
\begin{figure}[H]
\centering

\begin{tabular}{m{0.20\linewidth}|ccccc}
\hline
Layer path & multiply by 0.0 & multiply by 0.5 & default & multiply by 1.5 & multiply by 2.0 \\
\hline
time\_embed.2 & \includegraphics[width=0.145\linewidth]{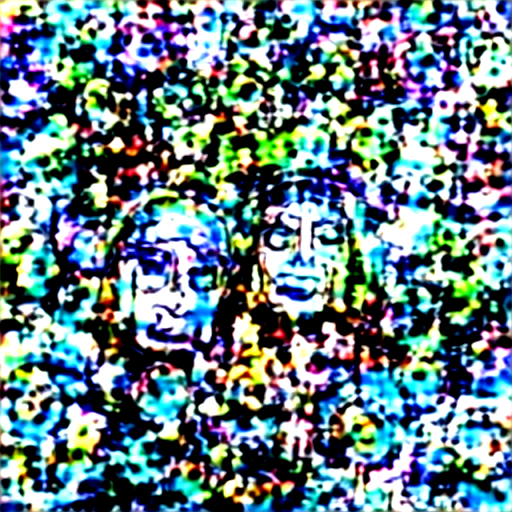} & \includegraphics[width=0.145\linewidth]{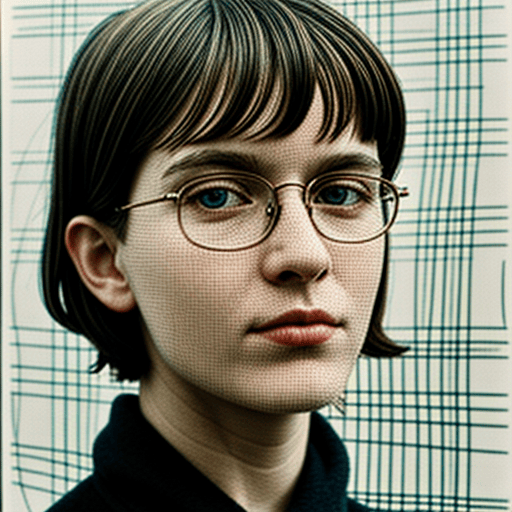} & \includegraphics[width=0.145\linewidth]{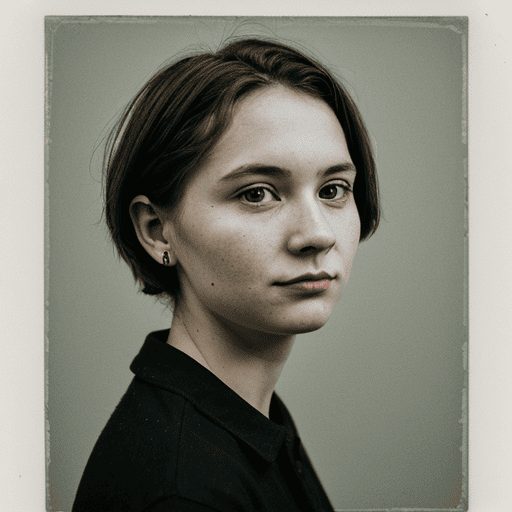} & \includegraphics[width=0.145\linewidth]{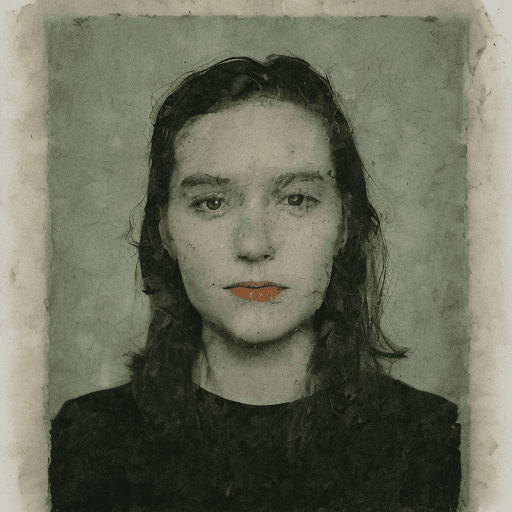} & \includegraphics[width=0.145\linewidth]{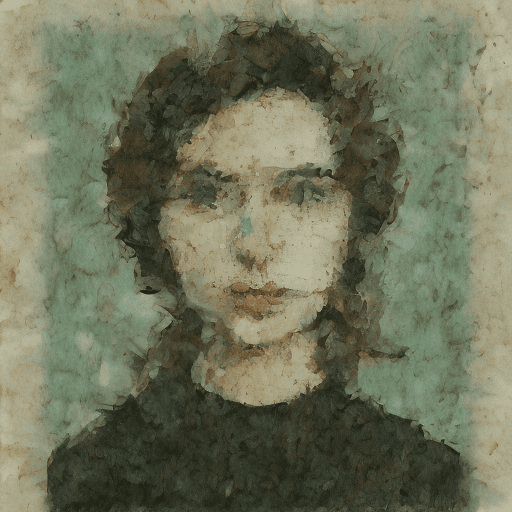} \\
input\_blocks.1.0 .in\_layers.0 & \includegraphics[width=0.145\linewidth]{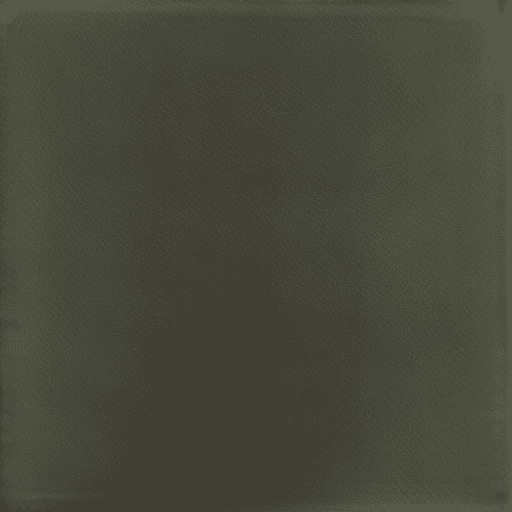} & \includegraphics[width=0.145\linewidth]{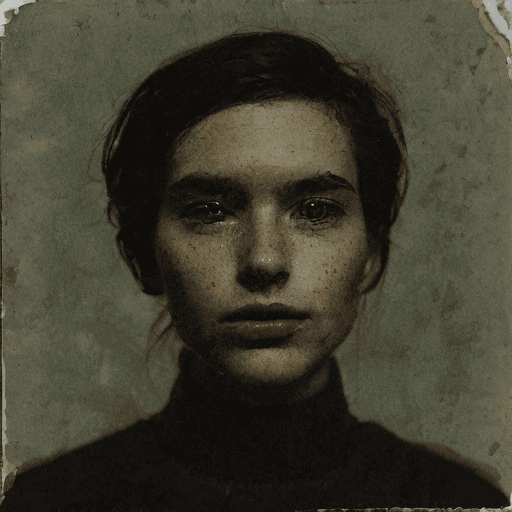} & \includegraphics[width=0.145\linewidth]{figs/multiple_paths/default_0.png} & \includegraphics[width=0.145\linewidth]{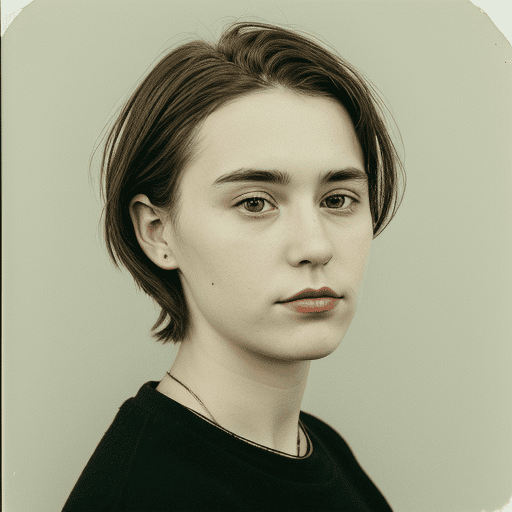} & \includegraphics[width=0.145\linewidth]{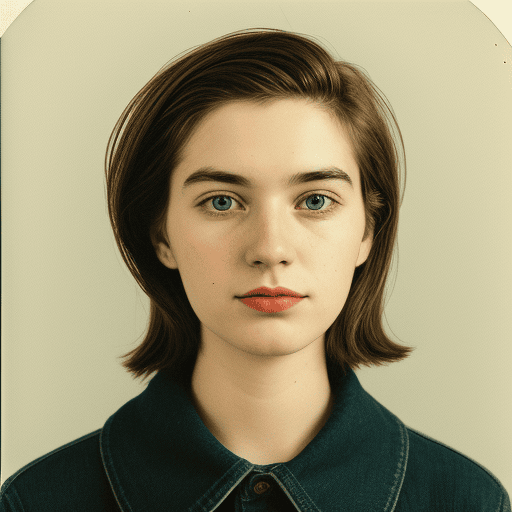} \\
input\_blocks.1.0 .out\_layers.2 & \includegraphics[width=0.145\linewidth]{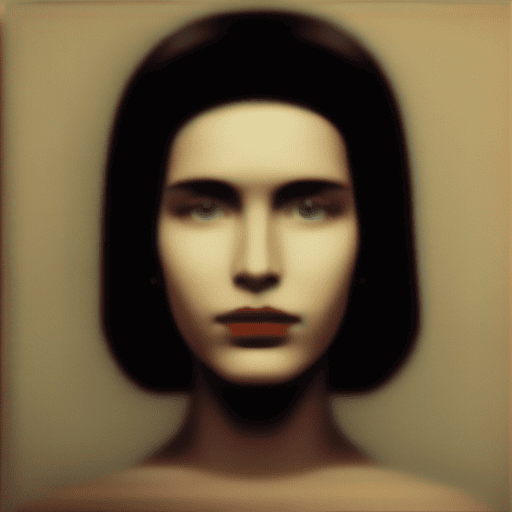} & \includegraphics[width=0.145\linewidth]{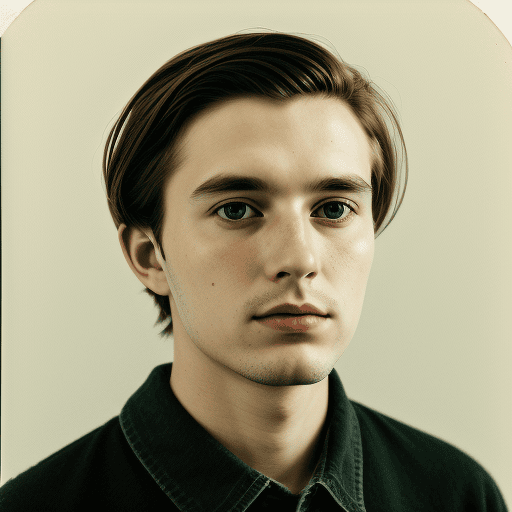} & \includegraphics[width=0.145\linewidth]{figs/multiple_paths/default_0.png} & \includegraphics[width=0.145\linewidth]{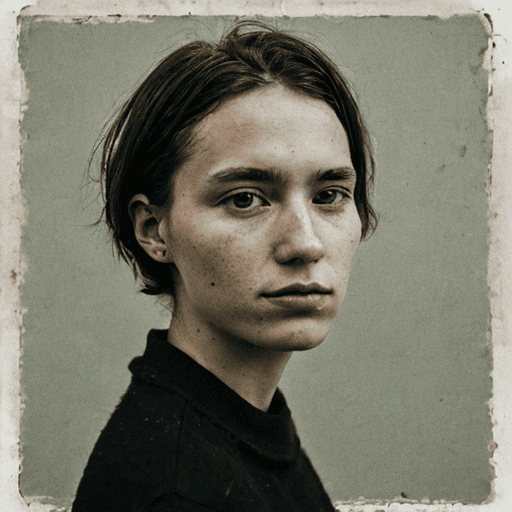} & \includegraphics[width=0.145\linewidth]{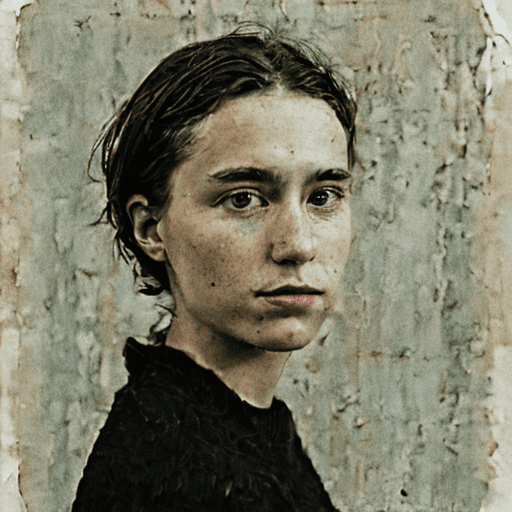} \\
input\_blocks.2.0 .skip\_connection & \includegraphics[width=0.145\linewidth]{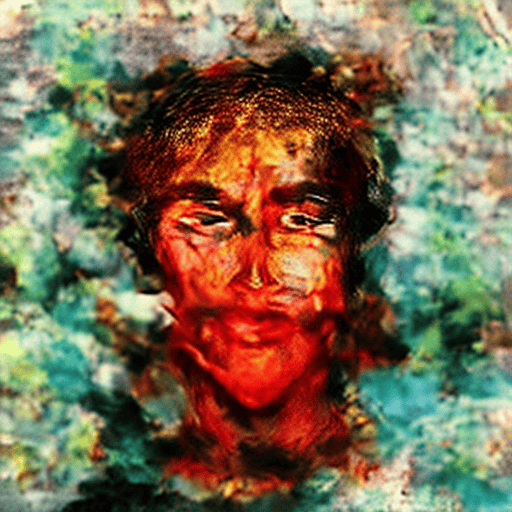} & \includegraphics[width=0.145\linewidth]{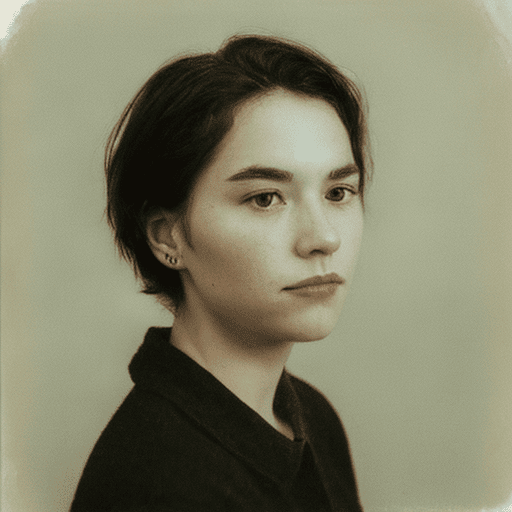} & \includegraphics[width=0.145\linewidth]{figs/multiple_paths/default_0.png} & \includegraphics[width=0.145\linewidth]{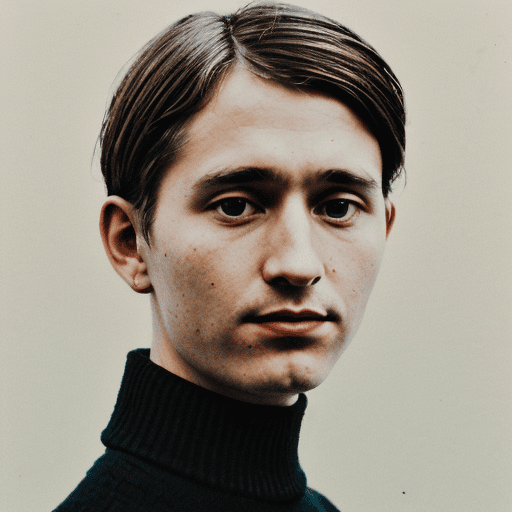} & \includegraphics[width=0.145\linewidth]{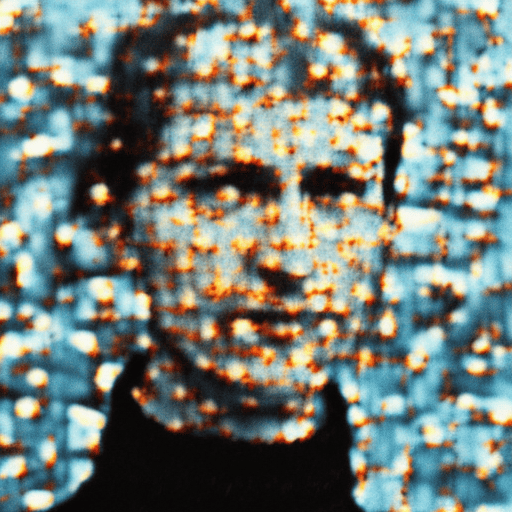} \\
input\_blocks.3.0.op & \includegraphics[width=0.145\linewidth]{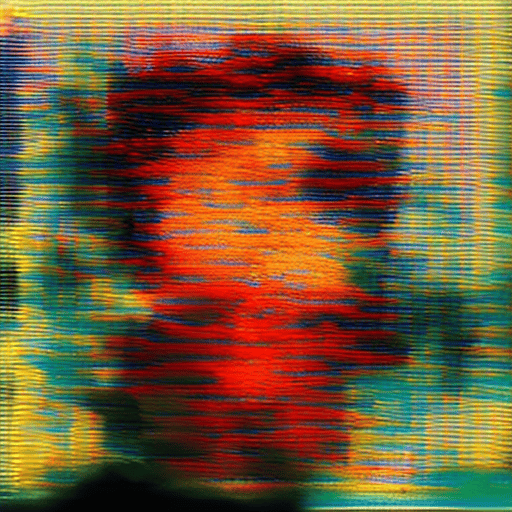} & \includegraphics[width=0.145\linewidth]{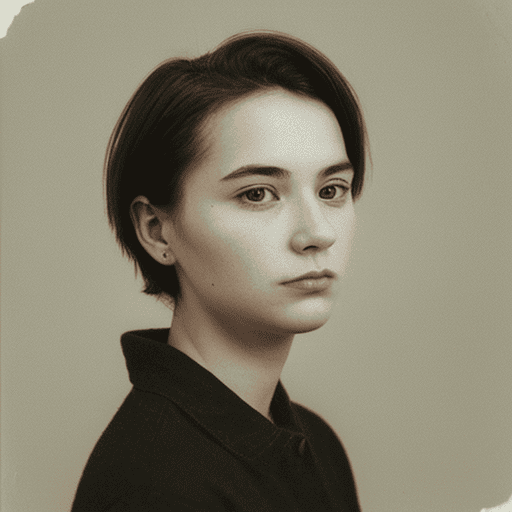} & \includegraphics[width=0.145\linewidth]{figs/multiple_paths/default_0.png} & \includegraphics[width=0.145\linewidth]{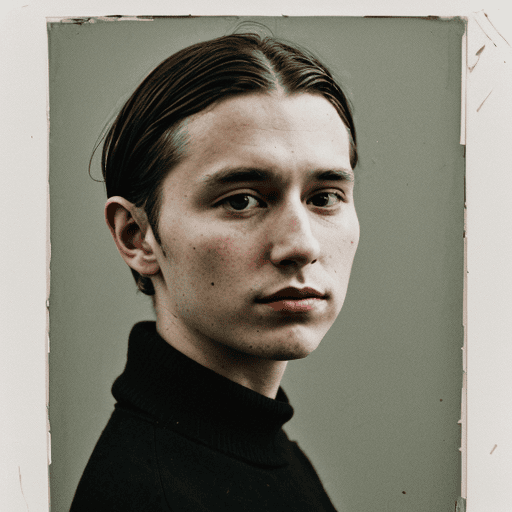} & \includegraphics[width=0.145\linewidth]{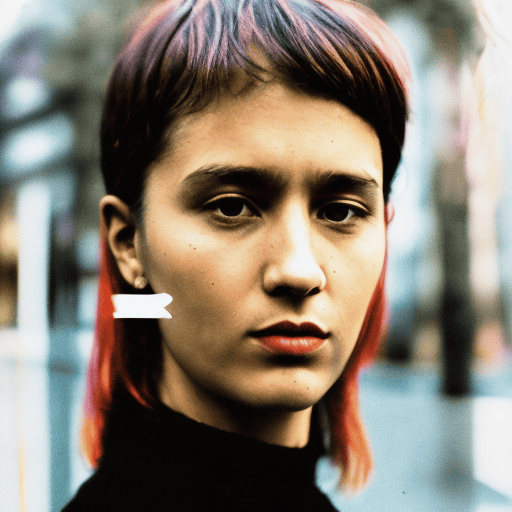} \\
input\_blocks.4.0 .skip\_connection & \includegraphics[width=0.145\linewidth]{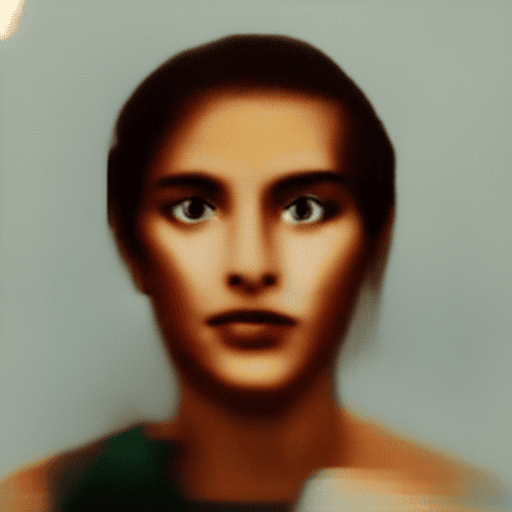} & \includegraphics[width=0.145\linewidth]{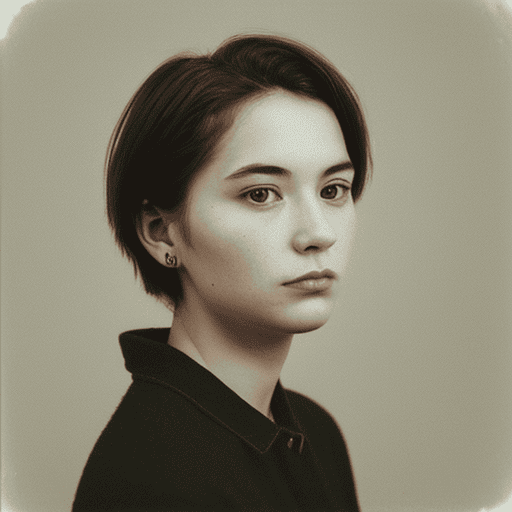} & \includegraphics[width=0.145\linewidth]{figs/multiple_paths/default_0.png} & \includegraphics[width=0.145\linewidth]{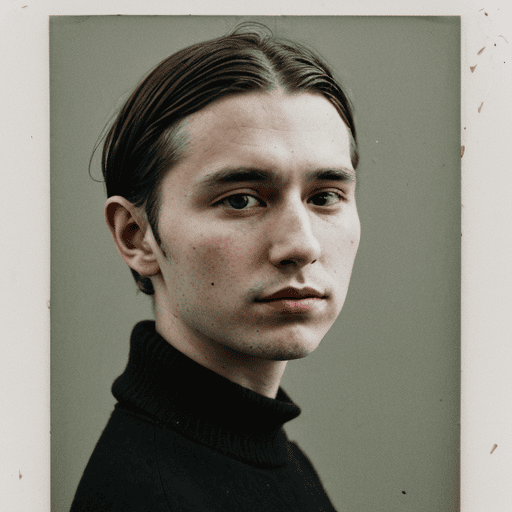} & \includegraphics[width=0.145\linewidth]{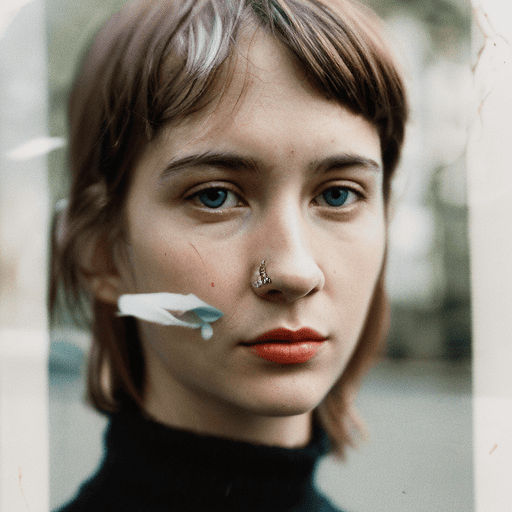} \\

\hline
\end{tabular}

\caption{One prompt; rows: bending layer path, columns: module\_args. Prompt: Analog style portrait of a person}
\label{fig:multiple_paths_1}
\end{figure}

\begin{figure}[H]
\centering

\begin{tabular}{p{0.28\linewidth}|ccccc}
\hline
Layer path & multiply by 0.0 & multiply by 0.5 & default & multiply by 1.5 & multiply by 2.0 \\
\hline
input\_blocks.5.1 .proj\_in & \includegraphics[width=0.145\linewidth]{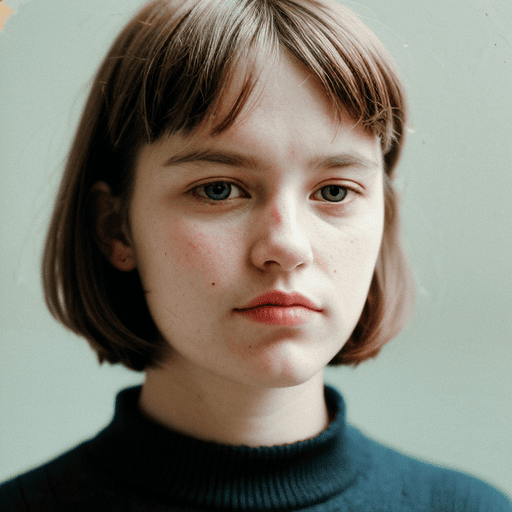} & \includegraphics[width=0.145\linewidth]{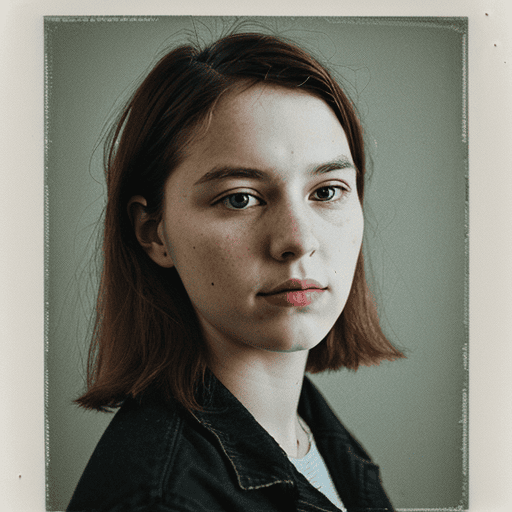} & \includegraphics[width=0.145\linewidth]{figs/multiple_paths/default_0.png} & \includegraphics[width=0.145\linewidth]{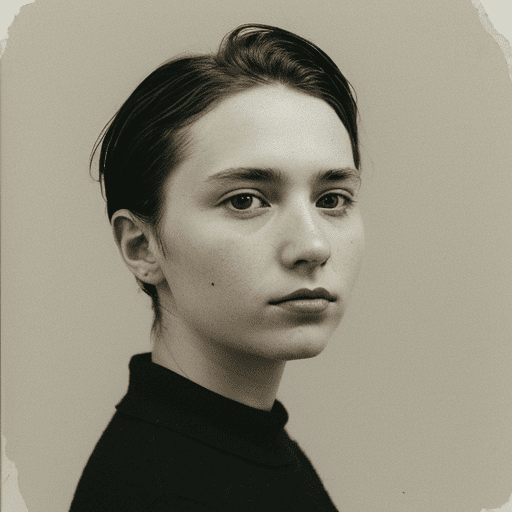} & \includegraphics[width=0.145\linewidth]{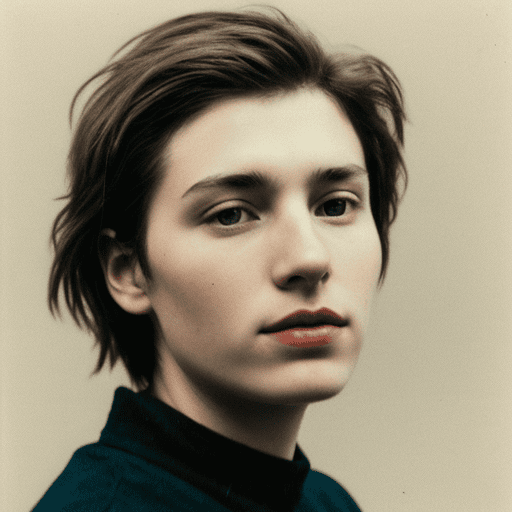} \\
input\_blocks.7.1 .proj\_in & \includegraphics[width=0.145\linewidth]{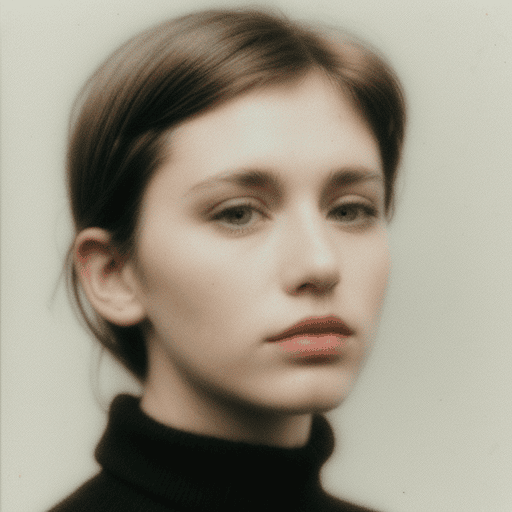} & \includegraphics[width=0.145\linewidth]{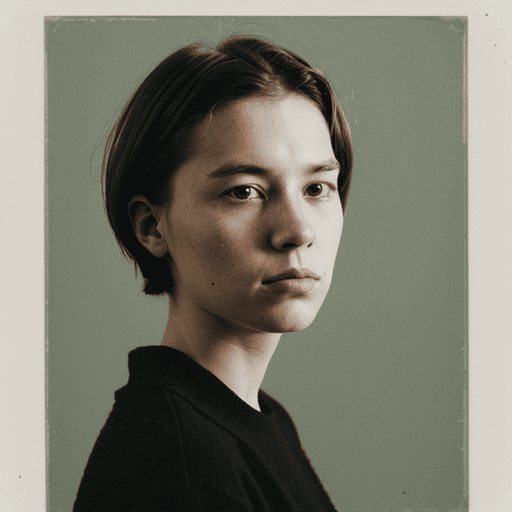} & \includegraphics[width=0.145\linewidth]{figs/multiple_paths/default_0.png} & \includegraphics[width=0.145\linewidth]{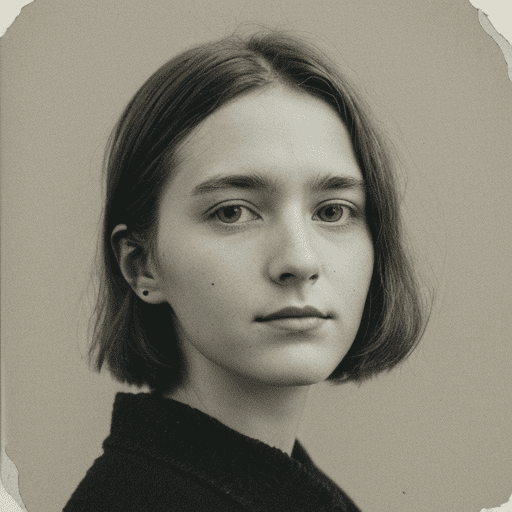} & \includegraphics[width=0.145\linewidth]{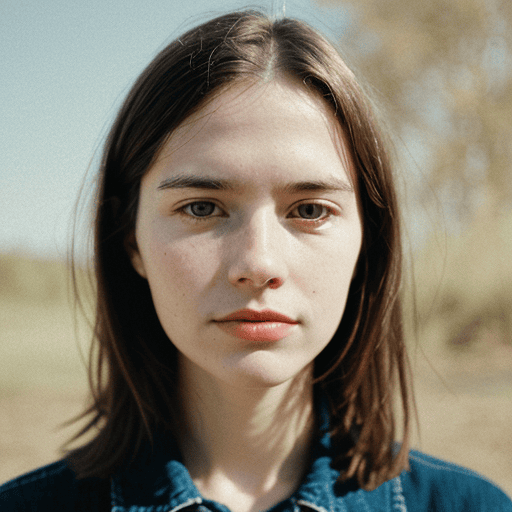} \\
input\_blocks.8.1 .transformer \_blocks.0.norm1 & \includegraphics[width=0.145\linewidth]{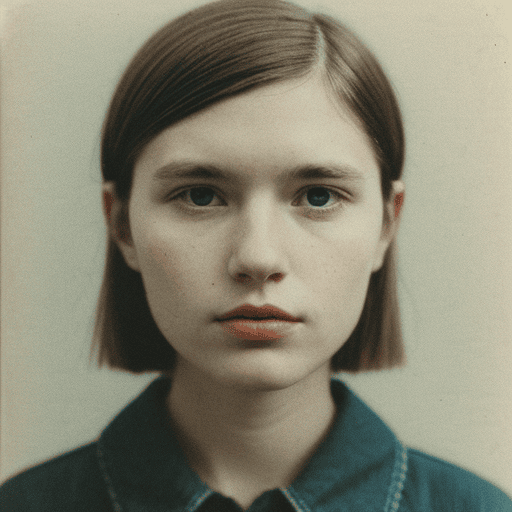} & \includegraphics[width=0.145\linewidth]{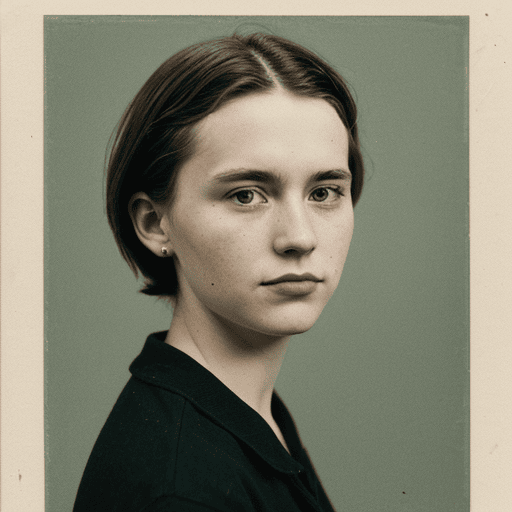} & \includegraphics[width=0.145\linewidth]{figs/multiple_paths/default_0.png} & \includegraphics[width=0.145\linewidth]{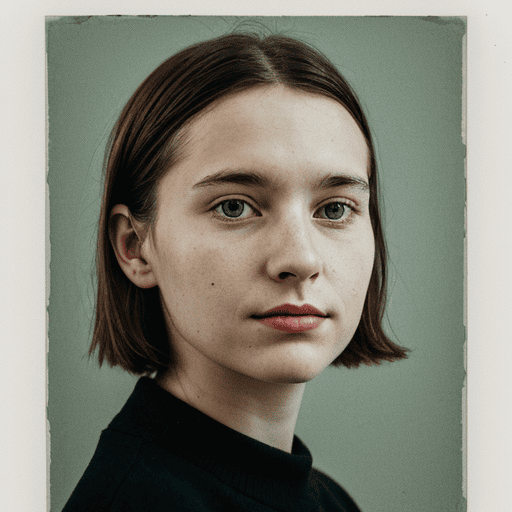} & \includegraphics[width=0.145\linewidth]{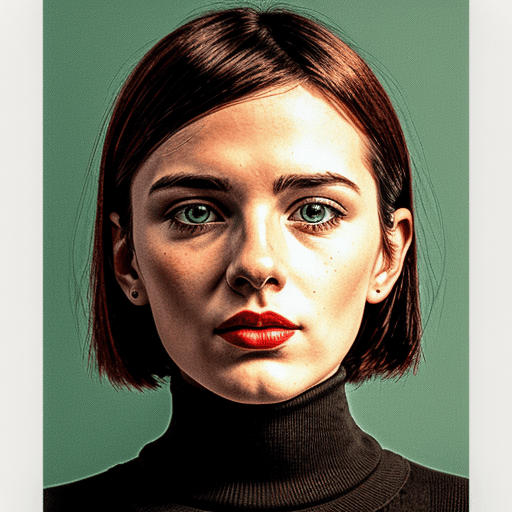} \\
middle\_block.1.transformer \_blocks.0.attn2.to\_out.0 & \includegraphics[width=0.145\linewidth]{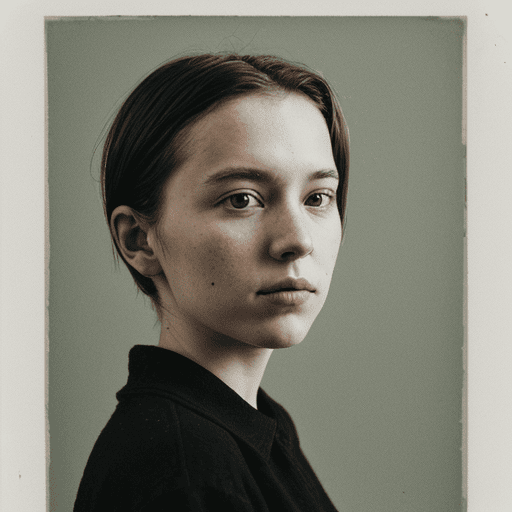} & \includegraphics[width=0.145\linewidth]{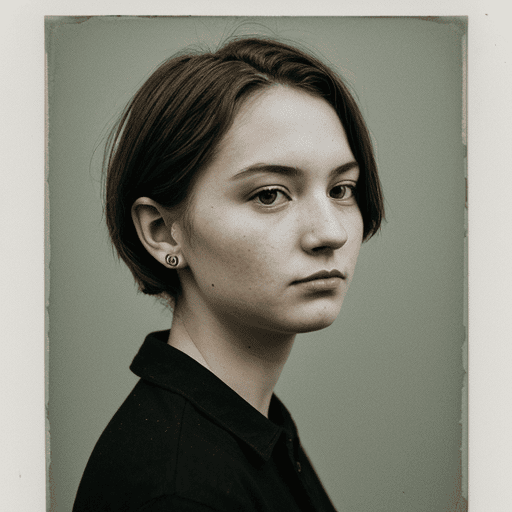} & \includegraphics[width=0.145\linewidth]{figs/multiple_paths/default_0.png} & \includegraphics[width=0.145\linewidth]{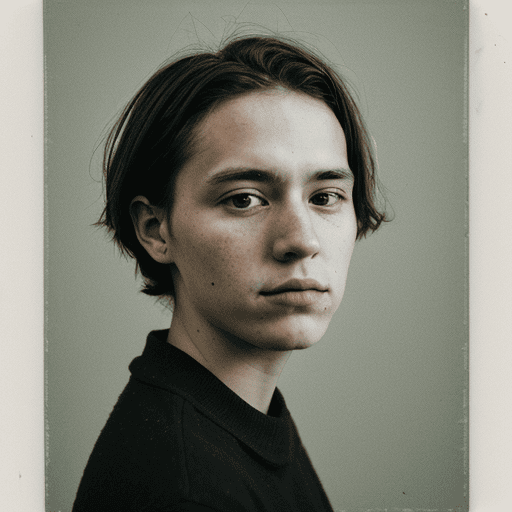} & \includegraphics[width=0.145\linewidth]{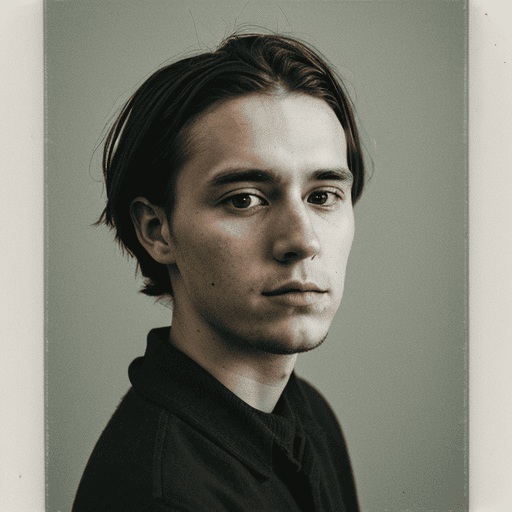} \\
output\_blocks.3.1.norm & \includegraphics[width=0.145\linewidth]{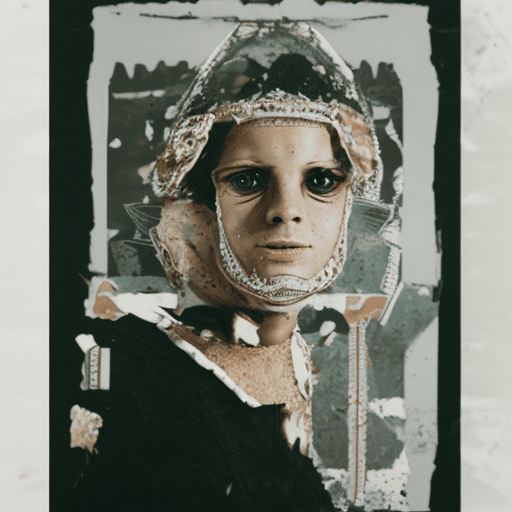} & \includegraphics[width=0.145\linewidth]{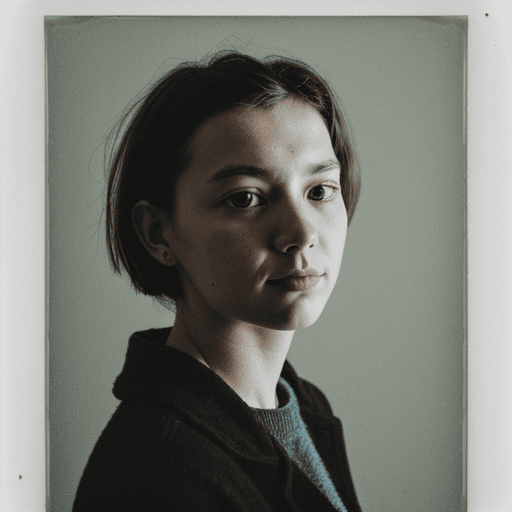} & \includegraphics[width=0.145\linewidth]{figs/multiple_paths/default_0.png} & \includegraphics[width=0.145\linewidth]{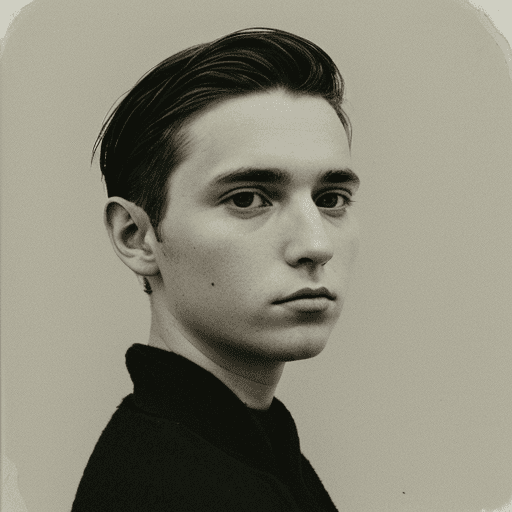} & \includegraphics[width=0.145\linewidth]{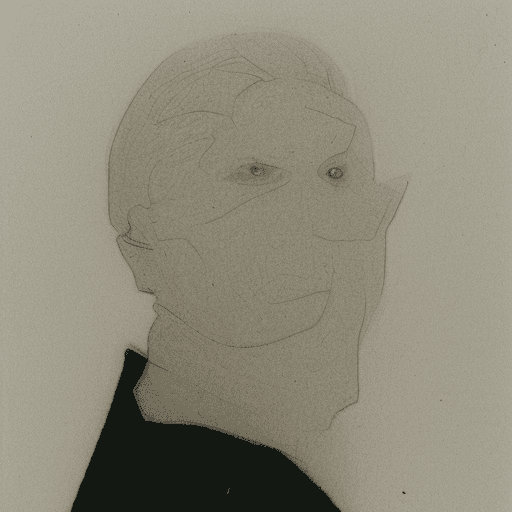} \\
output\_blocks.4.1 .transformer \_blocks.0.norm1 & \includegraphics[width=0.145\linewidth]{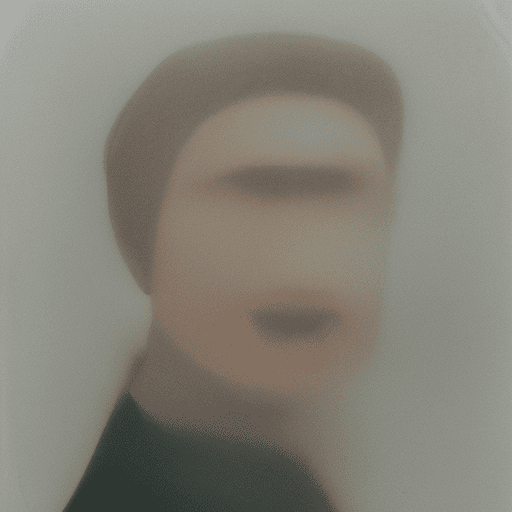} & \includegraphics[width=0.145\linewidth]{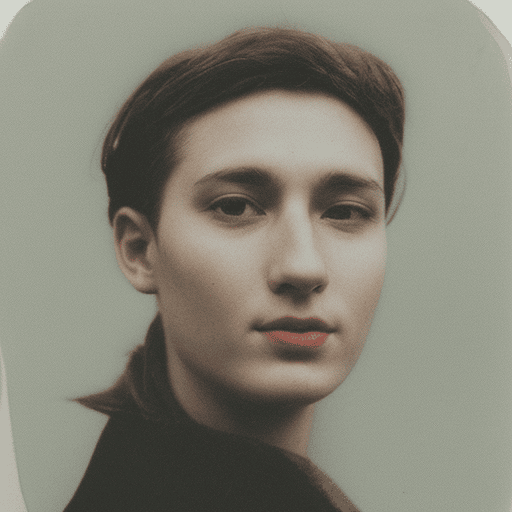} & \includegraphics[width=0.145\linewidth]{figs/multiple_paths/default_0.png} & \includegraphics[width=0.145\linewidth]{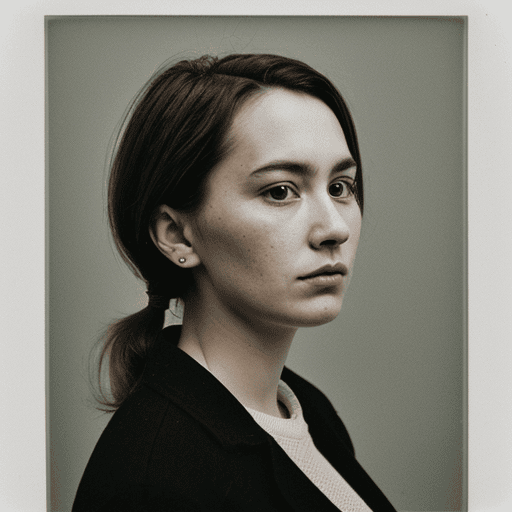} & \includegraphics[width=0.145\linewidth]{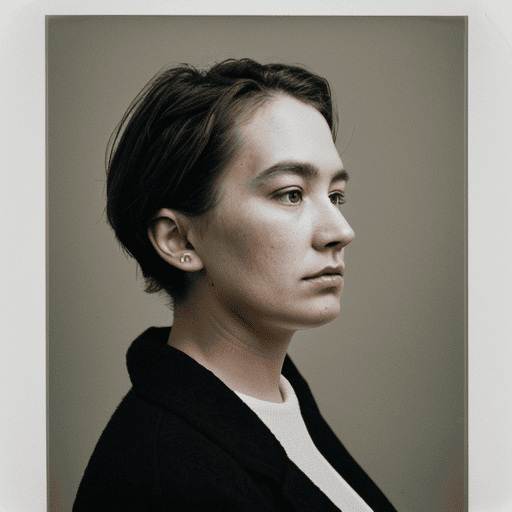} \\
output\_blocks.5.1 .transformer \_blocks.0.norm1 & \includegraphics[width=0.145\linewidth]{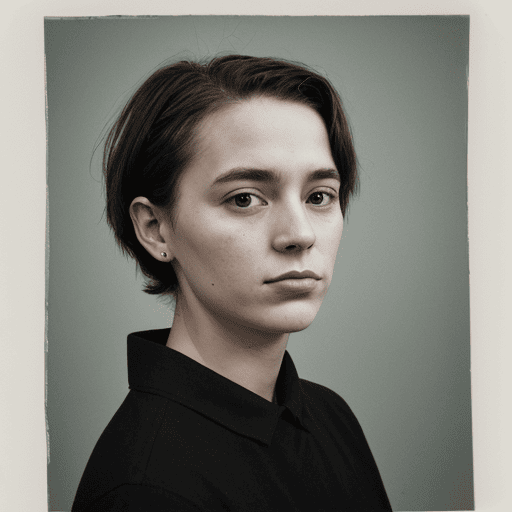} & \includegraphics[width=0.145\linewidth]{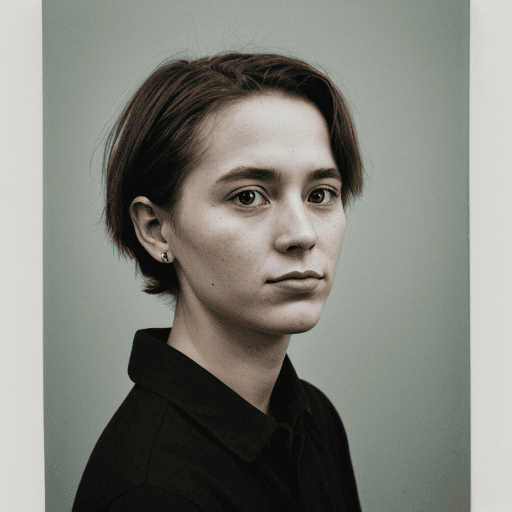} & \includegraphics[width=0.145\linewidth]{figs/multiple_paths/default_0.png} & \includegraphics[width=0.145\linewidth]{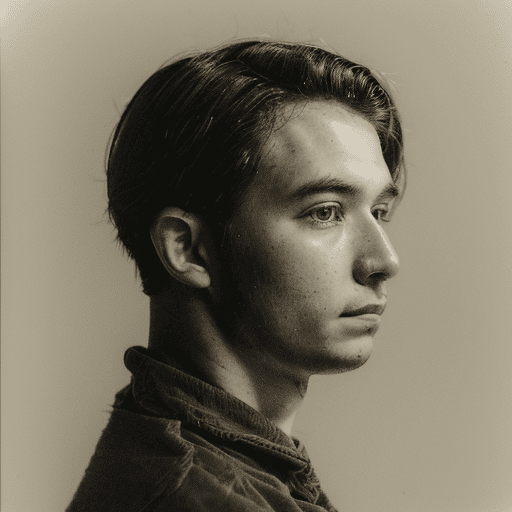} & \includegraphics[width=0.145\linewidth]{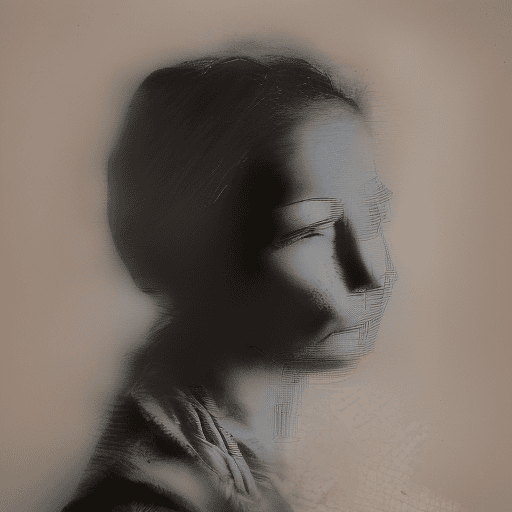} \\
output\_blocks.7.1 .transformer \_blocks.0.norm1 & \includegraphics[width=0.145\linewidth]{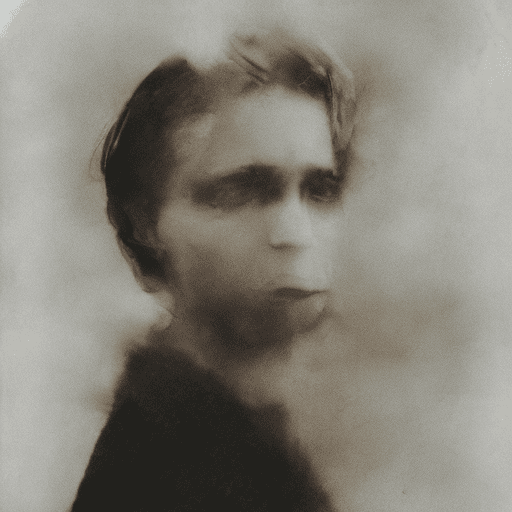} & \includegraphics[width=0.145\linewidth]{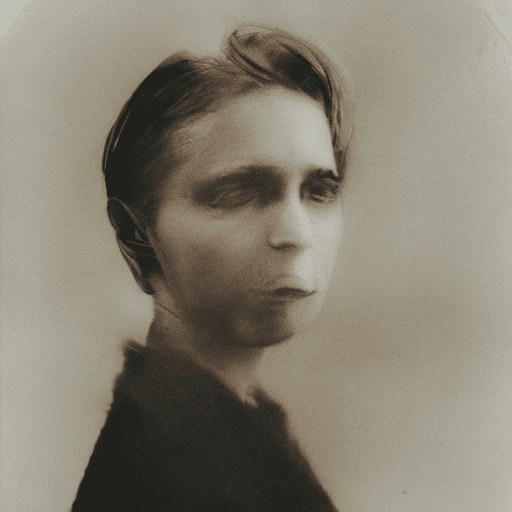} & \includegraphics[width=0.145\linewidth]{figs/multiple_paths/default_0.png} & \includegraphics[width=0.145\linewidth]{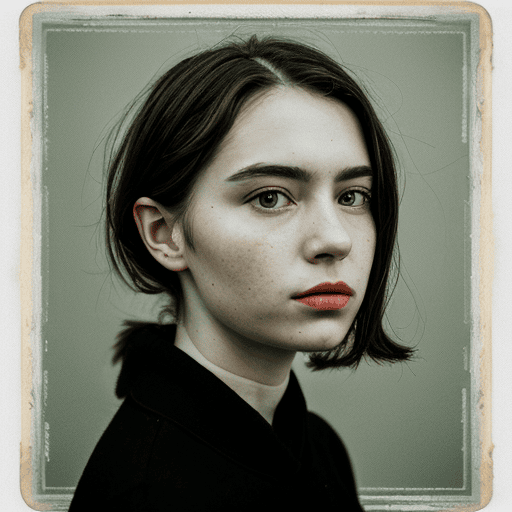} & \includegraphics[width=0.145\linewidth]{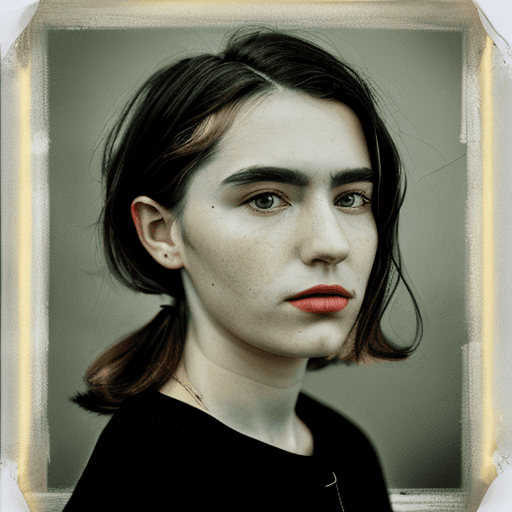} \\
\hline

\end{tabular}

\caption{One prompt; rows: bending layer path, columns: module\_args. Prompt: "Analog style portrait of a person"}
\label{fig:multiple_paths_2}
\end{figure}

Using the same prompt and seed, we bend different layers in the model using multiplication values of 0 (ablation), 0.5, 1.5 and 2.0. Figures ~\ref{fig:multiple_paths_1} and \ref{fig:multiple_paths_2} follow.

\subsubsection{Same Prompt, Different Seeds}
\label{sec:results_multiple_seeds}
\begin{figure}[H]
\centering

\medskip

\begin{tabular}{c|cccccc}
\hline
Multiply by / Seed & 42 & 0 & 123 & 456 & 789 & 786 \\
\hline

 0.0
& \includegraphics[width=0.130\linewidth]{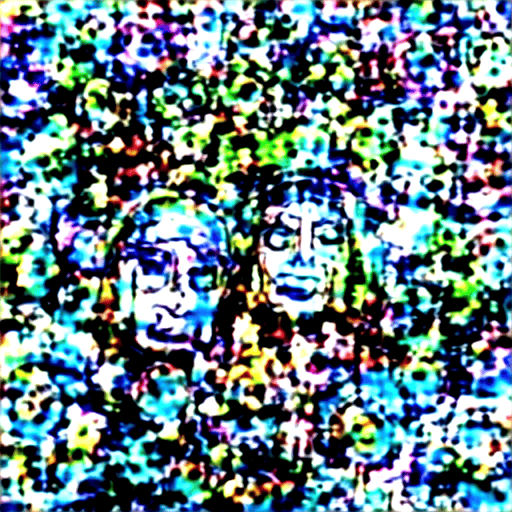}
& \includegraphics[width=0.130\linewidth]{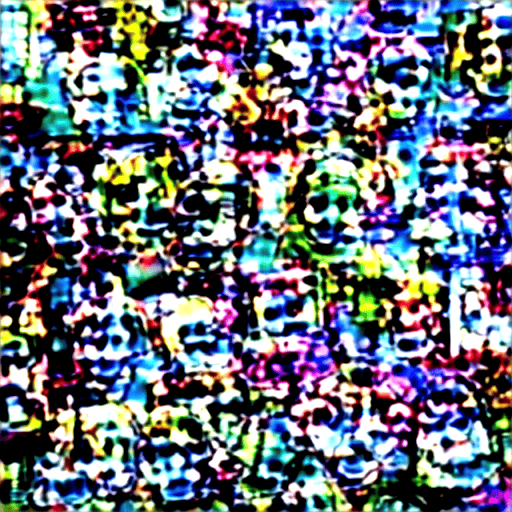}
& \includegraphics[width=0.130\linewidth]{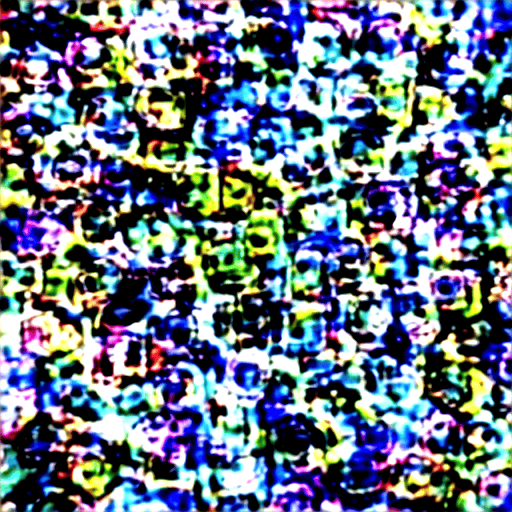}
& \includegraphics[width=0.130\linewidth]{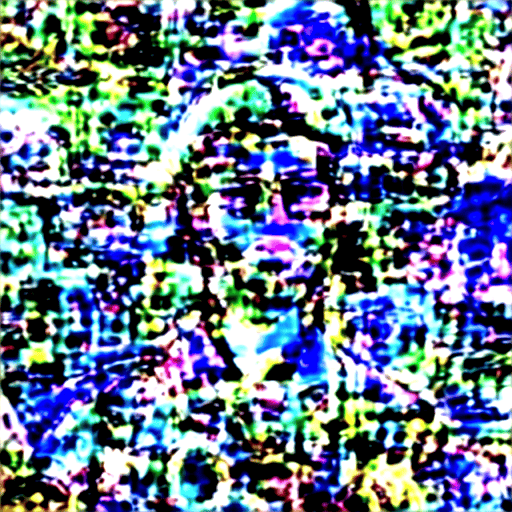}
& \includegraphics[width=0.130\linewidth]{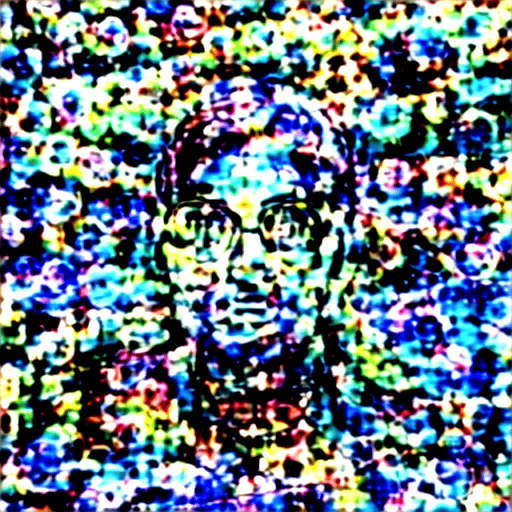}
& \includegraphics[width=0.130\linewidth]{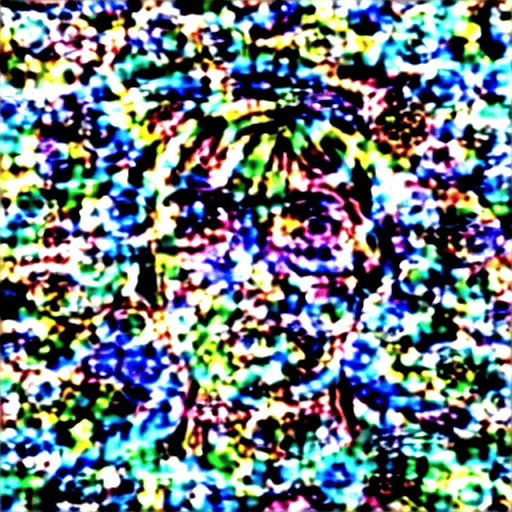}
\\

0.5
& \includegraphics[width=0.130\linewidth]{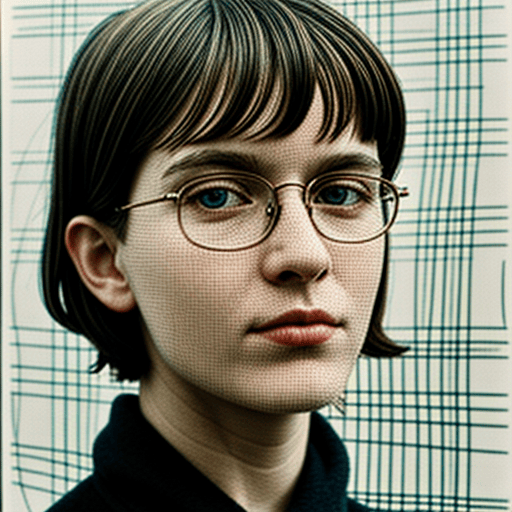}
& \includegraphics[width=0.130\linewidth]{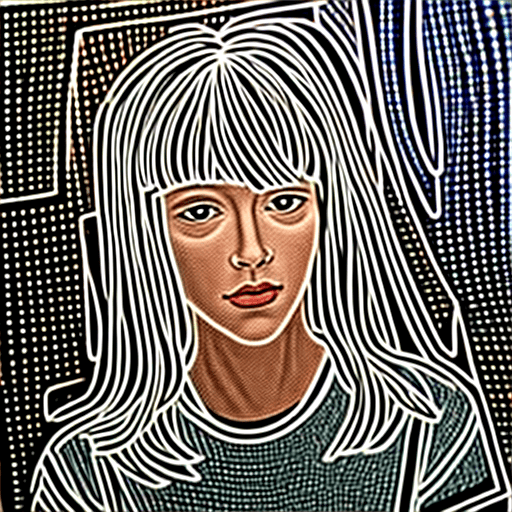}
& \includegraphics[width=0.130\linewidth]{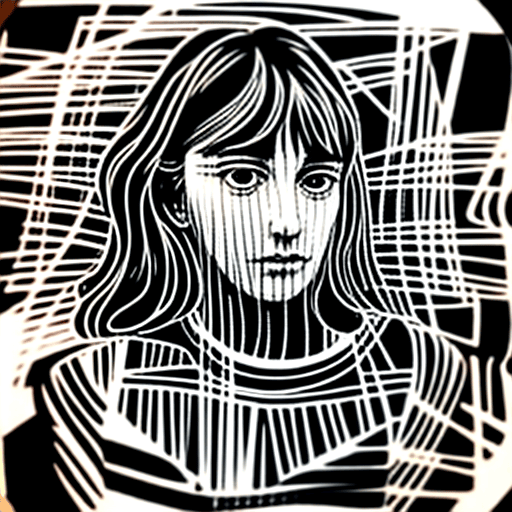}
& \includegraphics[width=0.130\linewidth]{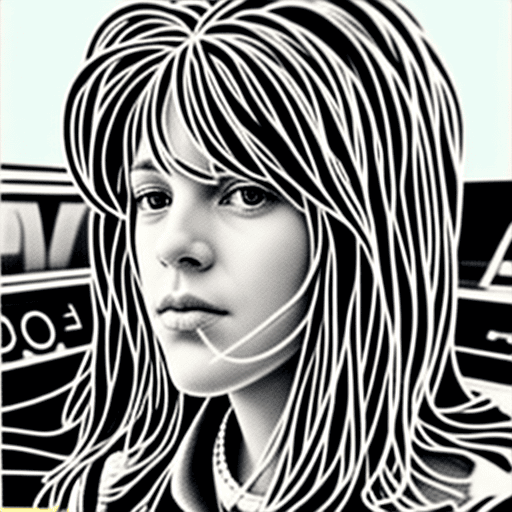}
& \includegraphics[width=0.130\linewidth]{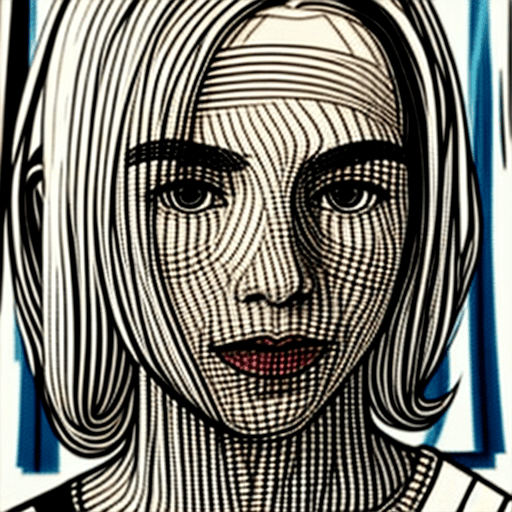}
& \includegraphics[width=0.130\linewidth]{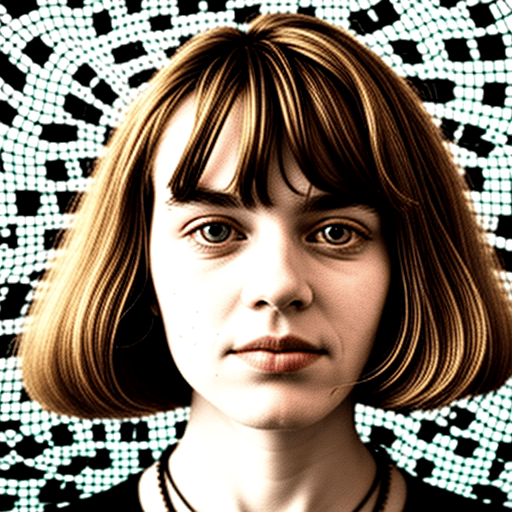}
\\

1.0
& \includegraphics[width=0.130\linewidth]{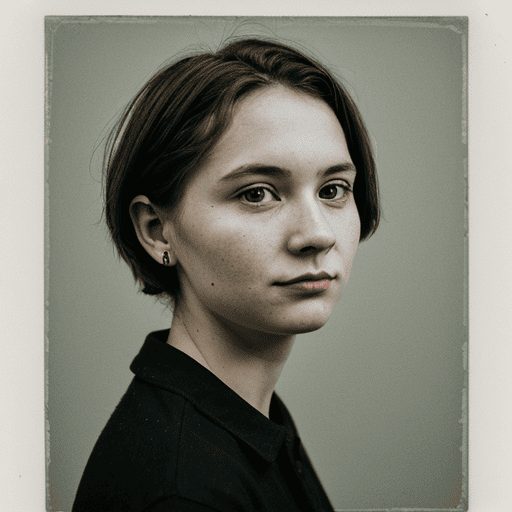}
& \includegraphics[width=0.130\linewidth]{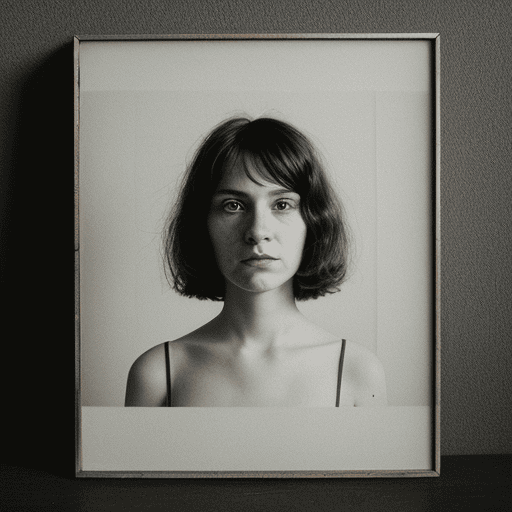}
& \includegraphics[width=0.130\linewidth]{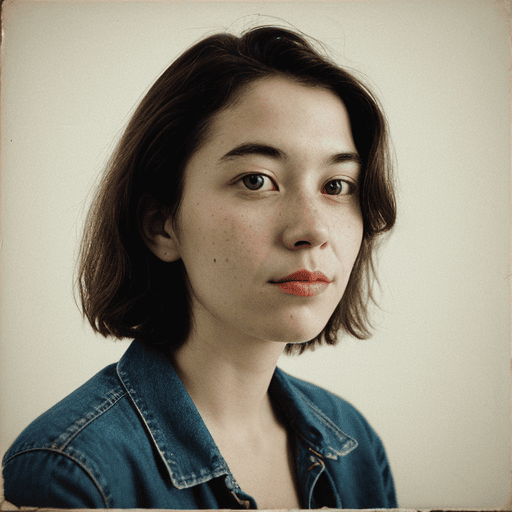}
& \includegraphics[width=0.130\linewidth]{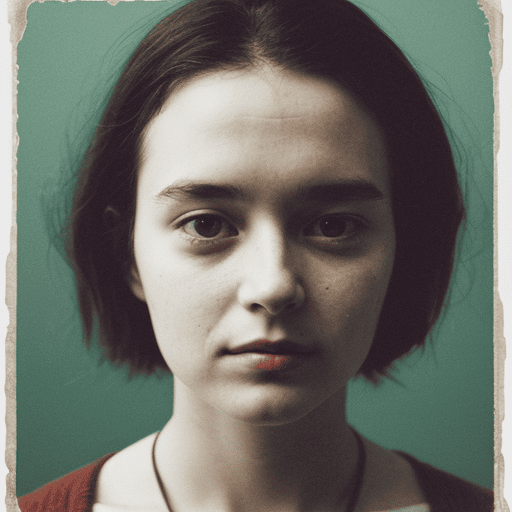}
& \includegraphics[width=0.130\linewidth]{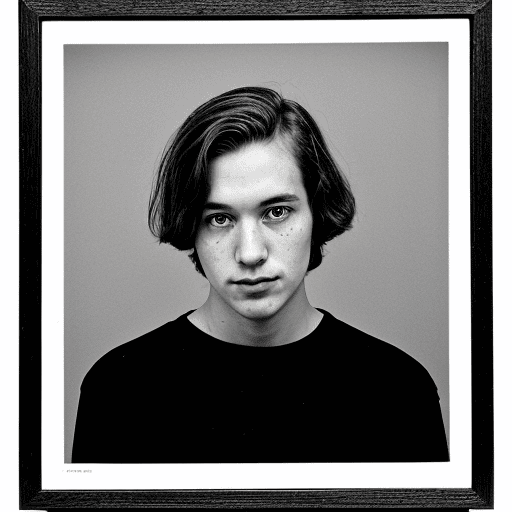}
& \includegraphics[width=0.130\linewidth]{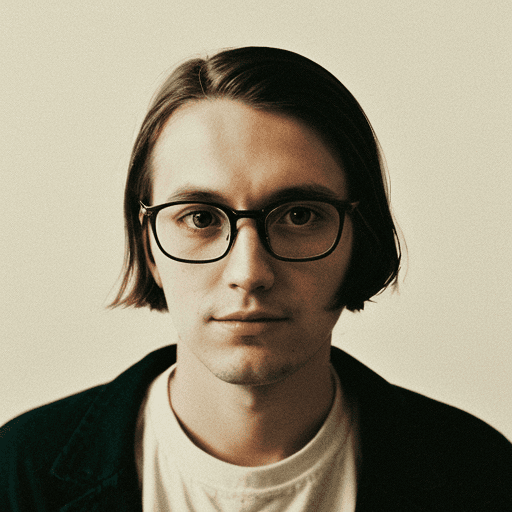}
\\

1.5
& \includegraphics[width=0.130\linewidth]{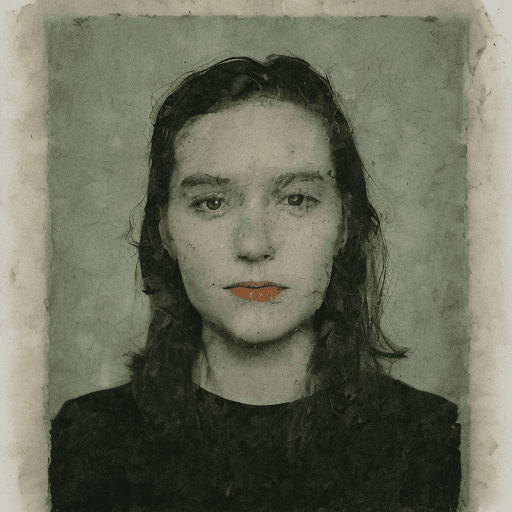}
& \includegraphics[width=0.130\linewidth]{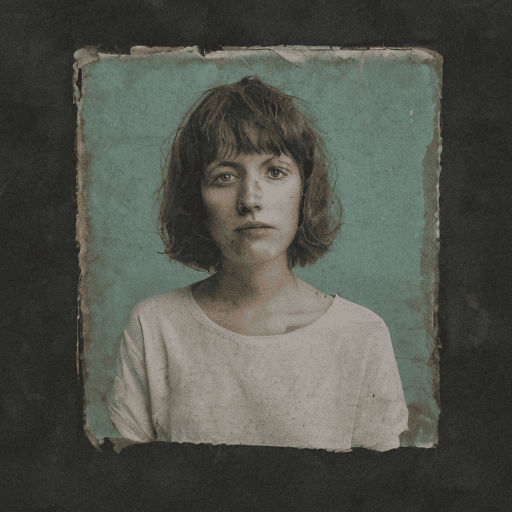}
& \includegraphics[width=0.130\linewidth]{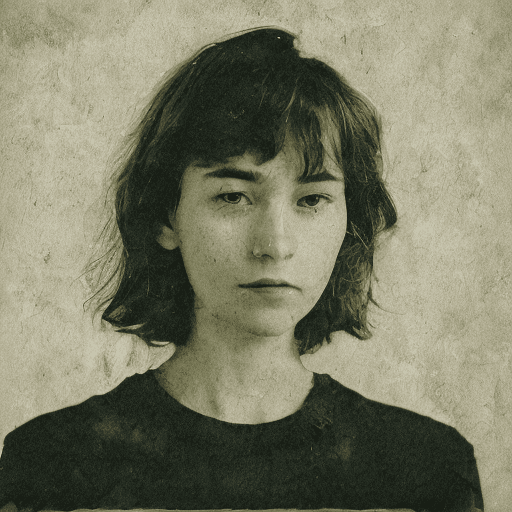}
& \includegraphics[width=0.130\linewidth]{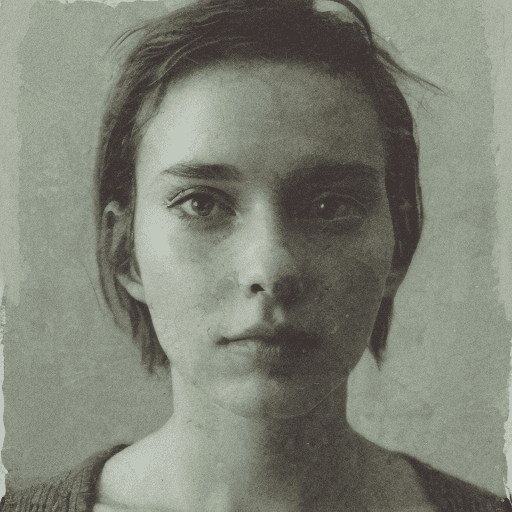}
& \includegraphics[width=0.130\linewidth]{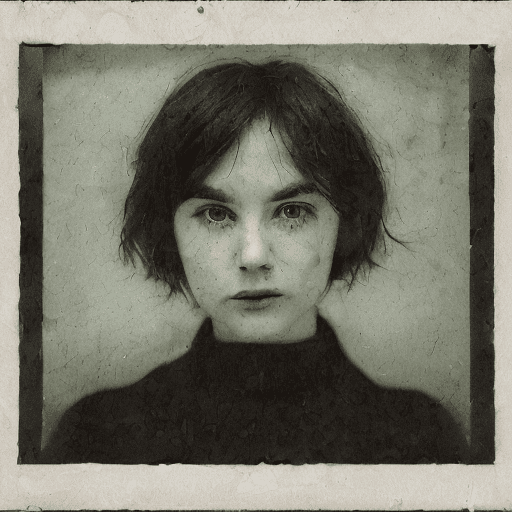}
& \includegraphics[width=0.130\linewidth]{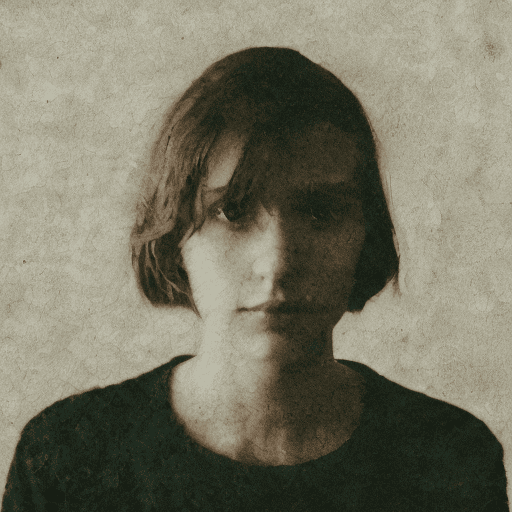}
\\

2.0
& \includegraphics[width=0.130\linewidth]{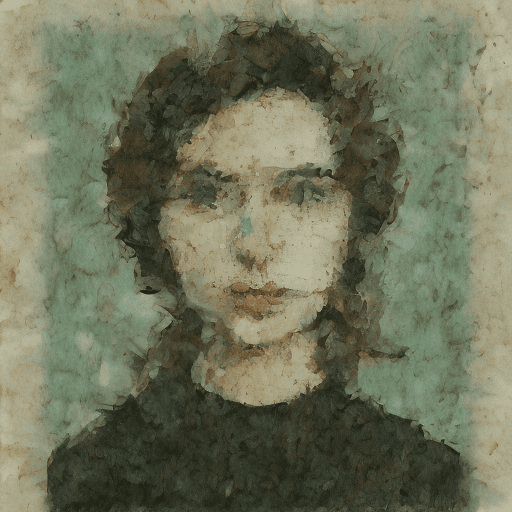}
& \includegraphics[width=0.130\linewidth]{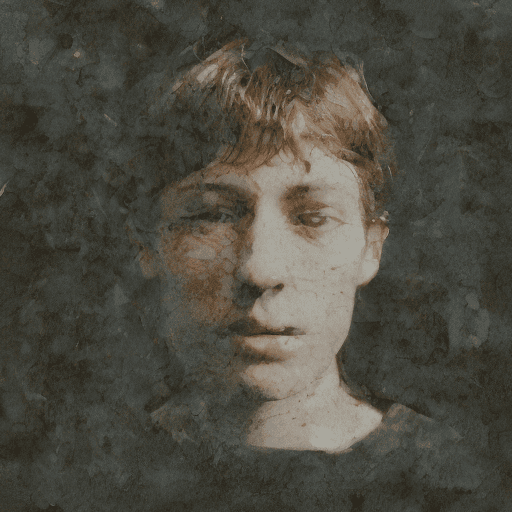}
& \includegraphics[width=0.130\linewidth]{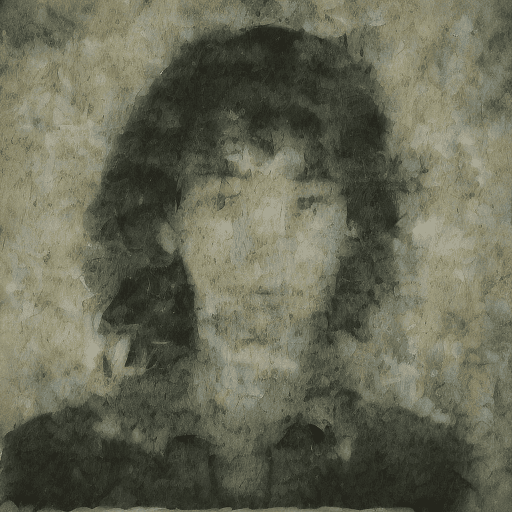}
& \includegraphics[width=0.130\linewidth]{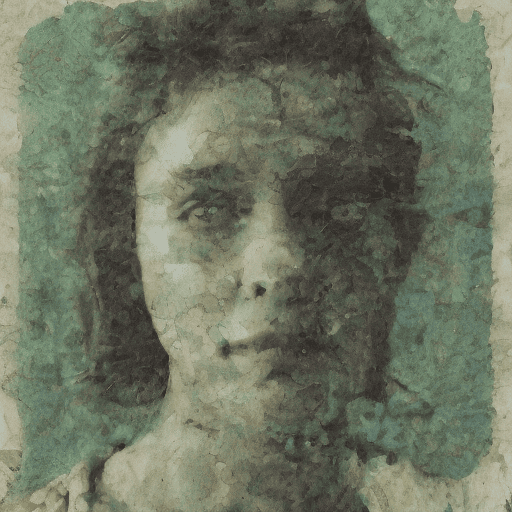}
& \includegraphics[width=0.130\linewidth]{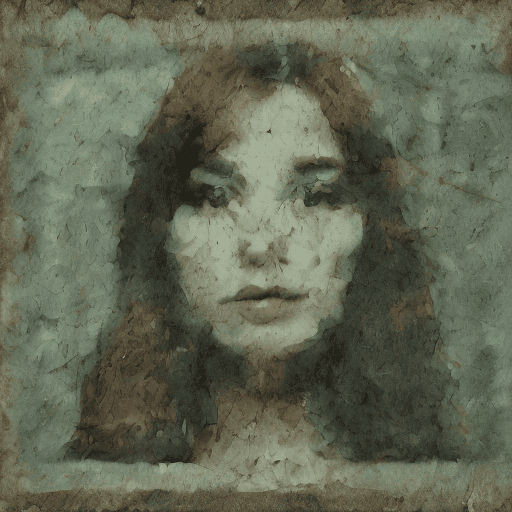}
& \includegraphics[width=0.130\linewidth]{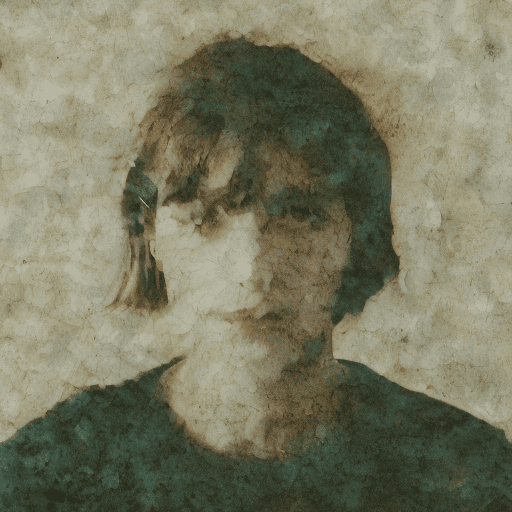}
\\
\hline
\end{tabular}

\caption{Bending \textit{time\_embed.2}; Same prompt, different seeds.}
\label{fig:multiple_seeds_1}

\end{figure}

\begin{figure}[h]
\centering

\medskip

\begin{tabular}{c|cccccc}
\hline
Multiply by / Seed & 42 & 0 & 123 & 456 & 789 & 786 \\
\hline

0.0
& \includegraphics[width=0.130\linewidth]{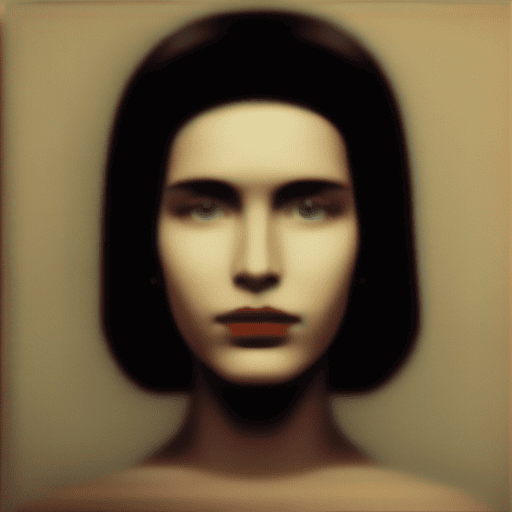}
& \includegraphics[width=0.130\linewidth]{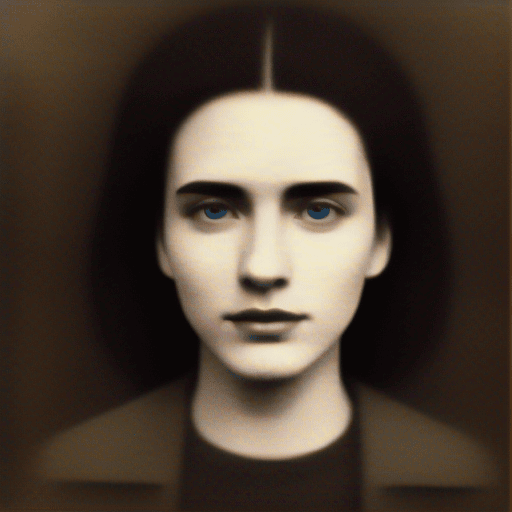}
& \includegraphics[width=0.130\linewidth]{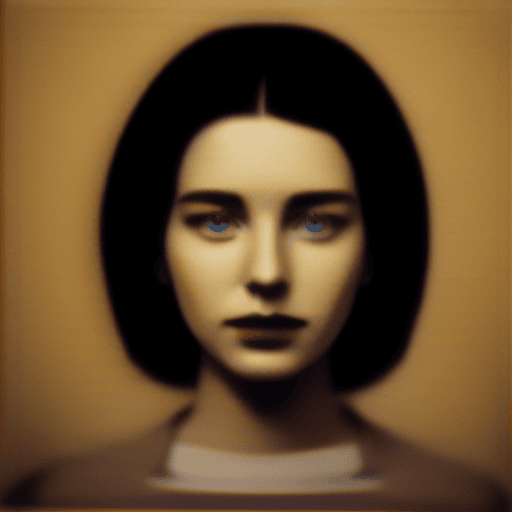}
& \includegraphics[width=0.130\linewidth]{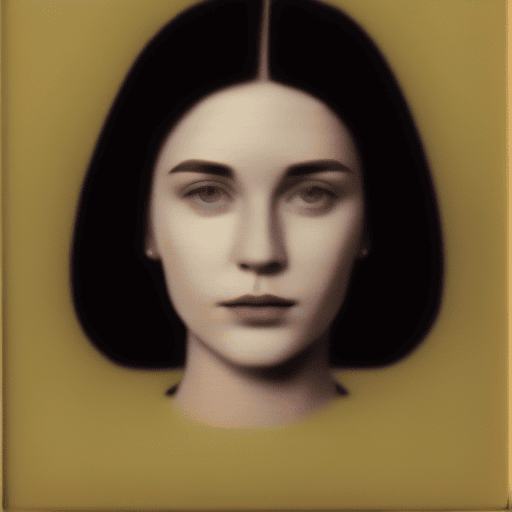}
& \includegraphics[width=0.130\linewidth]{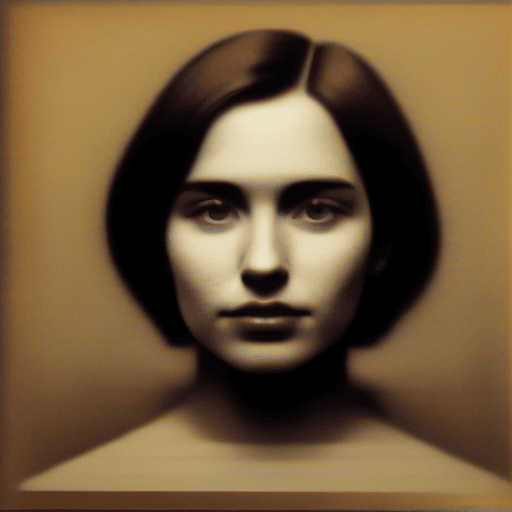}
& \includegraphics[width=0.130\linewidth]{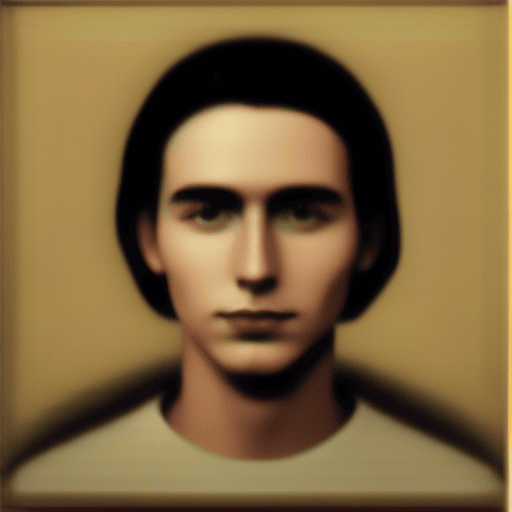}
\\

0.5
& \includegraphics[width=0.130\linewidth]{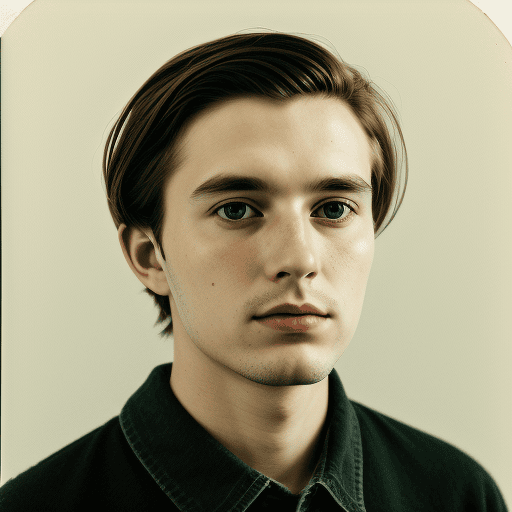}
& \includegraphics[width=0.130\linewidth]{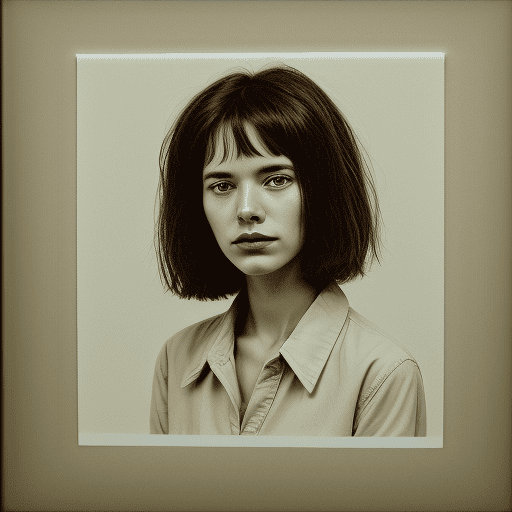}
& \includegraphics[width=0.130\linewidth]{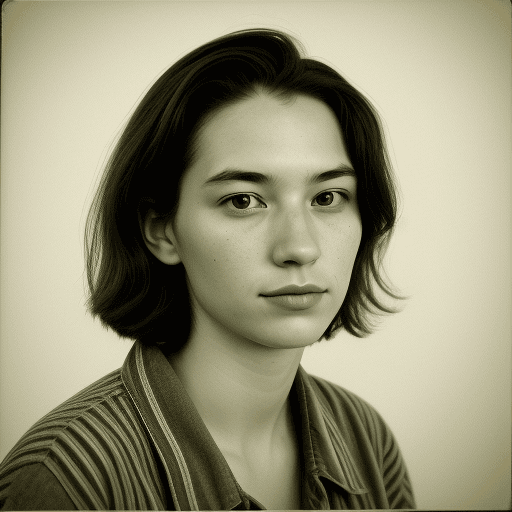}
& \includegraphics[width=0.130\linewidth]{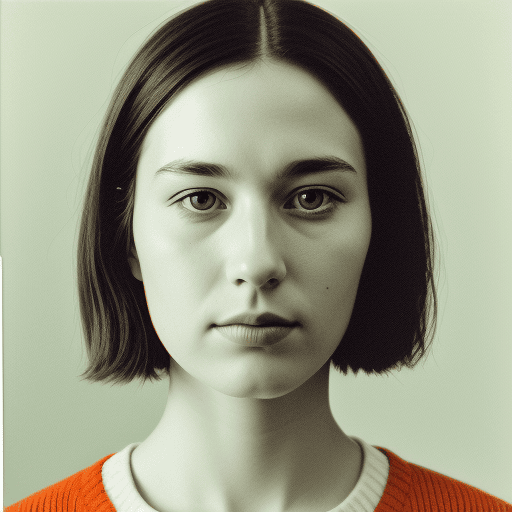}
& \includegraphics[width=0.130\linewidth]{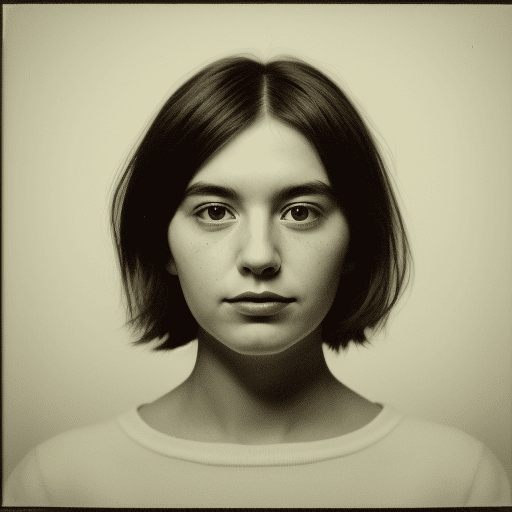}
& \includegraphics[width=0.130\linewidth]{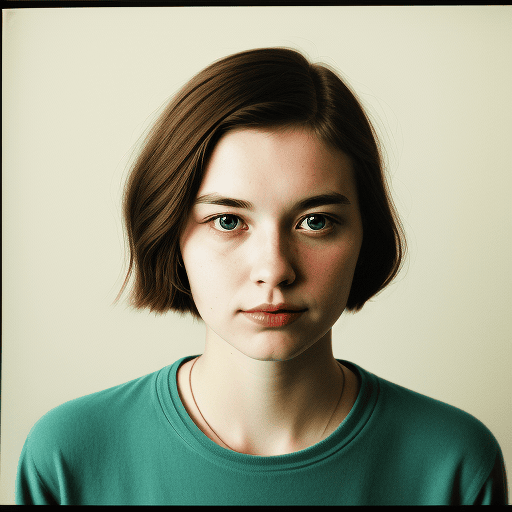}
\\

1.0
& \includegraphics[width=0.130\linewidth]{figs/multiple_seeds/default_0.png}
& \includegraphics[width=0.130\linewidth]{figs/multiple_seeds/default_1.png}
& \includegraphics[width=0.130\linewidth]{figs/multiple_seeds/default_2.png}
& \includegraphics[width=0.130\linewidth]{figs/multiple_seeds/default_3.png}
& \includegraphics[width=0.130\linewidth]{figs/multiple_seeds/default_4.png}
& \includegraphics[width=0.130\linewidth]{figs/multiple_seeds/default_5.png}
\\

1.5
& \includegraphics[width=0.130\linewidth]{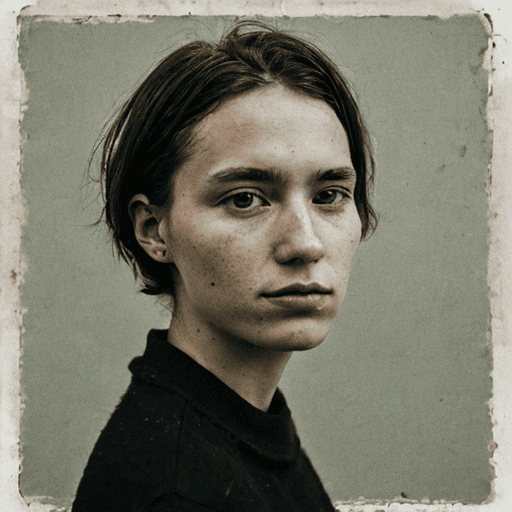}
& \includegraphics[width=0.130\linewidth]{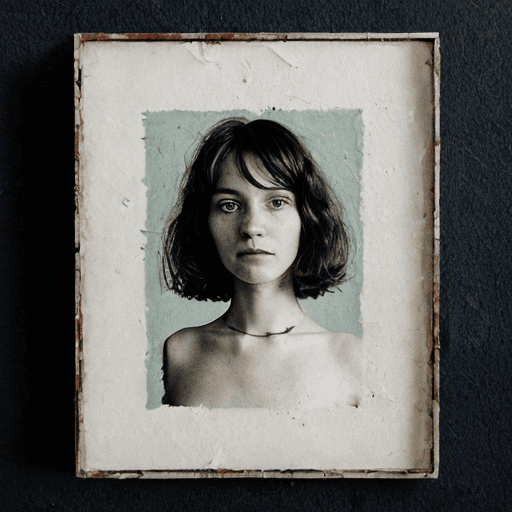}
& \includegraphics[width=0.130\linewidth]{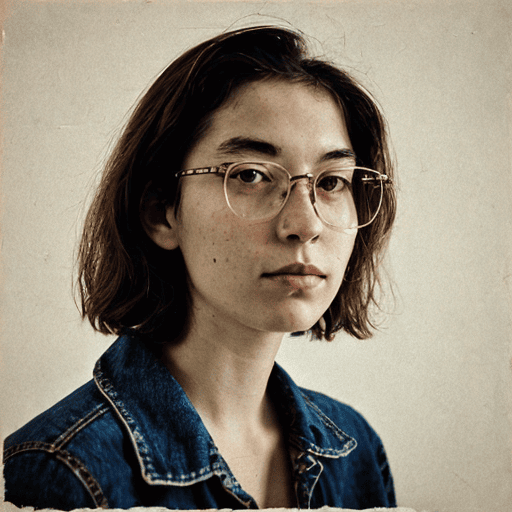}
& \includegraphics[width=0.130\linewidth]{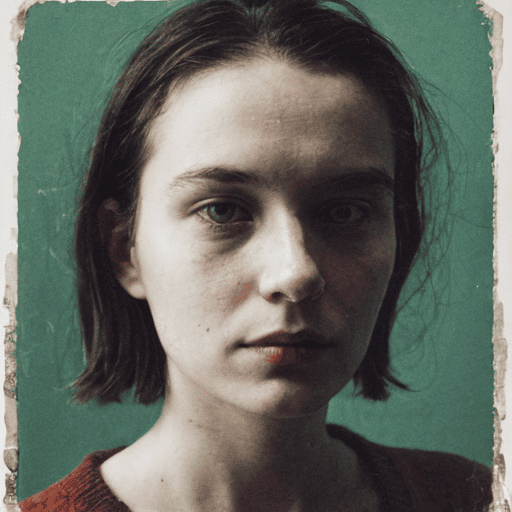}
& \includegraphics[width=0.130\linewidth]{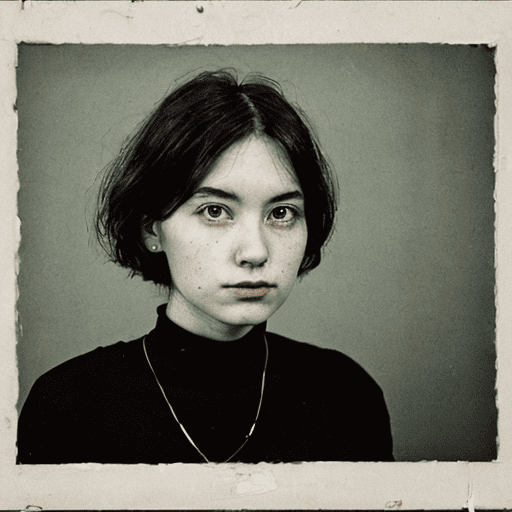}
& \includegraphics[width=0.130\linewidth]{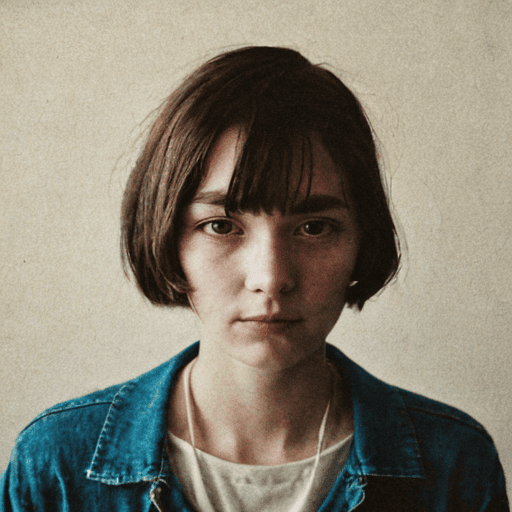}
\\

2.0
& \includegraphics[width=0.130\linewidth]{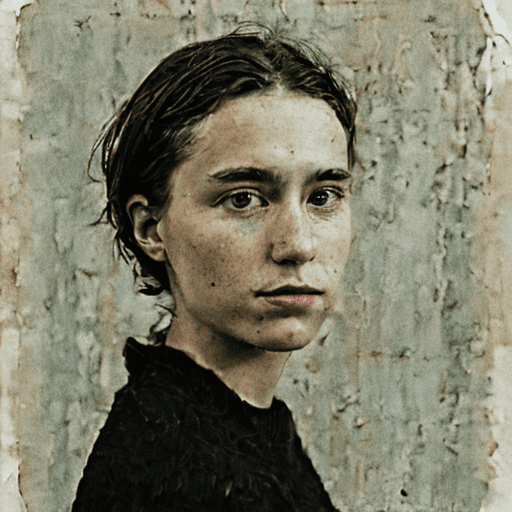}
& \includegraphics[width=0.130\linewidth]{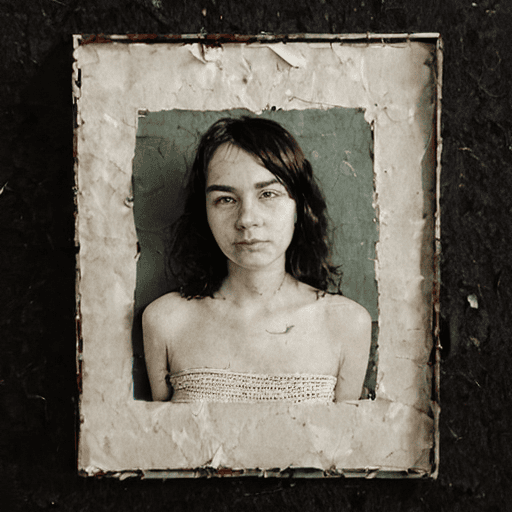}
& \includegraphics[width=0.130\linewidth]{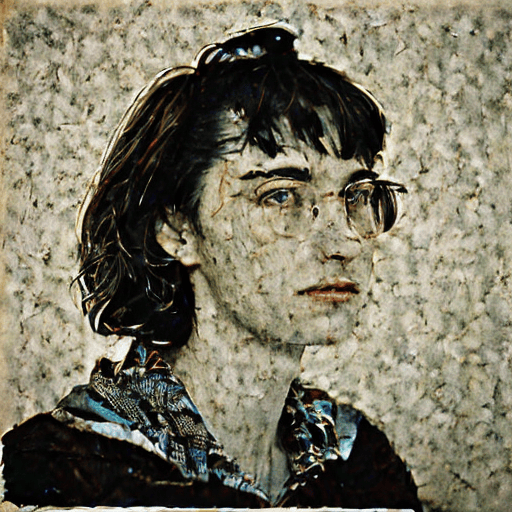}
& \includegraphics[width=0.130\linewidth]{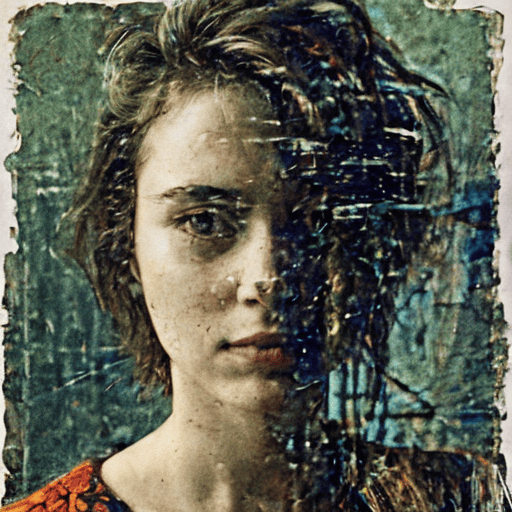}
& \includegraphics[width=0.130\linewidth]{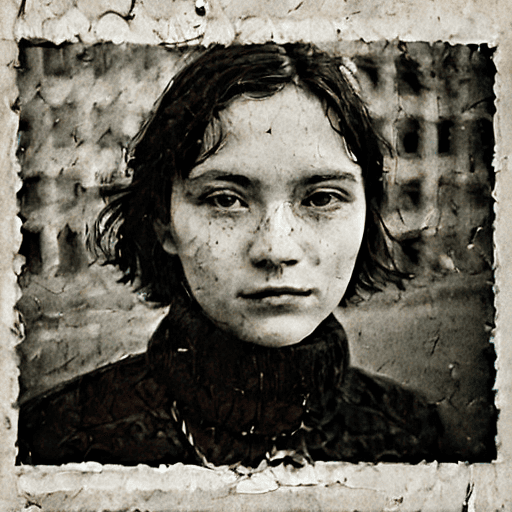}
& \includegraphics[width=0.130\linewidth]{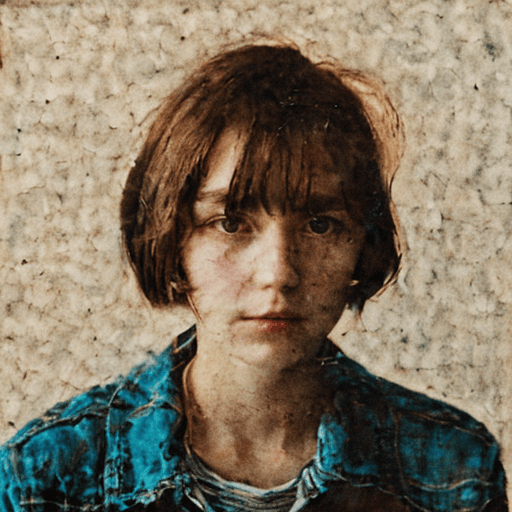}
\\
\hline
\end{tabular}

\caption{Bending \textit{input\_blocks.1.0.out\_layers.2}; Same prompt, different seeds.}
\label{fig:multiple_seeds_2}



\begin{tabular}{c|cccccc}
\hline
Multiply by / Seed & 42 & 0 & 123 & 456 & 789 & 786 \\
\hline

0.0
& \includegraphics[width=0.130\linewidth]{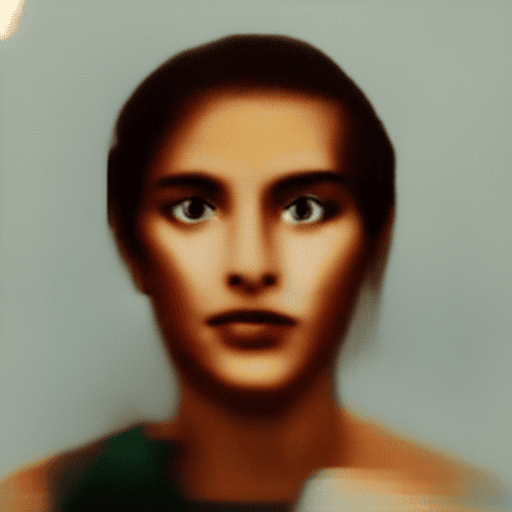}
& \includegraphics[width=0.130\linewidth]{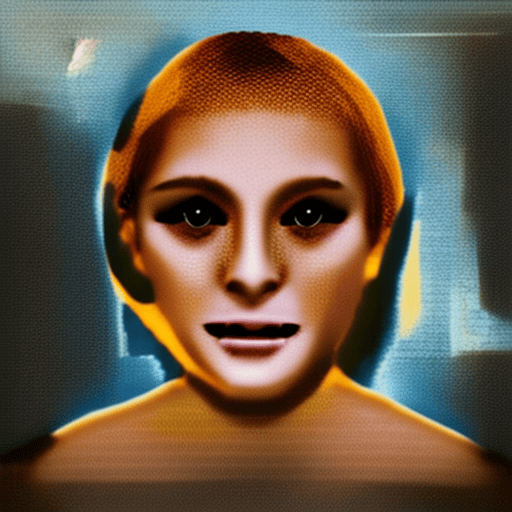}
& \includegraphics[width=0.130\linewidth]{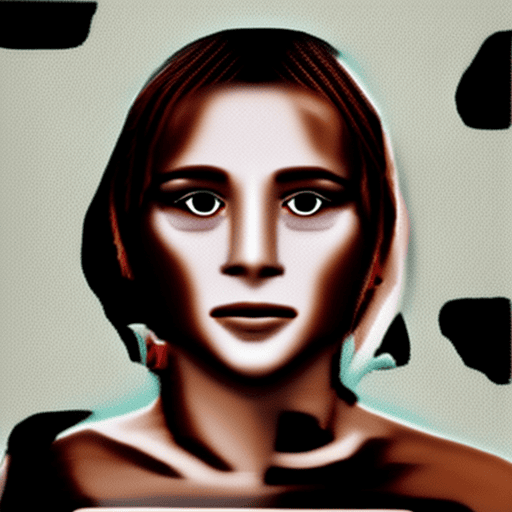}
& \includegraphics[width=0.130\linewidth]{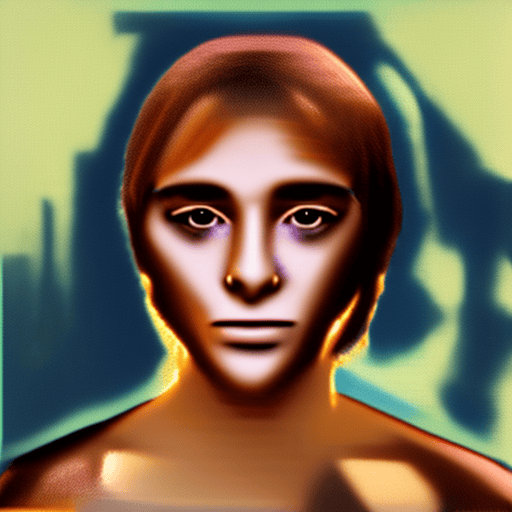}
& \includegraphics[width=0.130\linewidth]{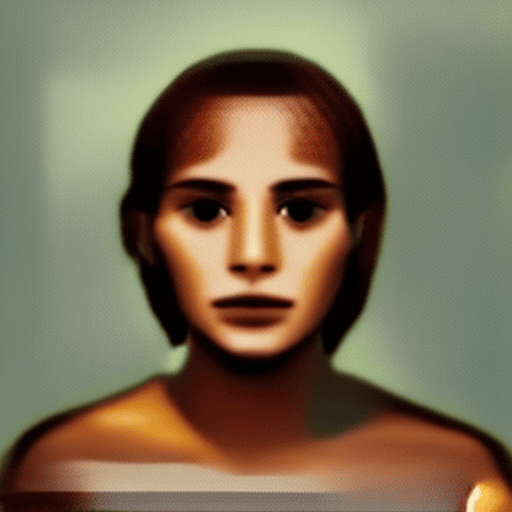}
& \includegraphics[width=0.130\linewidth]{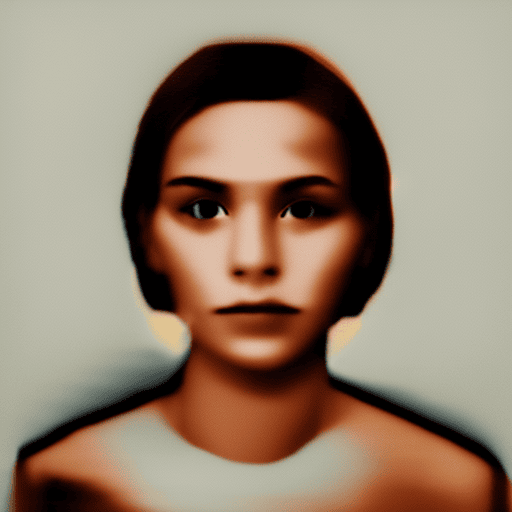}
\\

0.5
& \includegraphics[width=0.130\linewidth]{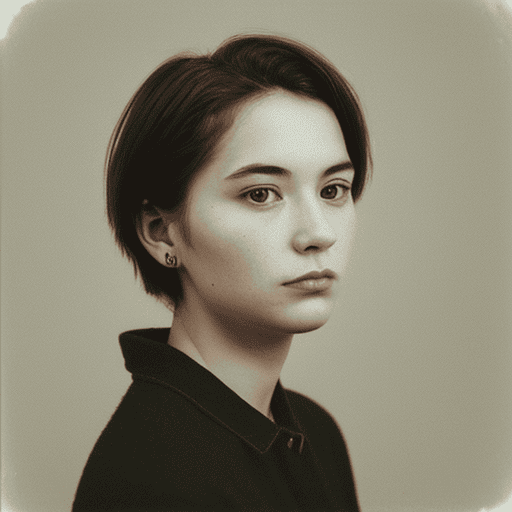}
& \includegraphics[width=0.130\linewidth]{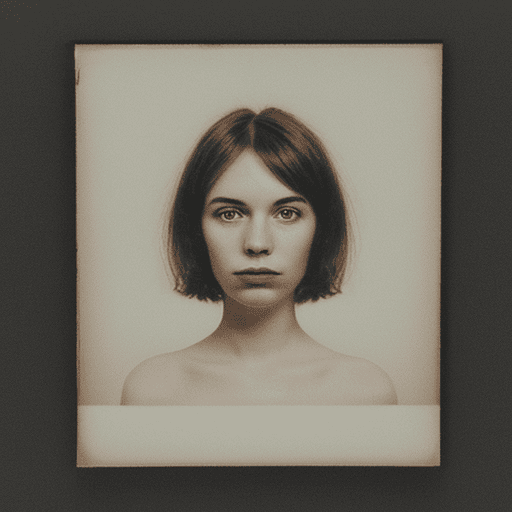}
& \includegraphics[width=0.130\linewidth]{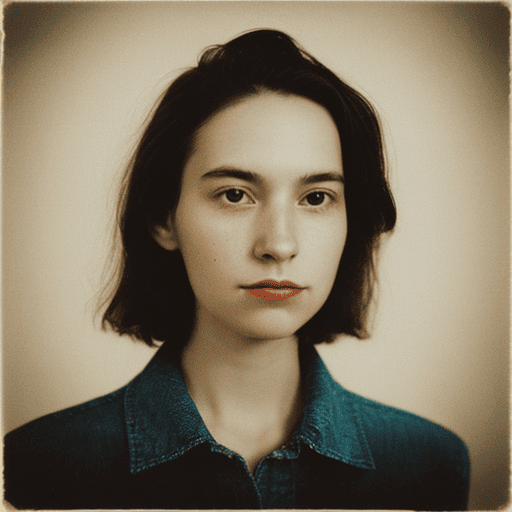}
& \includegraphics[width=0.130\linewidth]{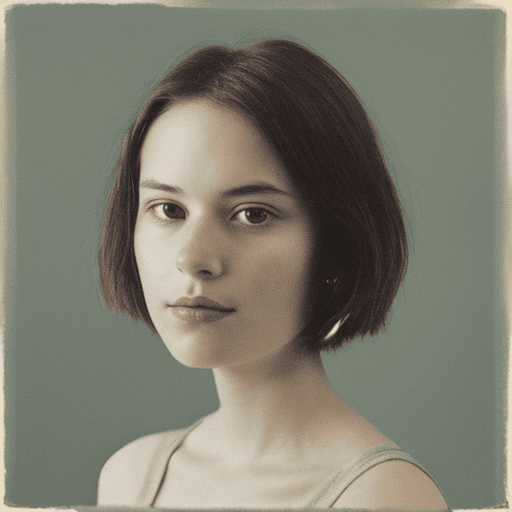}
& \includegraphics[width=0.130\linewidth]{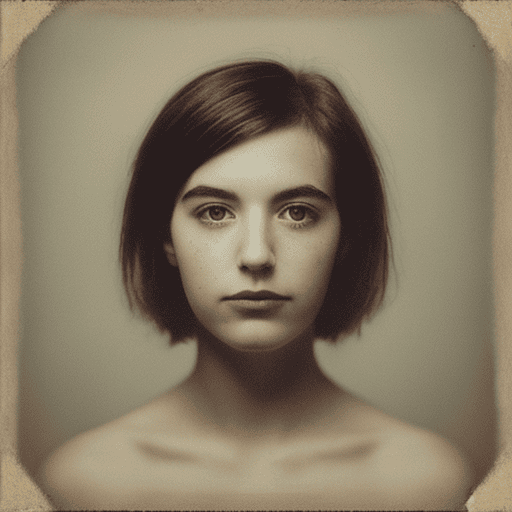}
& \includegraphics[width=0.130\linewidth]{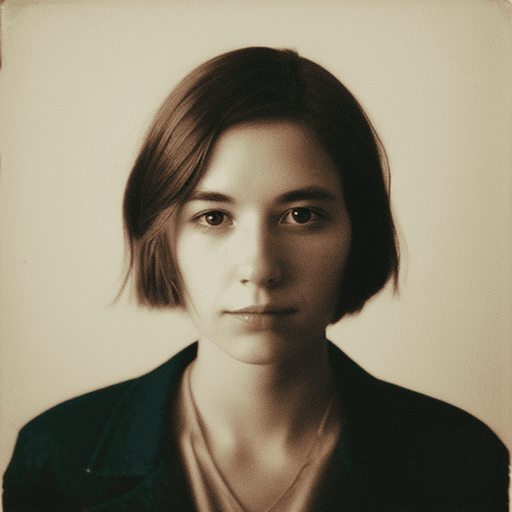}
\\

1.0
& \includegraphics[width=0.130\linewidth]{figs/multiple_seeds/default_0.png}
& \includegraphics[width=0.130\linewidth]{figs/multiple_seeds/default_1.png}
& \includegraphics[width=0.130\linewidth]{figs/multiple_seeds/default_2.png}
& \includegraphics[width=0.130\linewidth]{figs/multiple_seeds/default_3.png}
& \includegraphics[width=0.130\linewidth]{figs/multiple_seeds/default_4.png}
& \includegraphics[width=0.130\linewidth]{figs/multiple_seeds/default_5.png}
\\

1.5
& \includegraphics[width=0.130\linewidth]{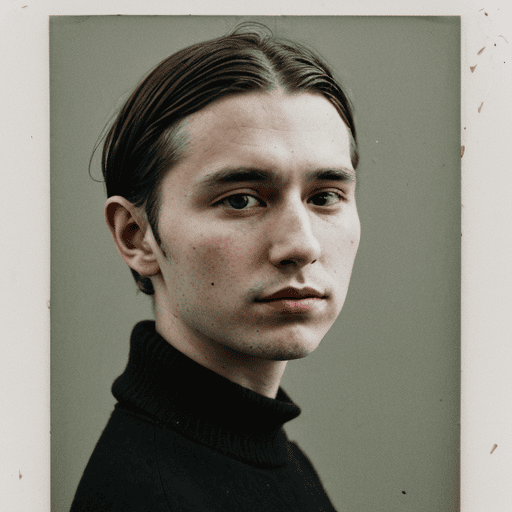}
& \includegraphics[width=0.130\linewidth]{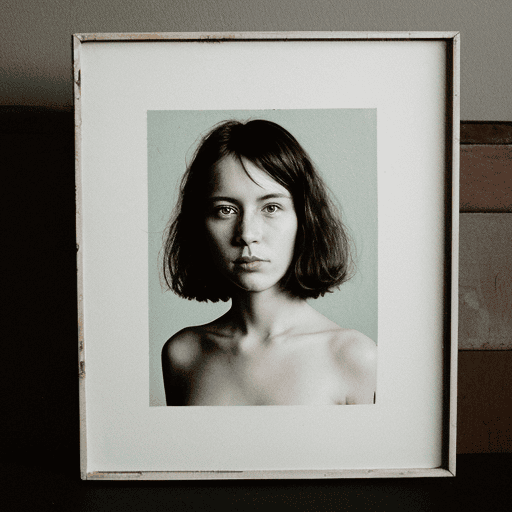}
& \includegraphics[width=0.130\linewidth]{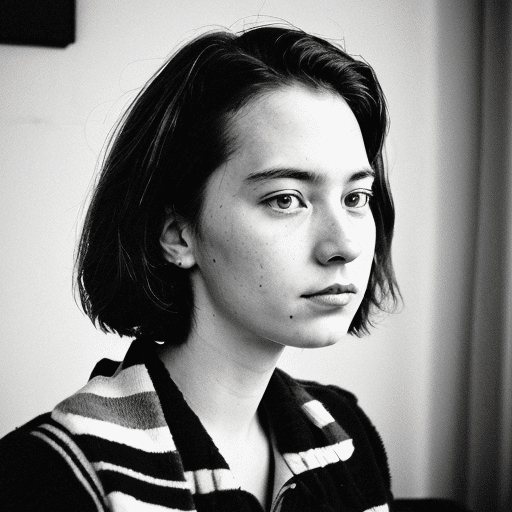}
& \includegraphics[width=0.130\linewidth]{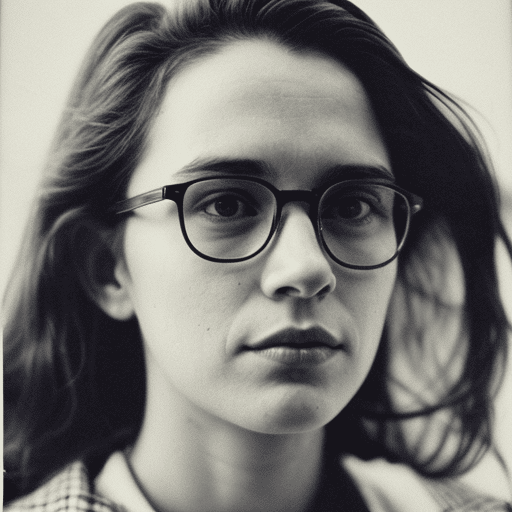}
& \includegraphics[width=0.130\linewidth]{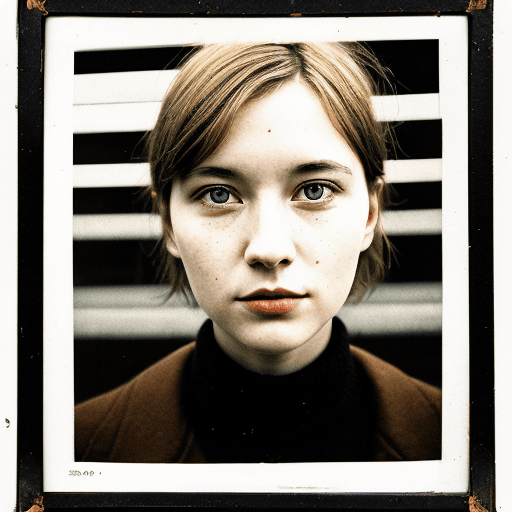}
& \includegraphics[width=0.130\linewidth]{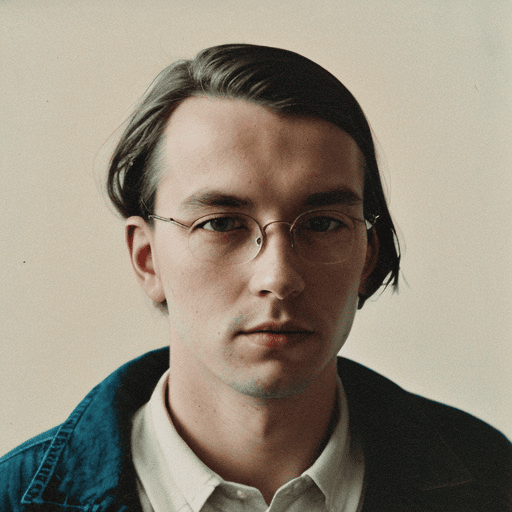}
\\

2.0
& \includegraphics[width=0.130\linewidth]{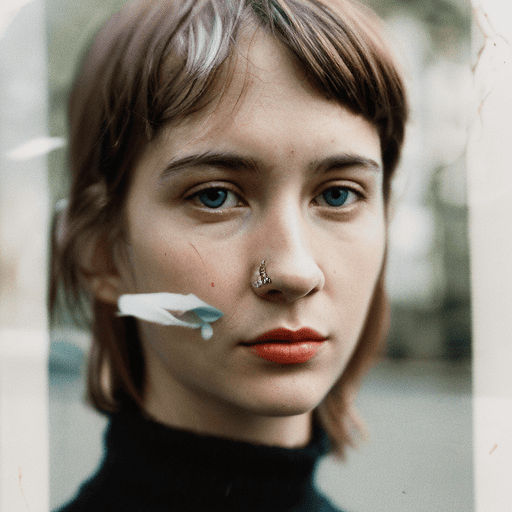}
& \includegraphics[width=0.130\linewidth]{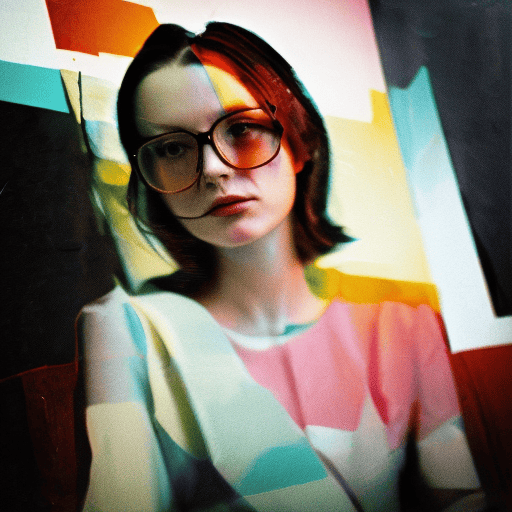}
& \includegraphics[width=0.130\linewidth]{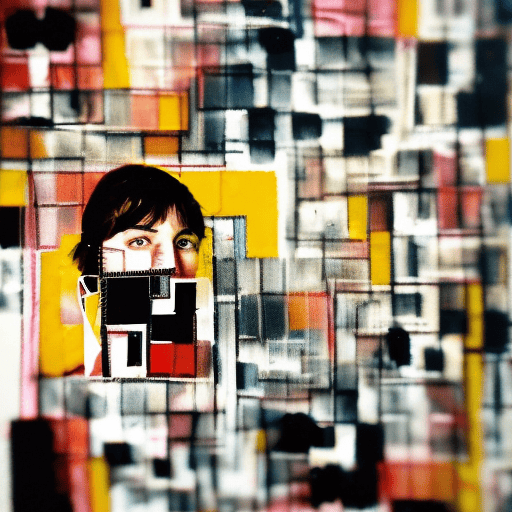}
& \includegraphics[width=0.130\linewidth]{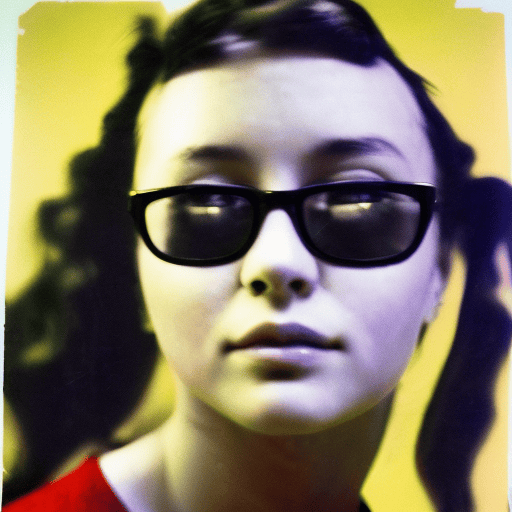}
& \includegraphics[width=0.130\linewidth]{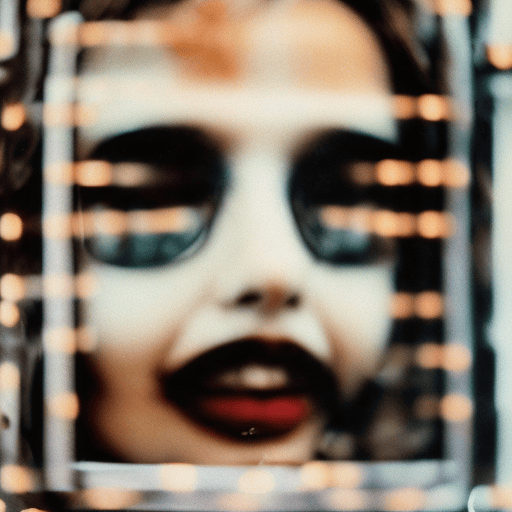}
& \includegraphics[width=0.130\linewidth]{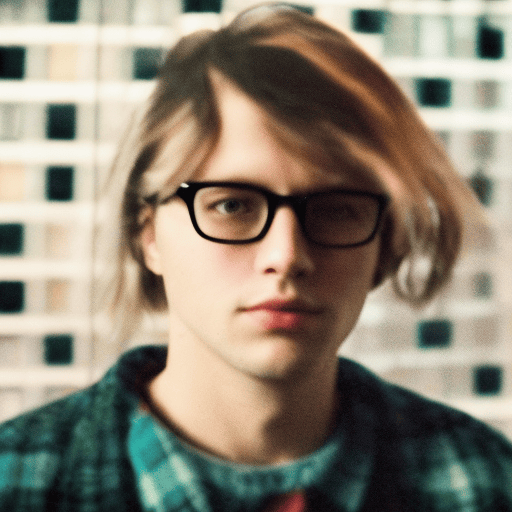}
\\
\hline
\end{tabular}

\caption{Bending \textit{input\_blocks.4.0.skip\_connection}; Same prompt, different seeds.}
\label{fig:multiple_seeds_3}
\end{figure}

\begin{figure}[H]
\centering


\begin{tabular}{c|cccccc}
\hline
Multiply by / Seed & 42 & 0 & 123 & 456 & 789 & 786 \\
\hline

0.0
& \includegraphics[width=0.130\linewidth]{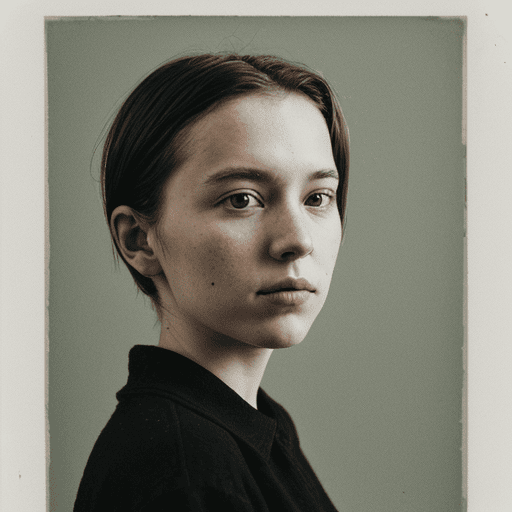}
& \includegraphics[width=0.130\linewidth]{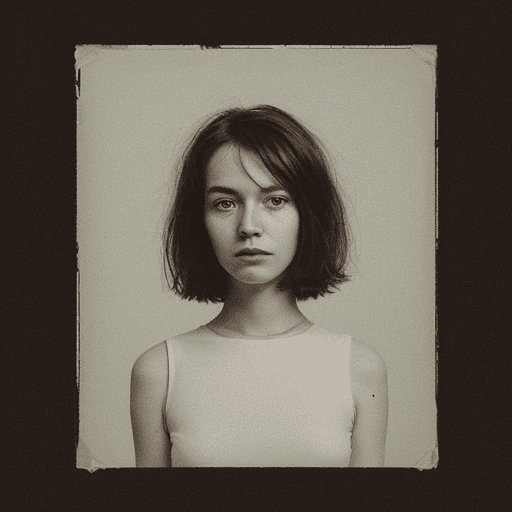}
& \includegraphics[width=0.130\linewidth]{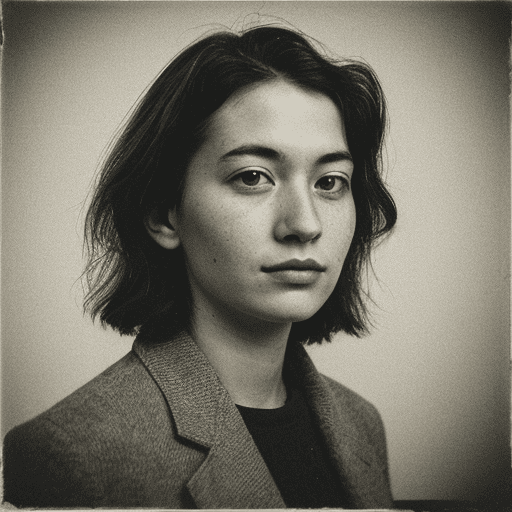}
& \includegraphics[width=0.130\linewidth]{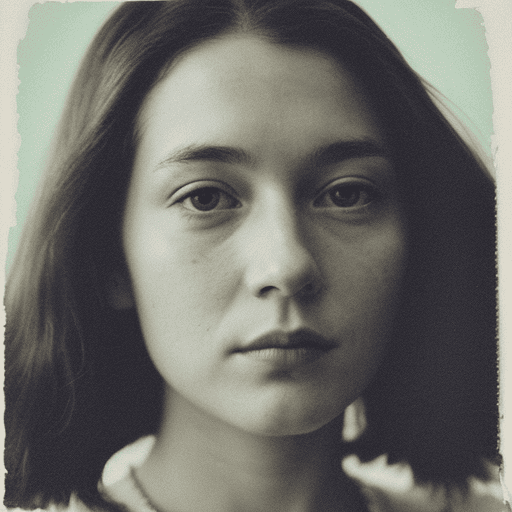}
& \includegraphics[width=0.130\linewidth]{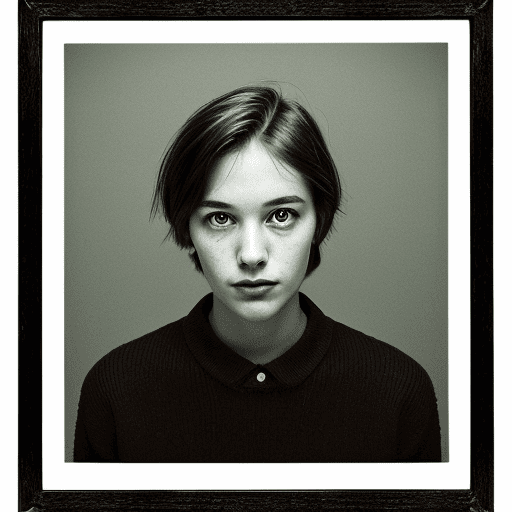}
& \includegraphics[width=0.130\linewidth]{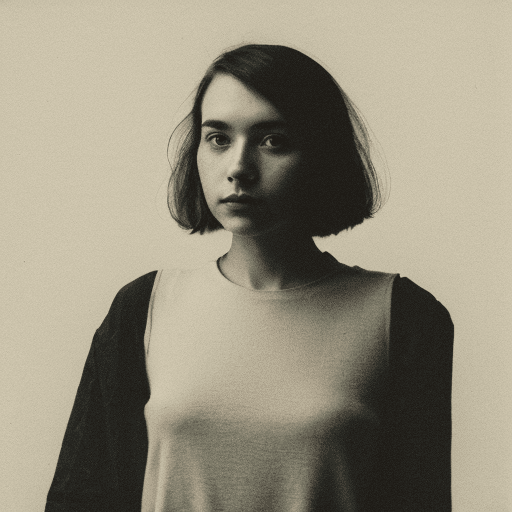}
\\

0.5
& \includegraphics[width=0.130\linewidth]{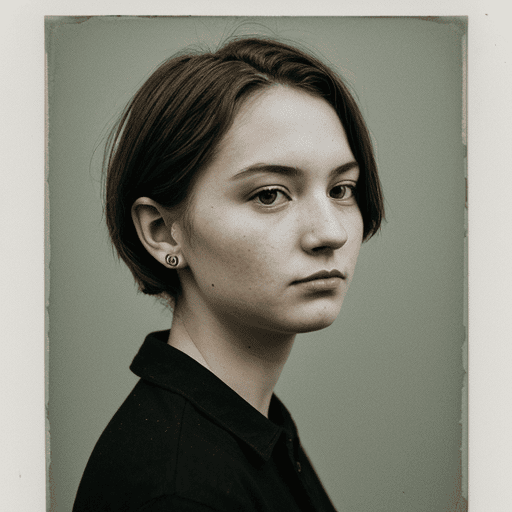}
& \includegraphics[width=0.130\linewidth]{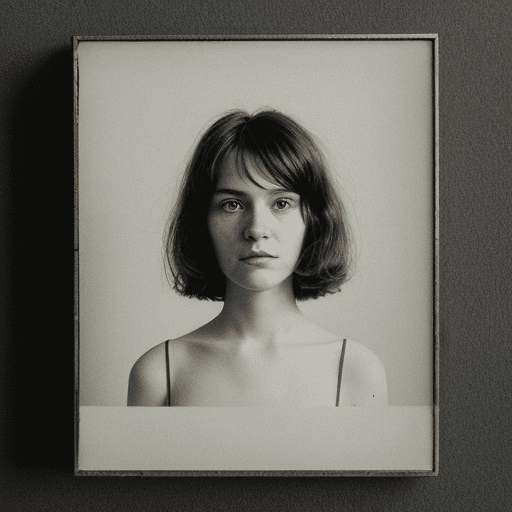}
& \includegraphics[width=0.130\linewidth]{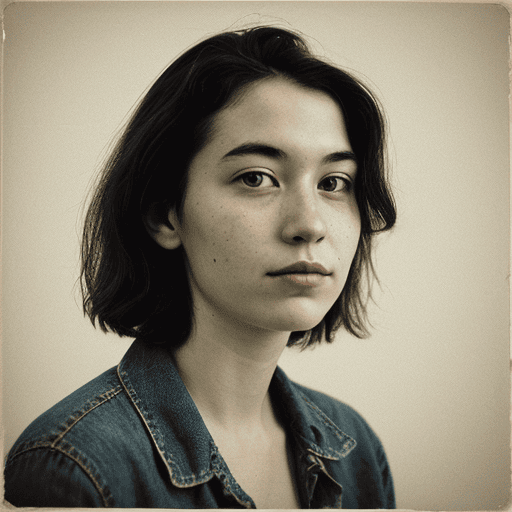}
& \includegraphics[width=0.130\linewidth]{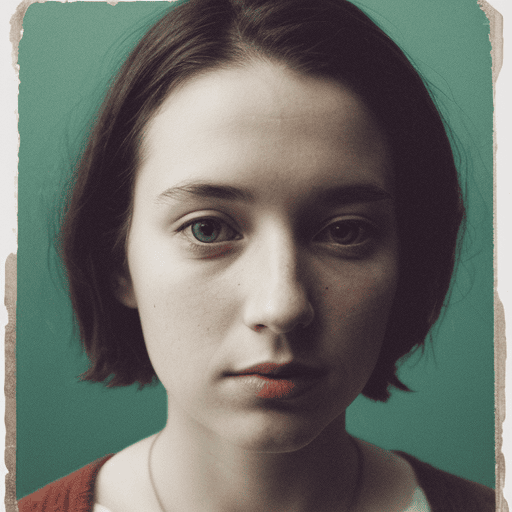}
& \includegraphics[width=0.130\linewidth]{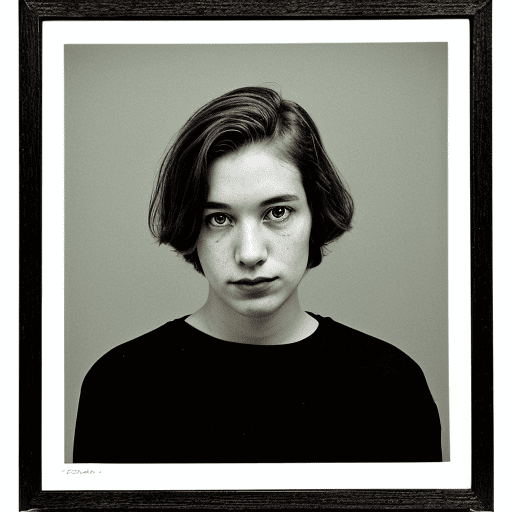}
& \includegraphics[width=0.130\linewidth]{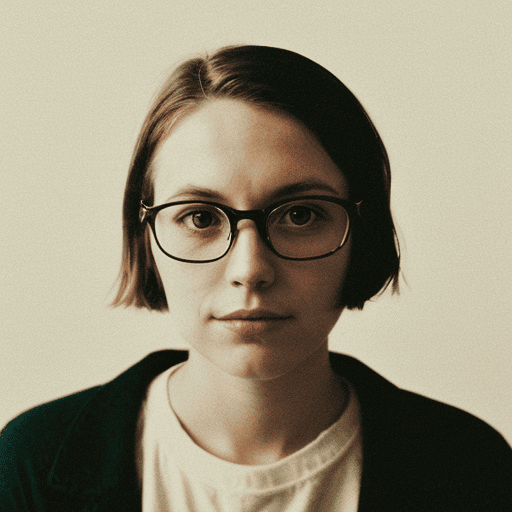}
\\

1.0
& \includegraphics[width=0.130\linewidth]{figs/multiple_seeds/default_0.png}
& \includegraphics[width=0.130\linewidth]{figs/multiple_seeds/default_1.png}
& \includegraphics[width=0.130\linewidth]{figs/multiple_seeds/default_2.png}
& \includegraphics[width=0.130\linewidth]{figs/multiple_seeds/default_3.png}
& \includegraphics[width=0.130\linewidth]{figs/multiple_seeds/default_4.png}
& \includegraphics[width=0.130\linewidth]{figs/multiple_seeds/default_5.png}
\\

1.5
& \includegraphics[width=0.130\linewidth]{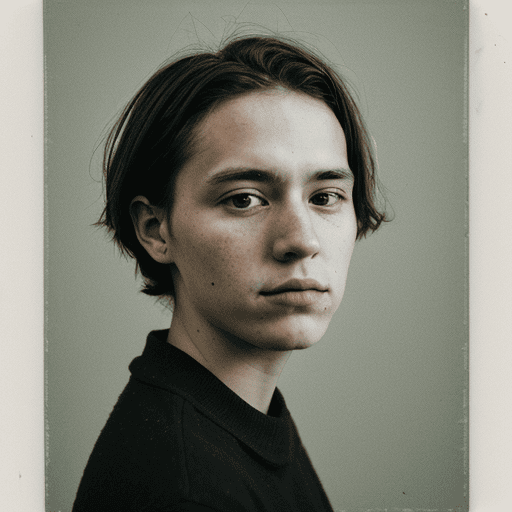}
& \includegraphics[width=0.130\linewidth]{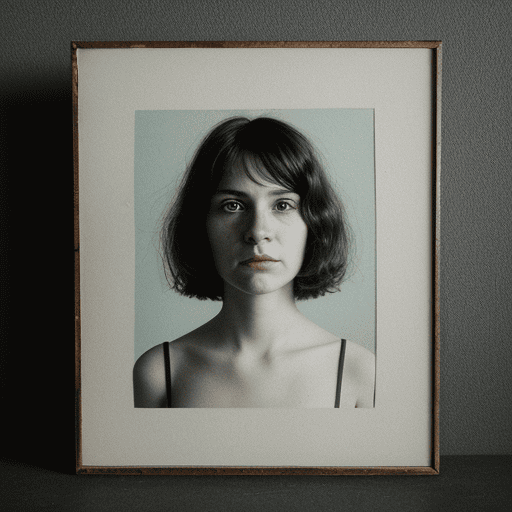}
& \includegraphics[width=0.130\linewidth]{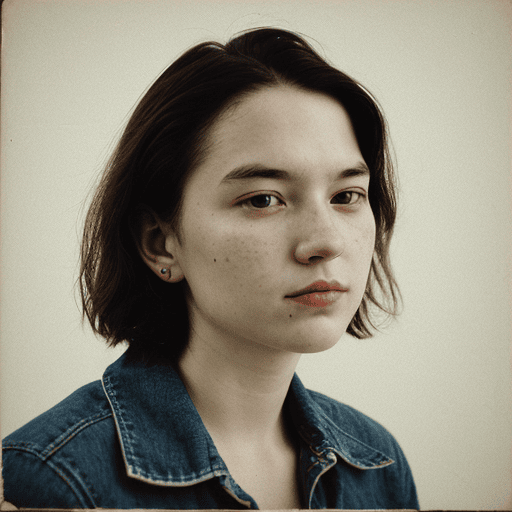}
& \includegraphics[width=0.130\linewidth]{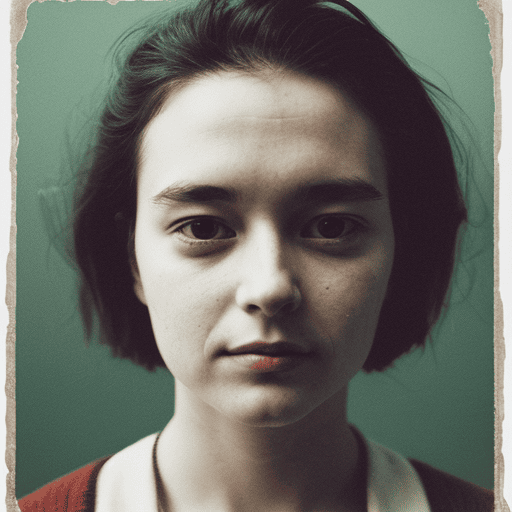}
& \includegraphics[width=0.130\linewidth]{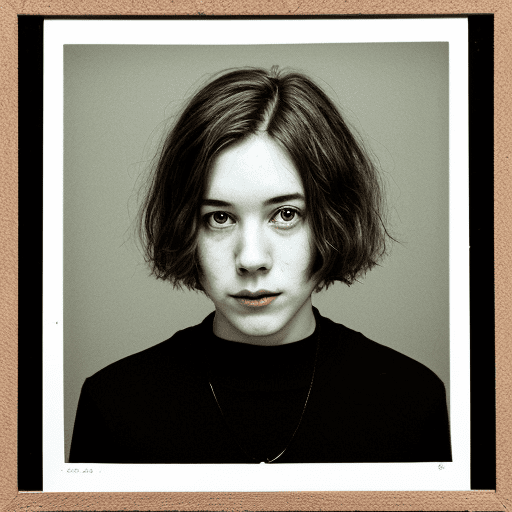}
& \includegraphics[width=0.130\linewidth]{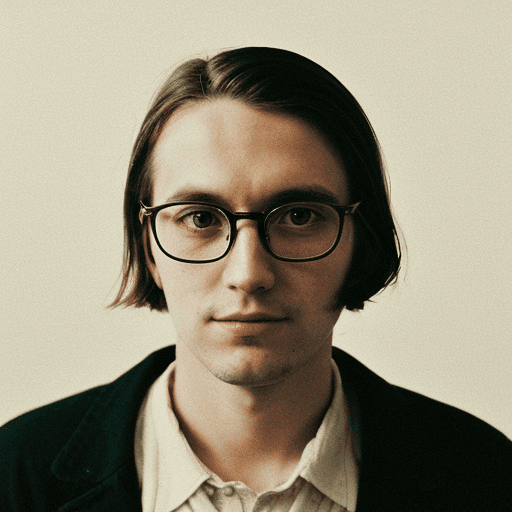}
\\

2.0
& \includegraphics[width=0.130\linewidth]{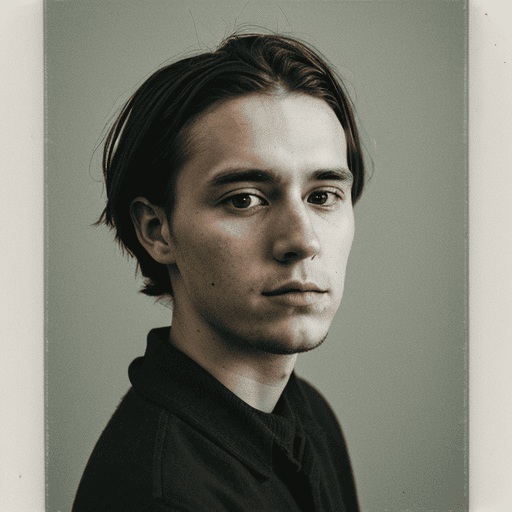}
& \includegraphics[width=0.130\linewidth]{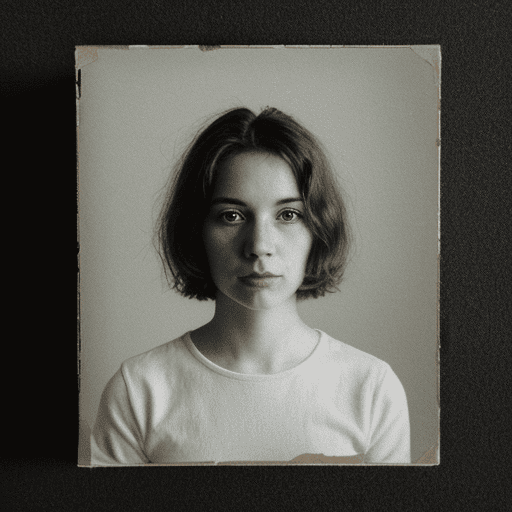}
& \includegraphics[width=0.130\linewidth]{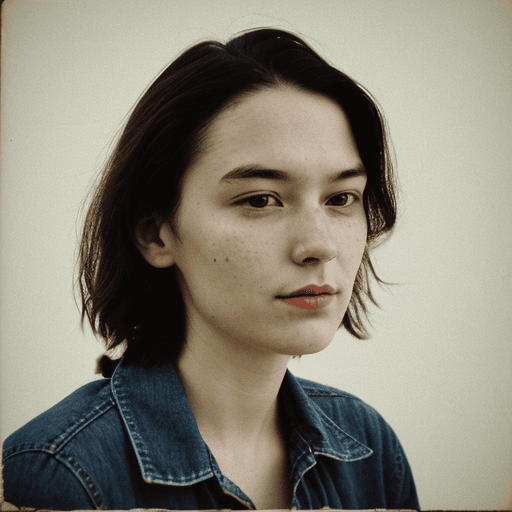}
& \includegraphics[width=0.130\linewidth]{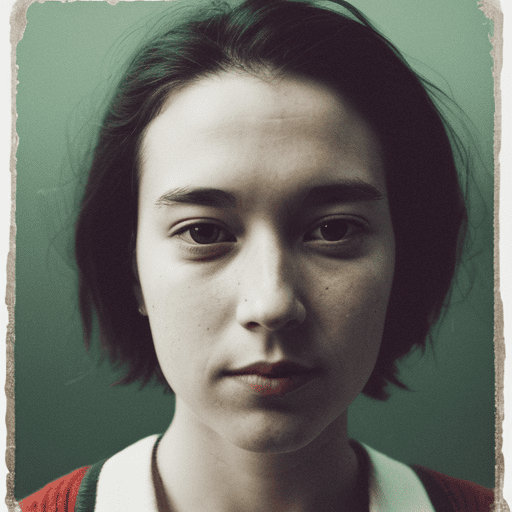}
& \includegraphics[width=0.130\linewidth]{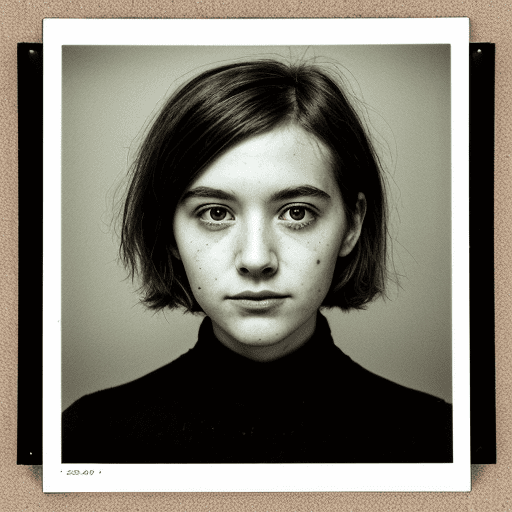}
& \includegraphics[width=0.130\linewidth]{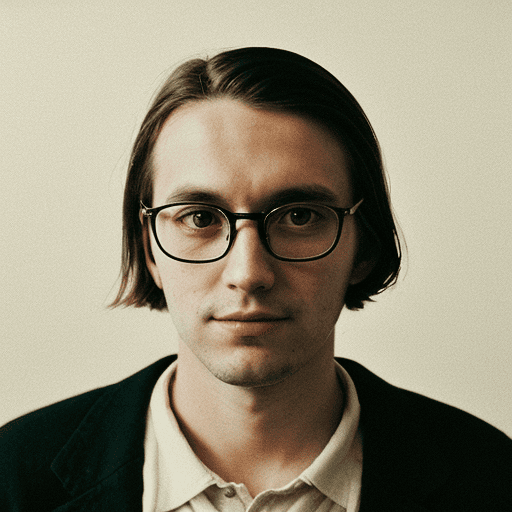}
\\
\hline
\end{tabular}

\caption{Bending  \textit{middle\_block.1.transformer\_blocks.0.attn2.to\_out}; Different seeds.}
\label{fig:multiple_seeds_4}



\begin{tabular}{c|cccccc}
\hline
Multiply by / Seed & 42 & 0 & 123 & 456 & 789 & 786 \\
\hline

0.0
& \includegraphics[width=0.130\linewidth]{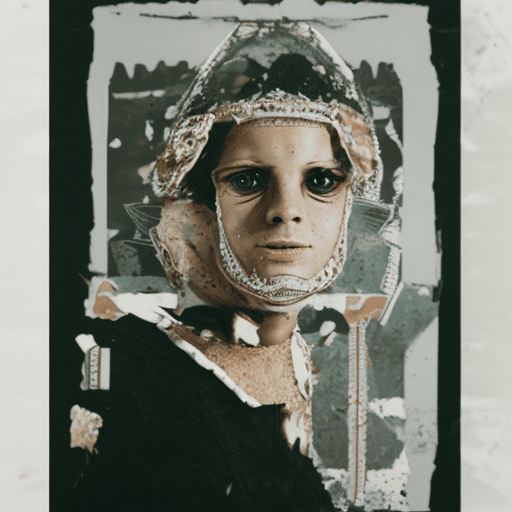}
& \includegraphics[width=0.130\linewidth]{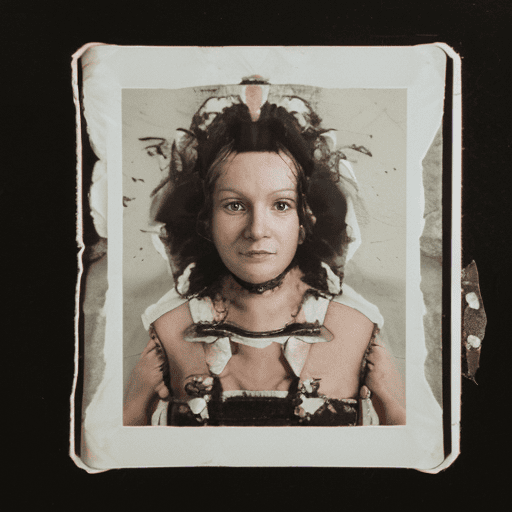}
& \includegraphics[width=0.130\linewidth]{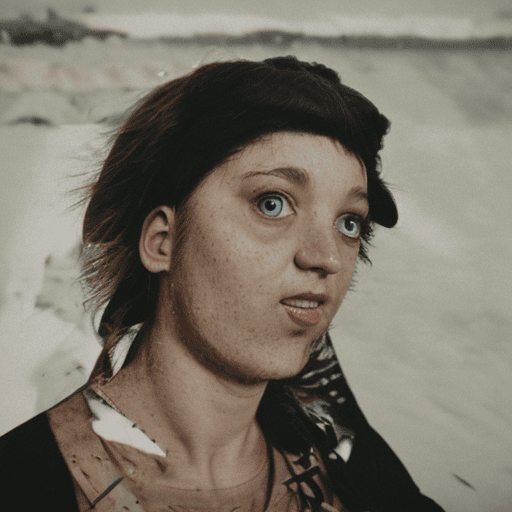}
& \includegraphics[width=0.130\linewidth]{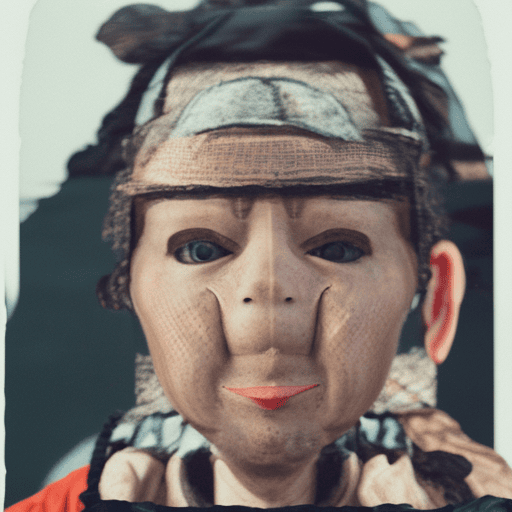}
& \includegraphics[width=0.130\linewidth]{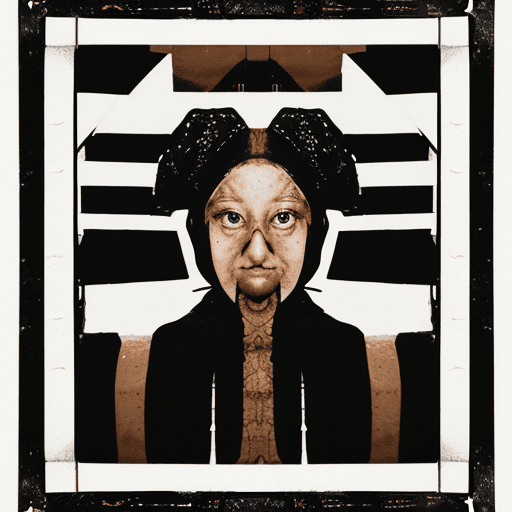}
& \includegraphics[width=0.130\linewidth]{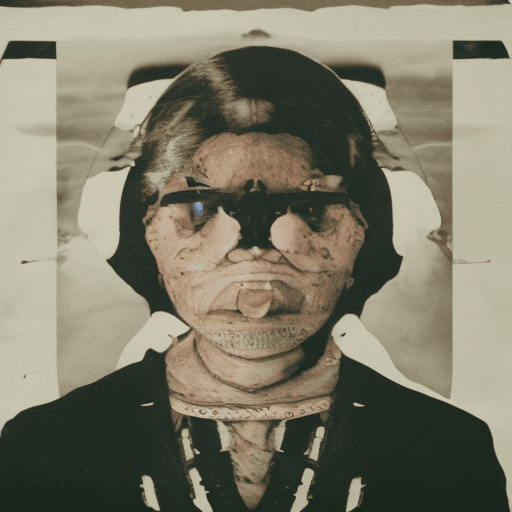}
\\

0.5
& \includegraphics[width=0.130\linewidth]{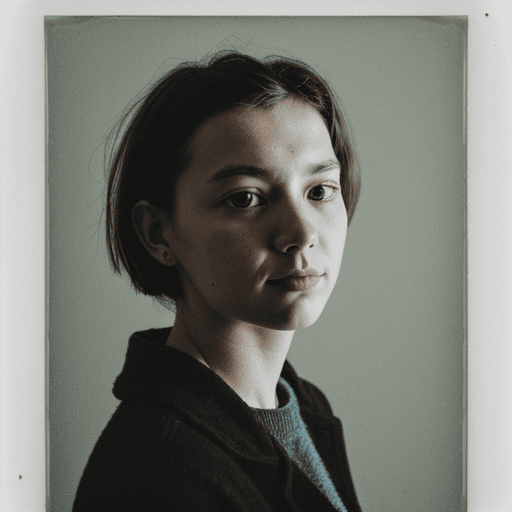}
& \includegraphics[width=0.130\linewidth]{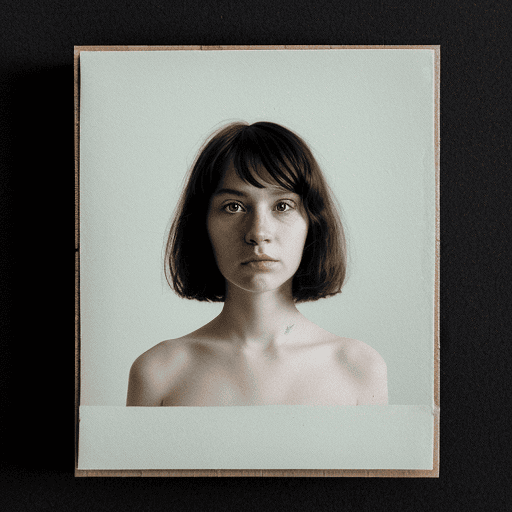}
& \includegraphics[width=0.130\linewidth]{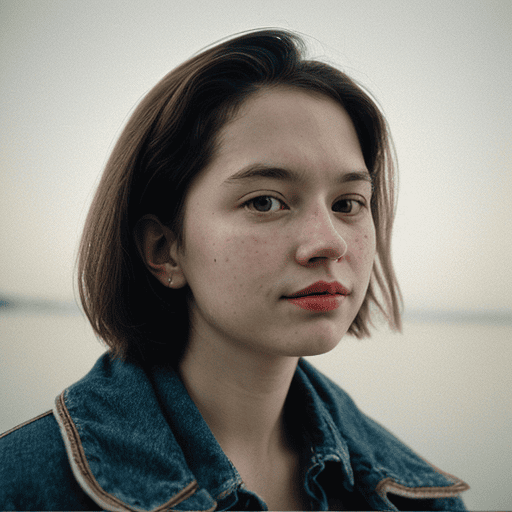}
& \includegraphics[width=0.130\linewidth]{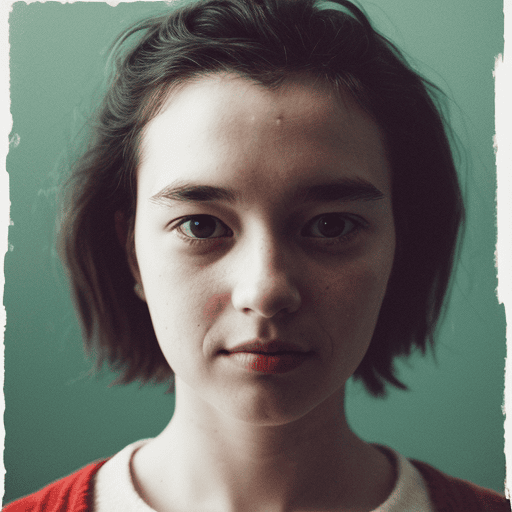}
& \includegraphics[width=0.130\linewidth]{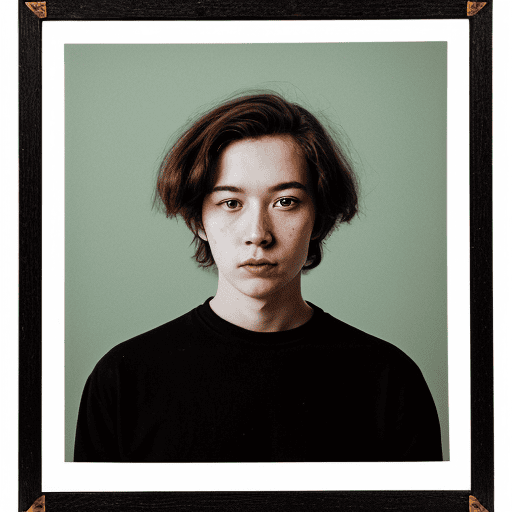}
& \includegraphics[width=0.130\linewidth]{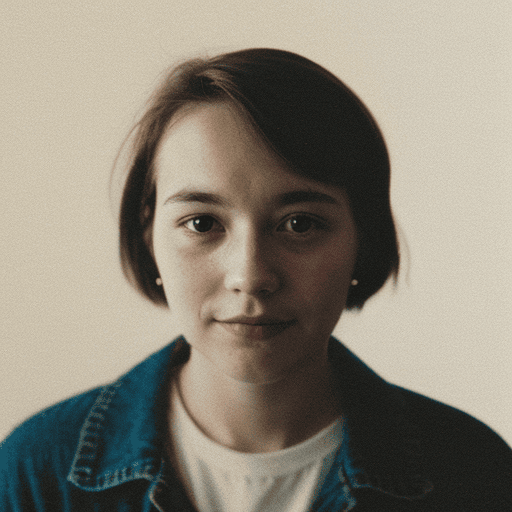}
\\

1.0
& \includegraphics[width=0.130\linewidth]{figs/multiple_seeds/default_0.png}
& \includegraphics[width=0.130\linewidth]{figs/multiple_seeds/default_1.png}
& \includegraphics[width=0.130\linewidth]{figs/multiple_seeds/default_2.png}
& \includegraphics[width=0.130\linewidth]{figs/multiple_seeds/default_3.png}
& \includegraphics[width=0.130\linewidth]{figs/multiple_seeds/default_4.png}
& \includegraphics[width=0.130\linewidth]{figs/multiple_seeds/default_5.png}
\\

1.5
& \includegraphics[width=0.130\linewidth]{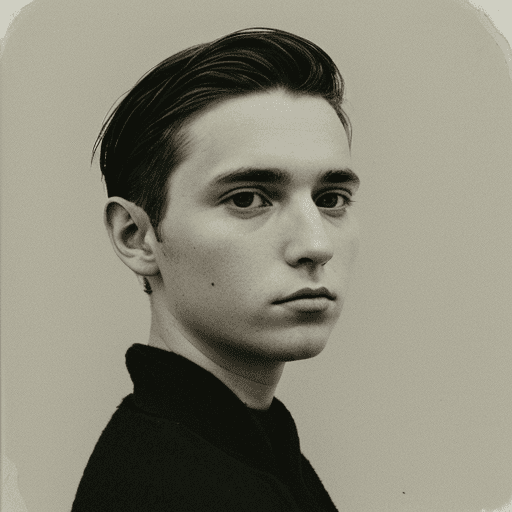}
& \includegraphics[width=0.130\linewidth]{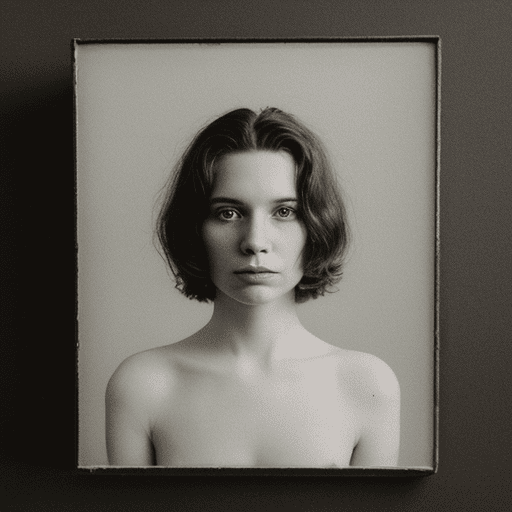}
& \includegraphics[width=0.130\linewidth]{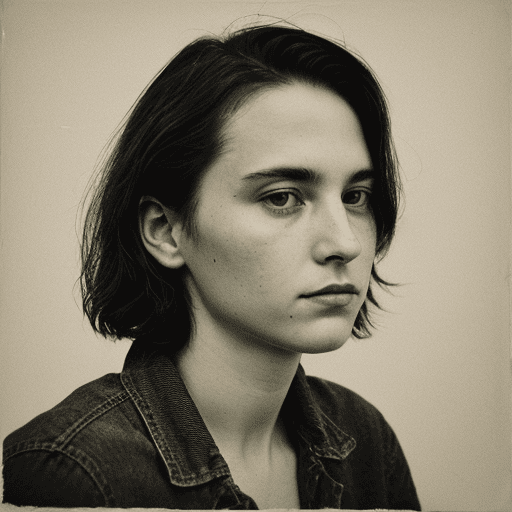}
& \includegraphics[width=0.130\linewidth]{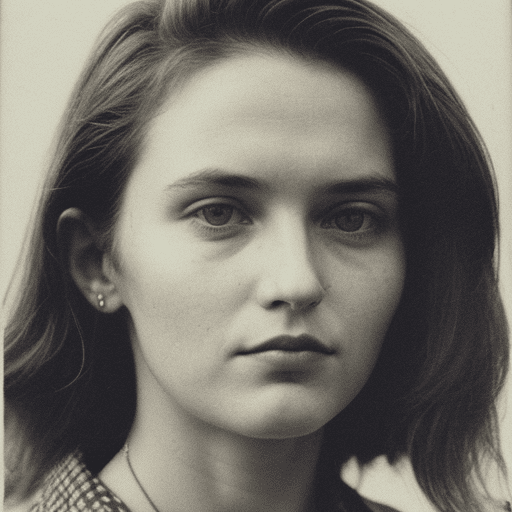}
& \includegraphics[width=0.130\linewidth]{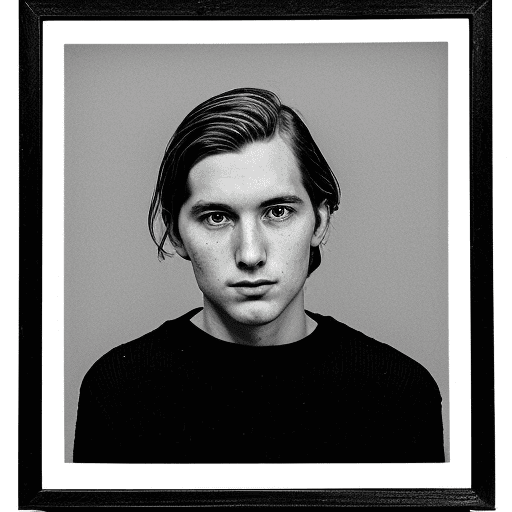}
& \includegraphics[width=0.130\linewidth]{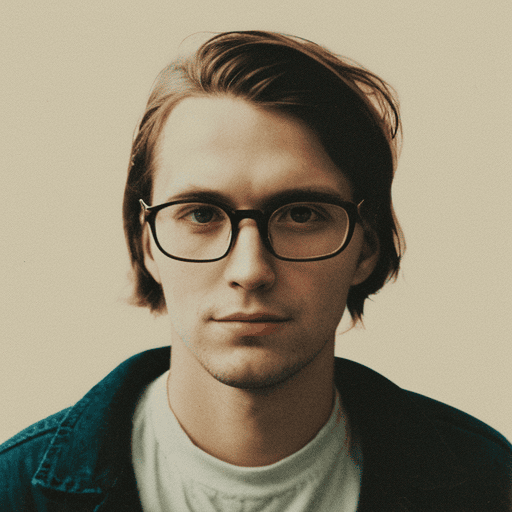}
\\

2.0
& \includegraphics[width=0.130\linewidth]{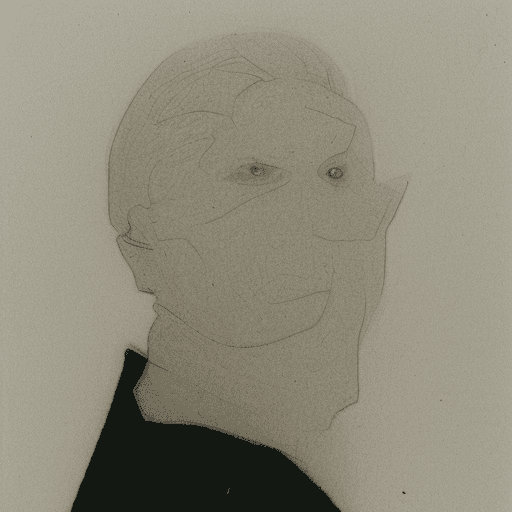}
& \includegraphics[width=0.130\linewidth]{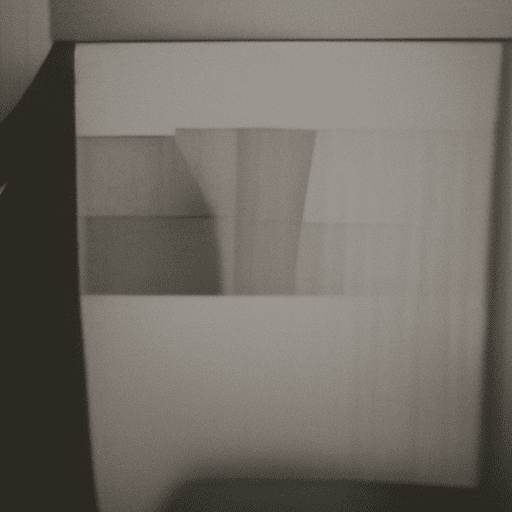}
& \includegraphics[width=0.130\linewidth]{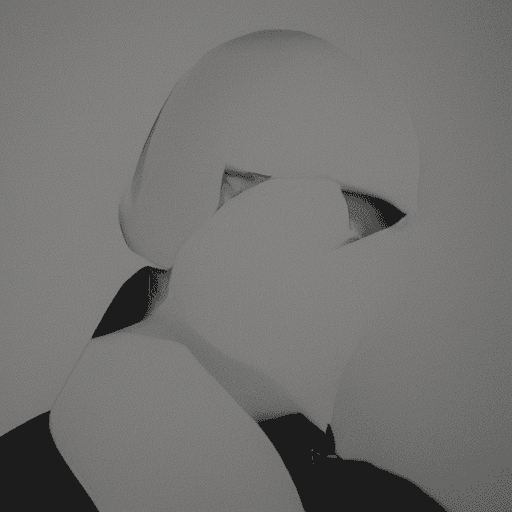}
& \includegraphics[width=0.130\linewidth]{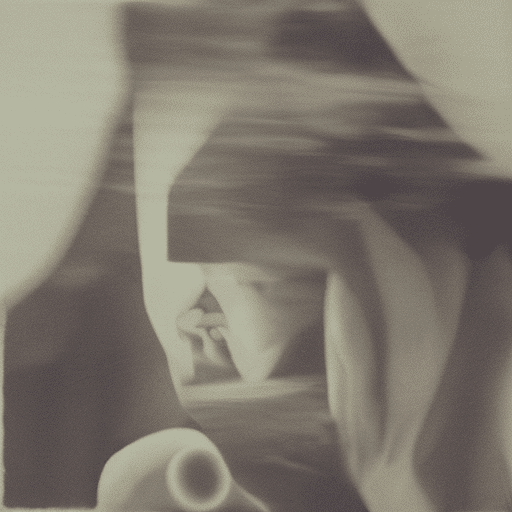}
& \includegraphics[width=0.130\linewidth]{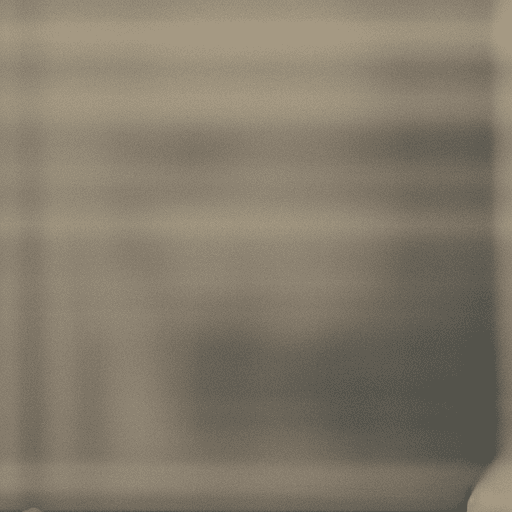}
& \includegraphics[width=0.130\linewidth]{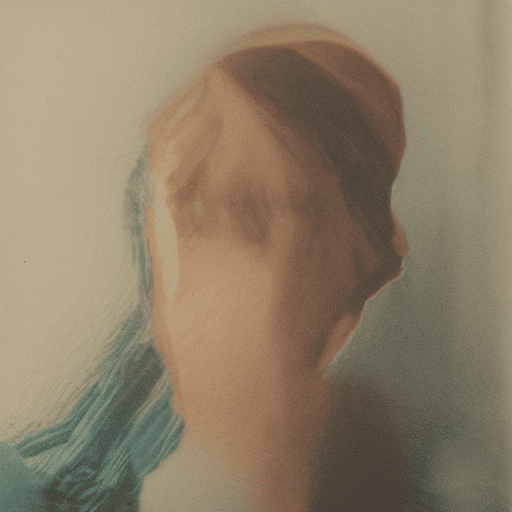}
\\
\hline
\end{tabular}

\caption{Bending \textit{output\_blocks.3.1.norm}; Same prompt, different seeds.}
\label{fig:multiple_seeds_5}
\end{figure}

Using the same prompt, "Analog style portrait of a person", we examine whether the same bending effects are consistent across different seeds. We repeat the experiment for select layers, hand-picked to span input, middle and output blocks in the UNet. The first row (seed 42) is the one used across the rest of the paper. Figures~\ref{fig:multiple_seeds_1}, ~\ref{fig:multiple_seeds_2}, ~\ref{fig:multiple_seeds_3}, ~\ref{fig:multiple_seeds_4}, ~\ref{fig:multiple_seeds_5} follow.

\clearpage

\subsubsection{Same Seed, Different Prompts}
\label{sec:results_multiple_prompts}
\begin{figure}[h]
\centering

\medskip

\begin{tabular}{c|cccccc}

\hline
Multiply & Prompt 1 & Prompt 2 & Prompt 3 & Prompt 4 & Prompt 5 & Prompt 6 \\
\hline
0.0
& \includegraphics[width=0.140\linewidth]{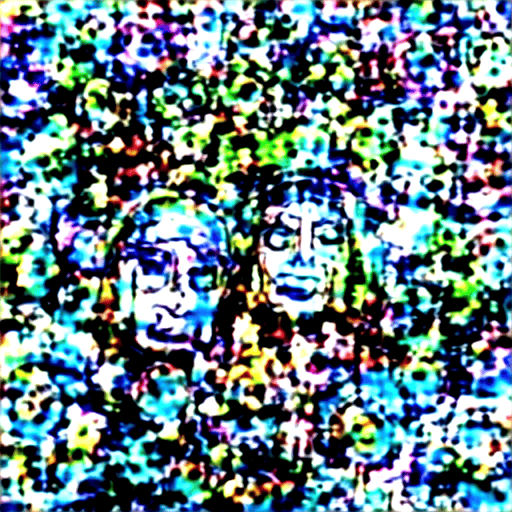}
& \includegraphics[width=0.140\linewidth]{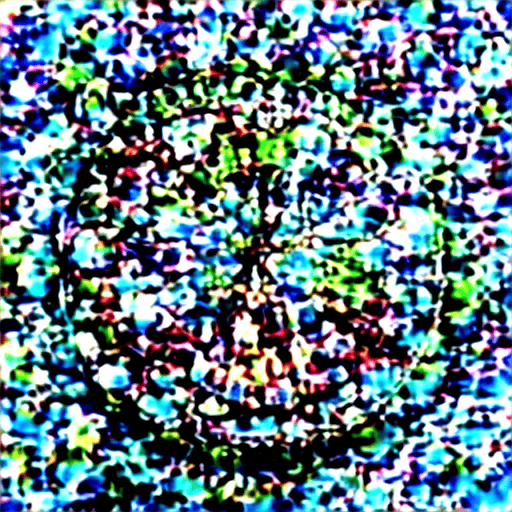}
& \includegraphics[width=0.140\linewidth]{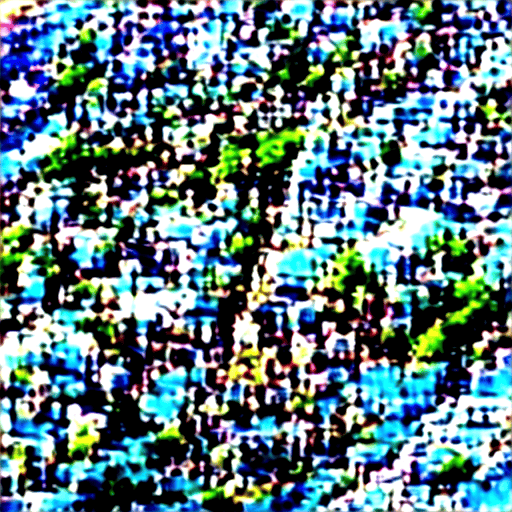}
& \includegraphics[width=0.140\linewidth]{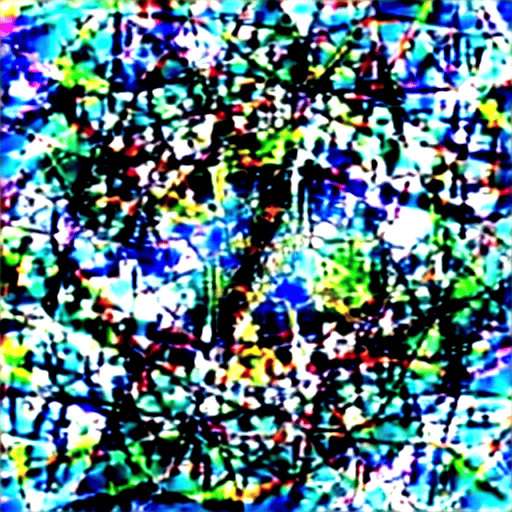}
& \includegraphics[width=0.140\linewidth]{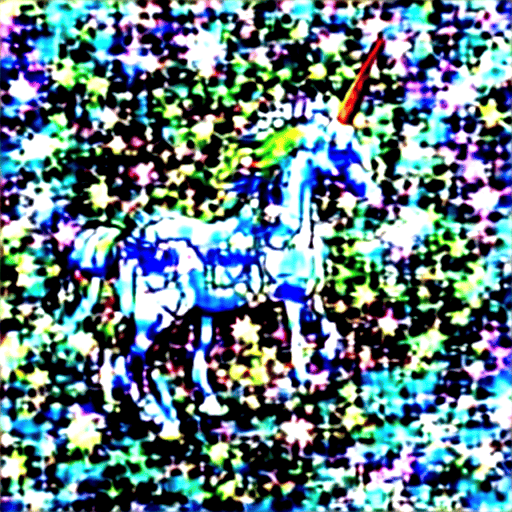}
& \includegraphics[width=0.140\linewidth]{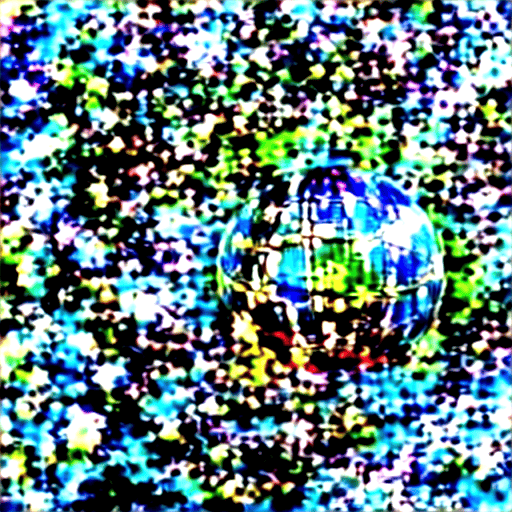}
\\

0.5
& \includegraphics[width=0.140\linewidth]{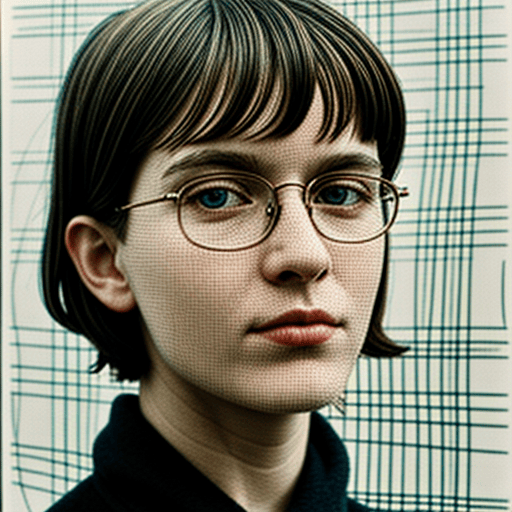}
& \includegraphics[width=0.140\linewidth]{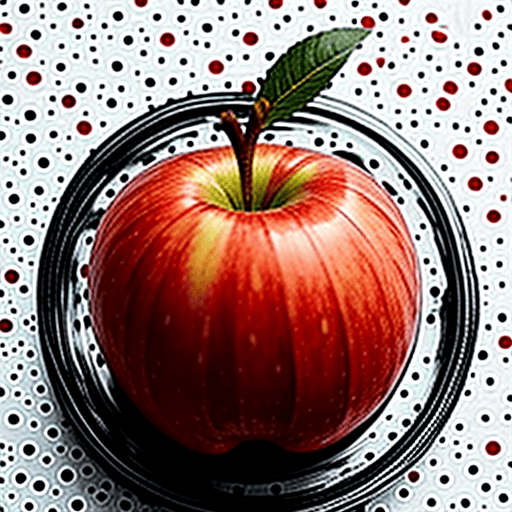}
& \includegraphics[width=0.140\linewidth]{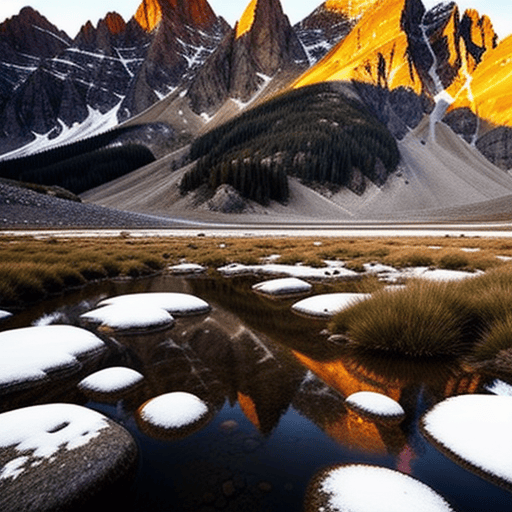}
& \includegraphics[width=0.140\linewidth]{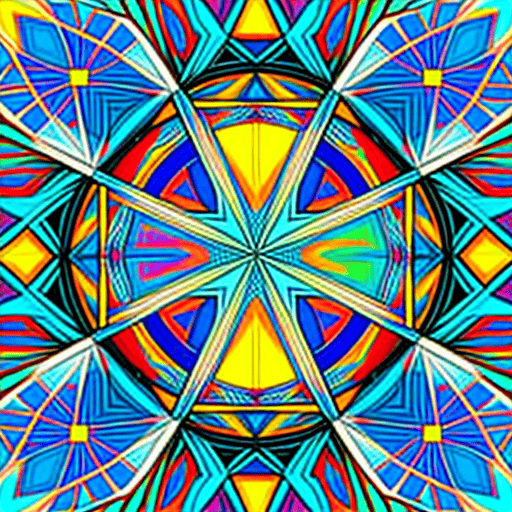}
& \includegraphics[width=0.140\linewidth]{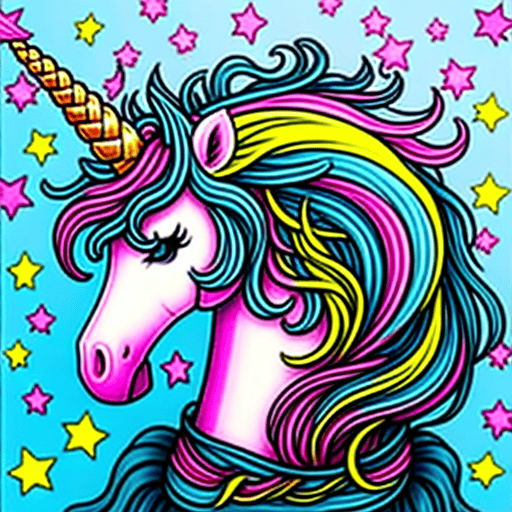}
& \includegraphics[width=0.140\linewidth]{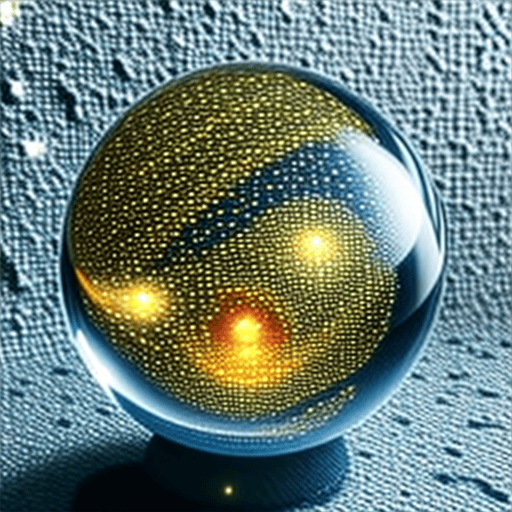}
\\

1
& \includegraphics[width=0.140\linewidth]{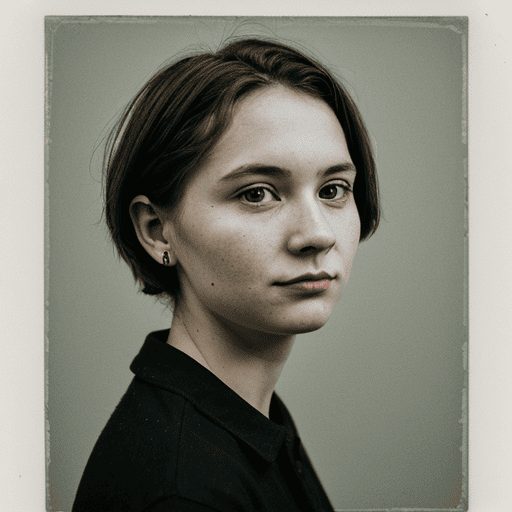}
& \includegraphics[width=0.140\linewidth]{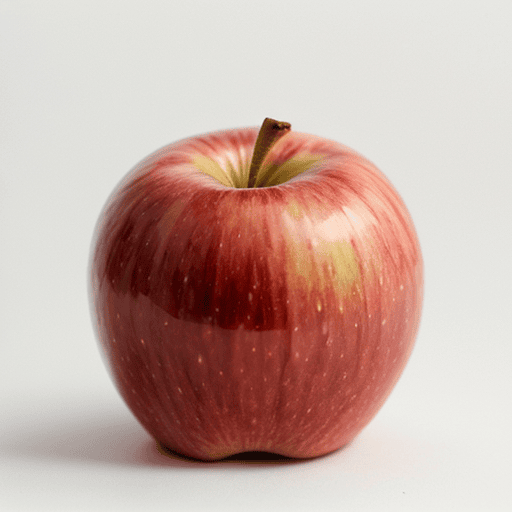}
& \includegraphics[width=0.140\linewidth]{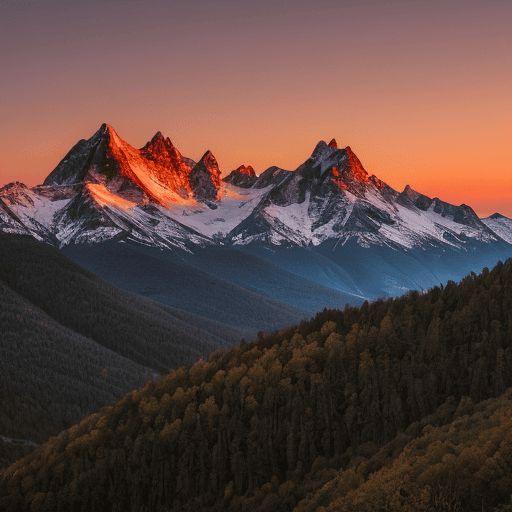}
& \includegraphics[width=0.140\linewidth]{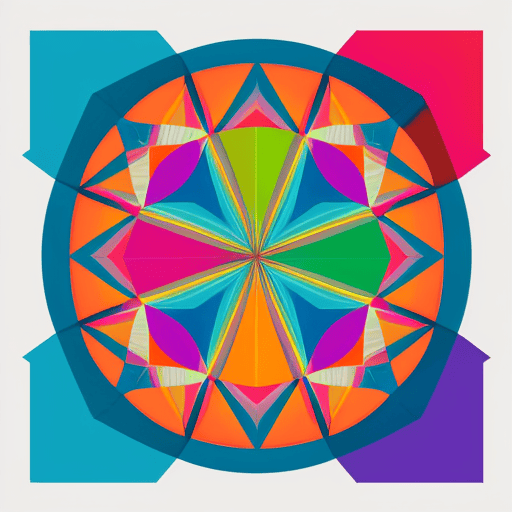}
& \includegraphics[width=0.140\linewidth]{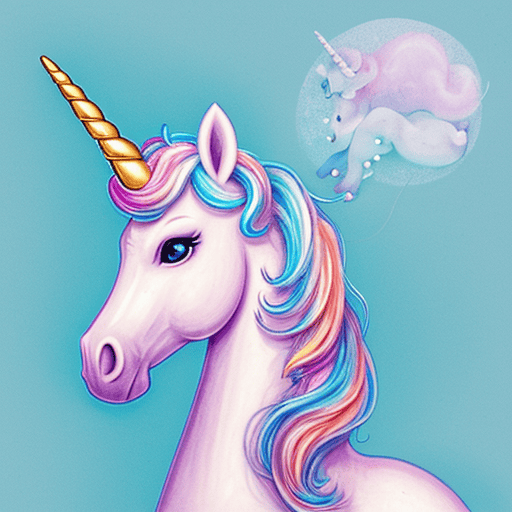}
& \includegraphics[width=0.140\linewidth]{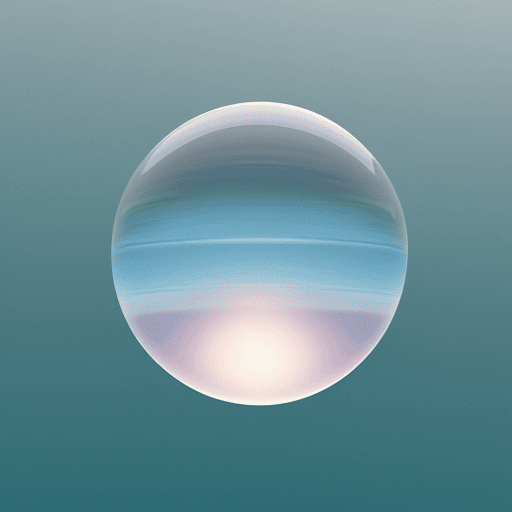}
\\

1.5
& \includegraphics[width=0.140\linewidth]{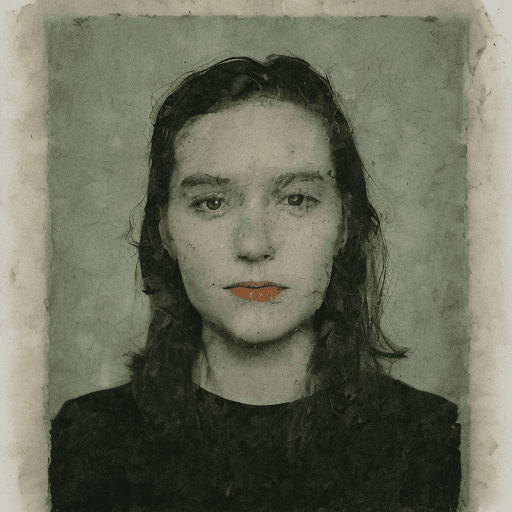}
& \includegraphics[width=0.140\linewidth]{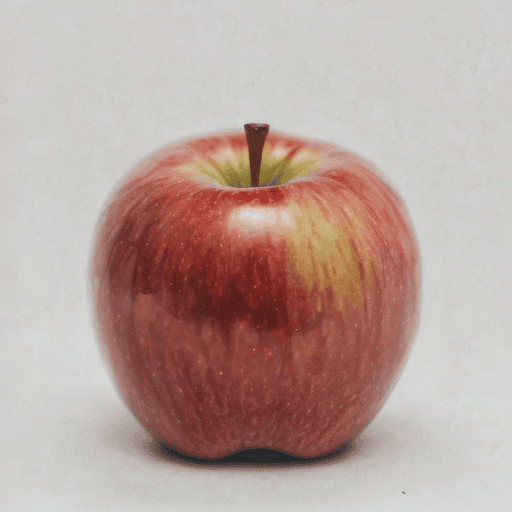}
& \includegraphics[width=0.140\linewidth]{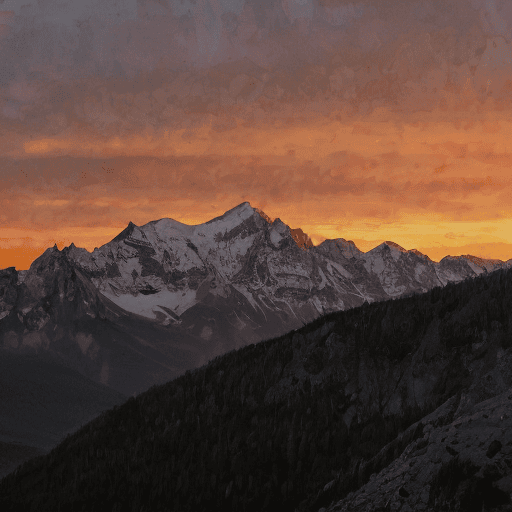}
& \includegraphics[width=0.140\linewidth]{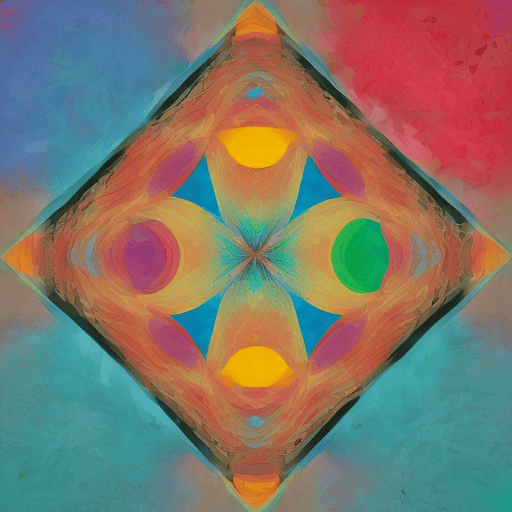}
& \includegraphics[width=0.140\linewidth]{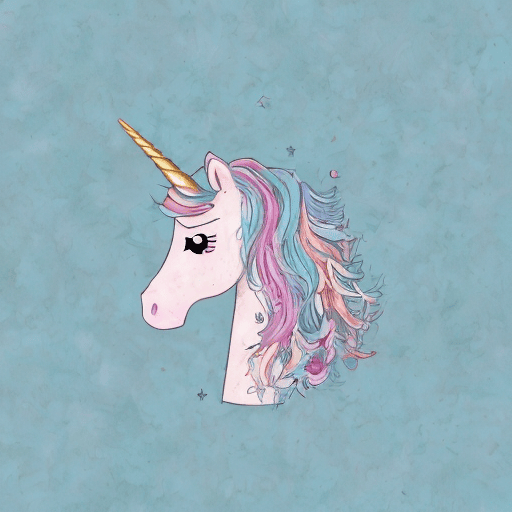}
& \includegraphics[width=0.140\linewidth]{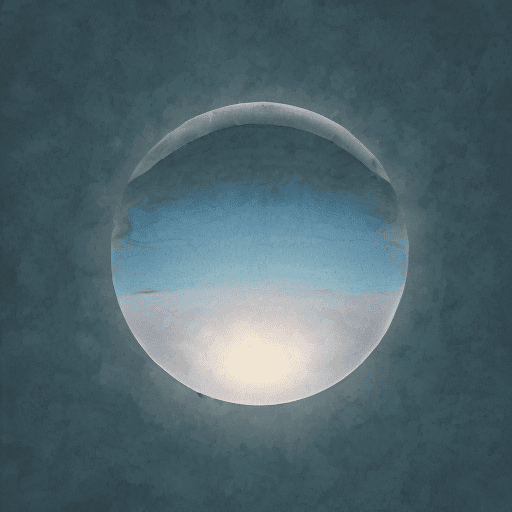}
\\

2.0
& \includegraphics[width=0.140\linewidth]{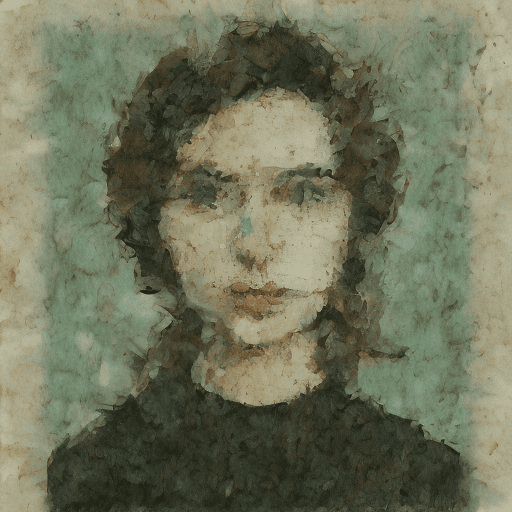}
& \includegraphics[width=0.140\linewidth]{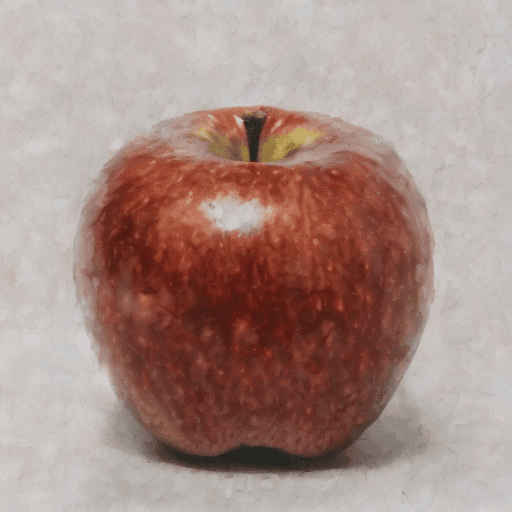}
& \includegraphics[width=0.140\linewidth]{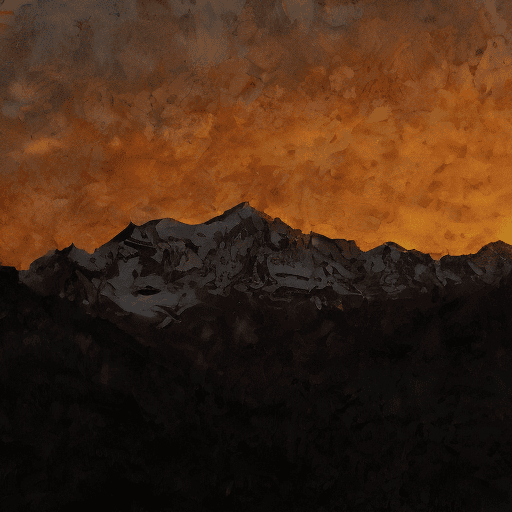}
& \includegraphics[width=0.140\linewidth]{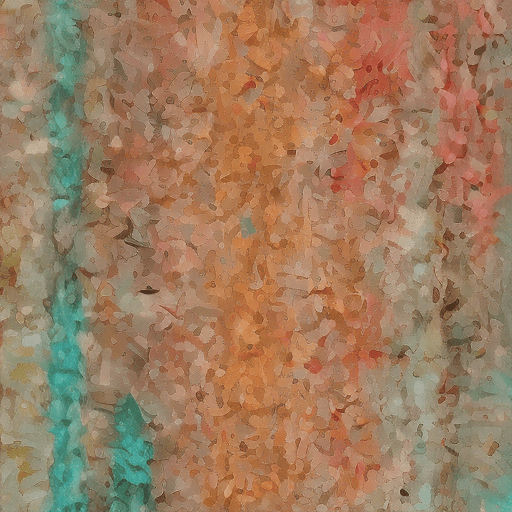}
& \includegraphics[width=0.140\linewidth]{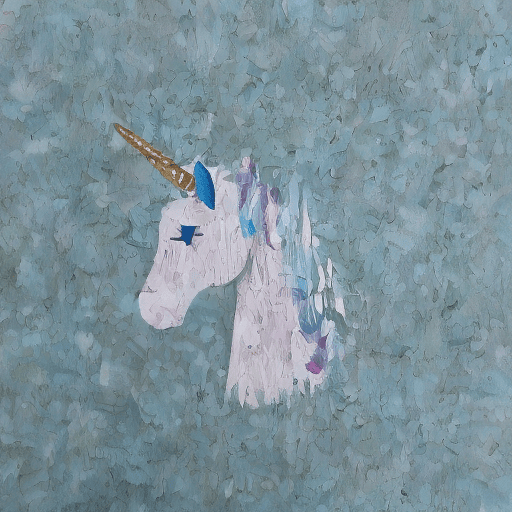}
& \includegraphics[width=0.140\linewidth]{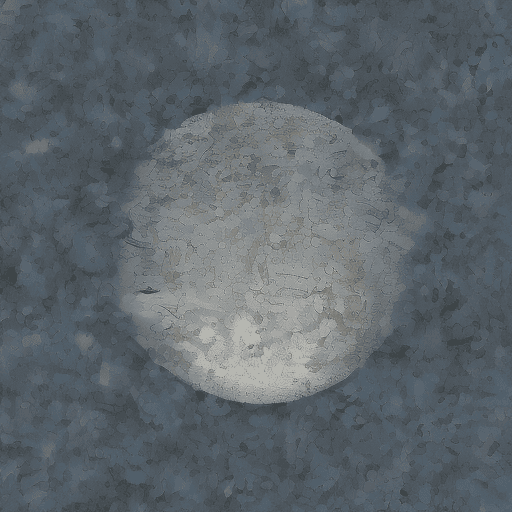}
\\
\hline
\end{tabular}

\caption{Results of bending the layer \textit{time\_embed.2}; Different prompts, same seed.}
\label{fig:multiple_prompts_1}

\end{figure}

\begin{figure}[h]
\centering

\medskip

\begin{tabular}{c|cccccc}
\hline
Multiply & Prompt 1 & Prompt 2 & Prompt 3 & Prompt 4 & Prompt 5 & Prompt 6 \\
\hline

0.0
& \includegraphics[width=0.140\linewidth]{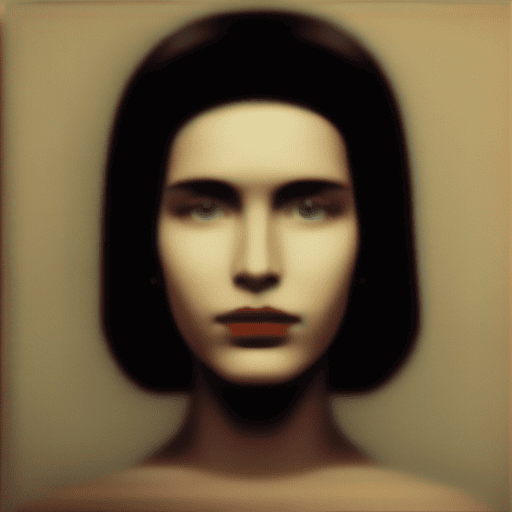}
& \includegraphics[width=0.140\linewidth]{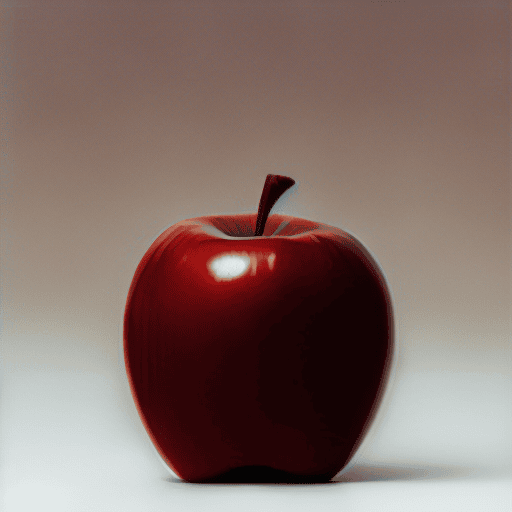}
& \includegraphics[width=0.140\linewidth]{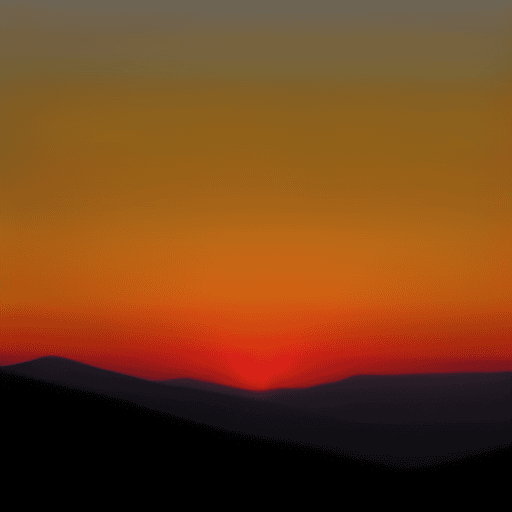}
& \includegraphics[width=0.140\linewidth]{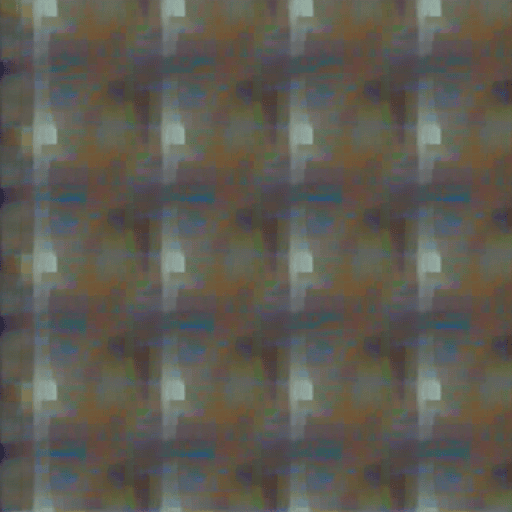}
& \includegraphics[width=0.140\linewidth]{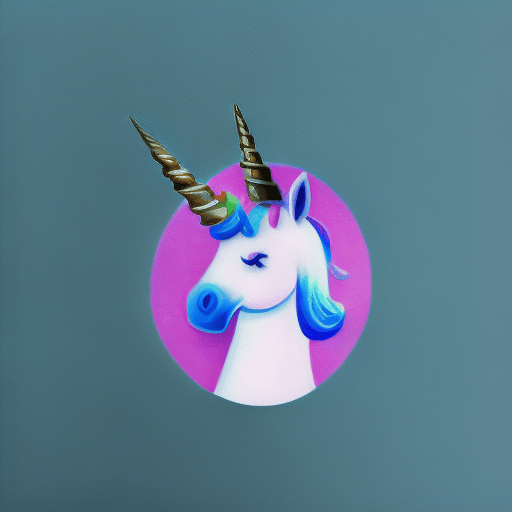}
& \includegraphics[width=0.140\linewidth]{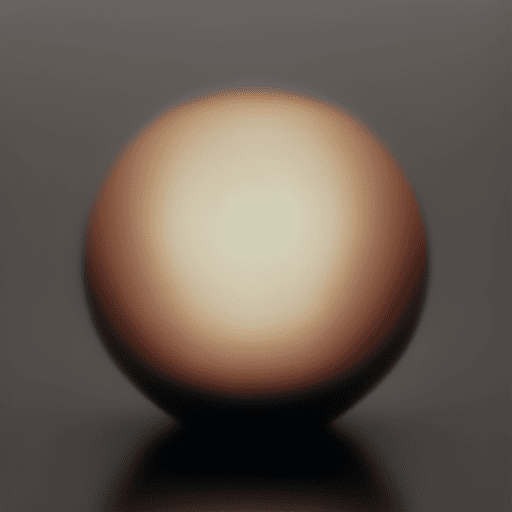}
\\

0.5
& \includegraphics[width=0.140\linewidth]{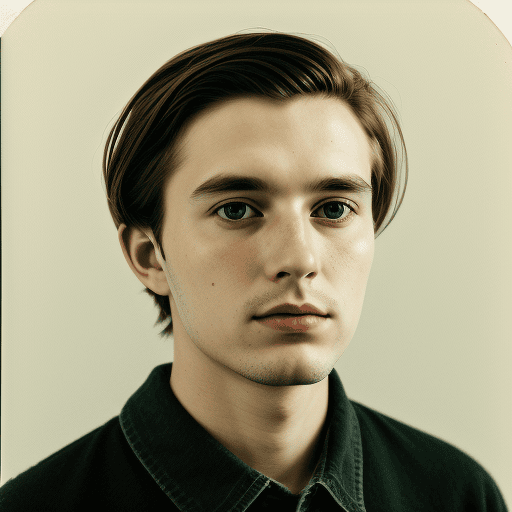}
& \includegraphics[width=0.140\linewidth]{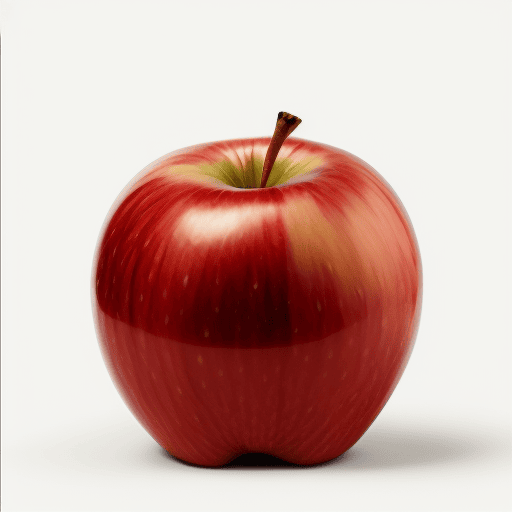}
& \includegraphics[width=0.140\linewidth]{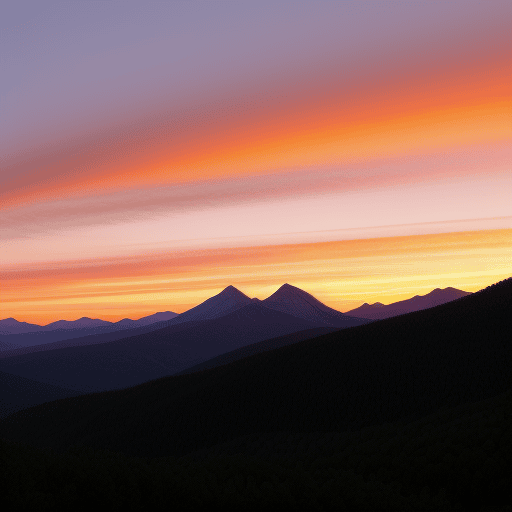}
& \includegraphics[width=0.140\linewidth]{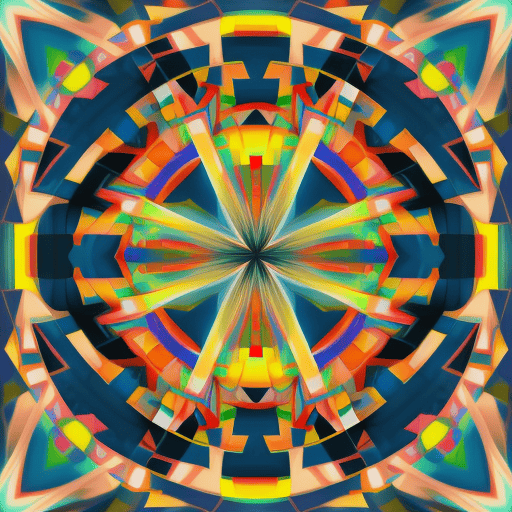}
& \includegraphics[width=0.140\linewidth]{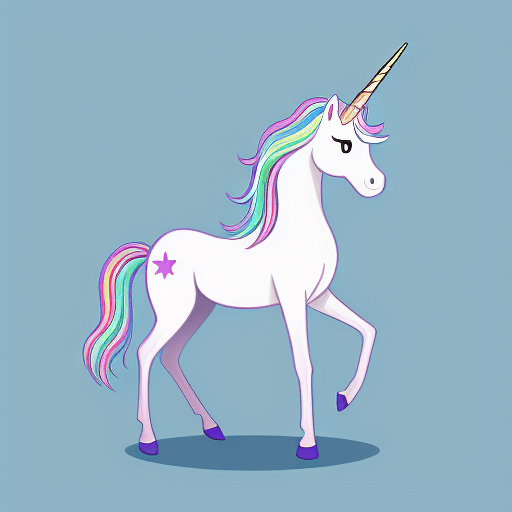}
& \includegraphics[width=0.140\linewidth]{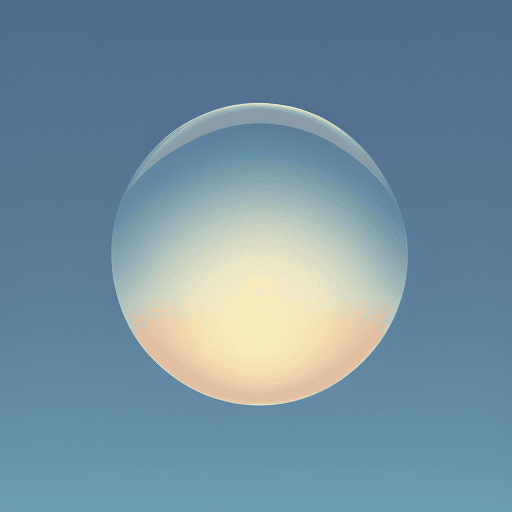}
\\

default
& \includegraphics[width=0.140\linewidth]{figs/multiple_prompts/default_0.png}
& \includegraphics[width=0.140\linewidth]{figs/multiple_prompts/default_1.png}
& \includegraphics[width=0.140\linewidth]{figs/multiple_prompts/default_3.png}
& \includegraphics[width=0.140\linewidth]{figs/multiple_prompts/default_4.png}
& \includegraphics[width=0.140\linewidth]{figs/multiple_prompts/default_5.png}
& \includegraphics[width=0.140\linewidth]{figs/multiple_prompts/default_6.png}
\\

1.5
& \includegraphics[width=0.140\linewidth]{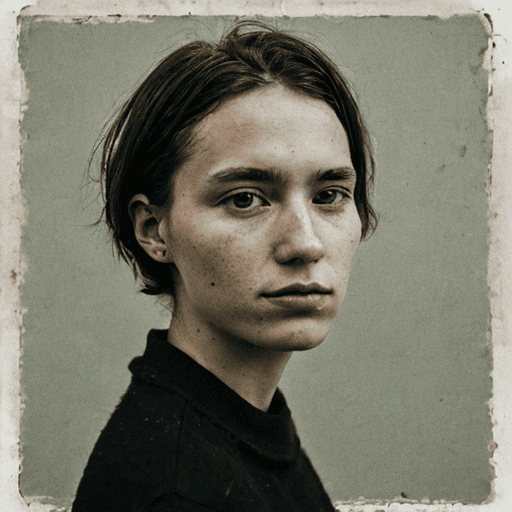}
& \includegraphics[width=0.140\linewidth]{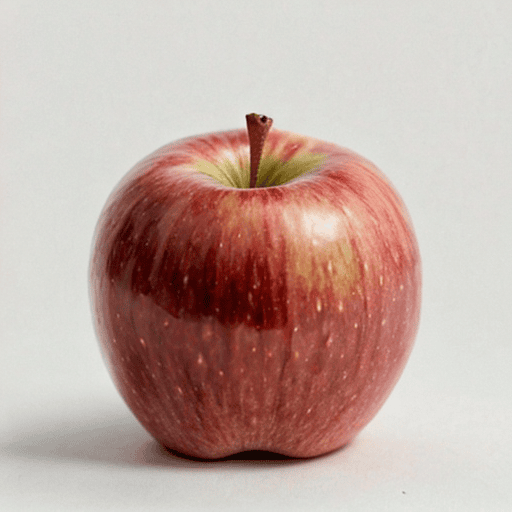}
& \includegraphics[width=0.140\linewidth]{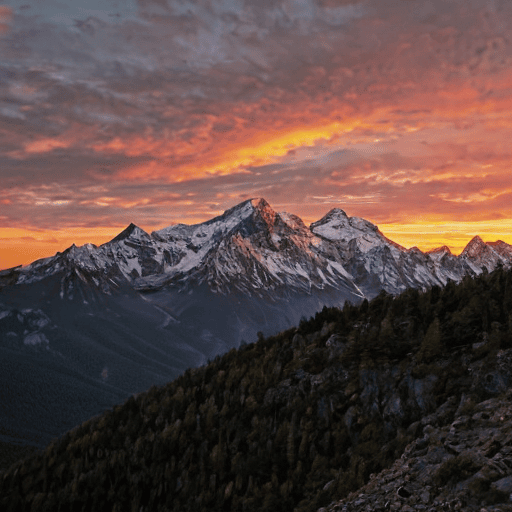}
& \includegraphics[width=0.140\linewidth]{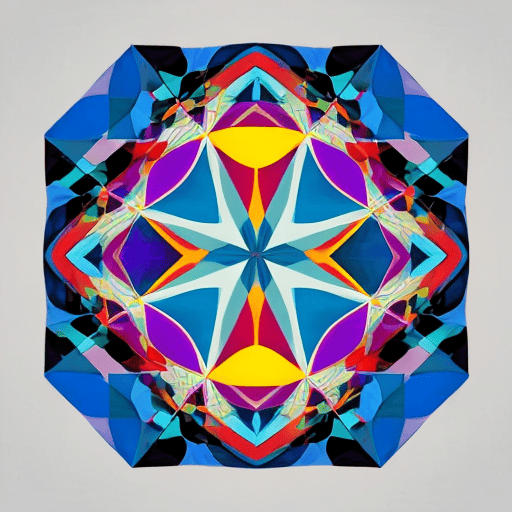}
& \includegraphics[width=0.140\linewidth]{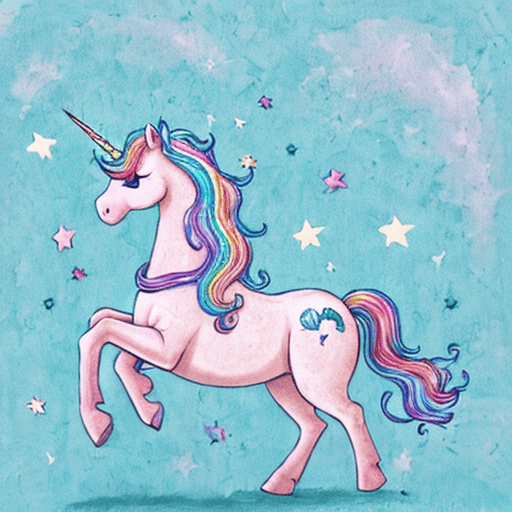}
& \includegraphics[width=0.140\linewidth]{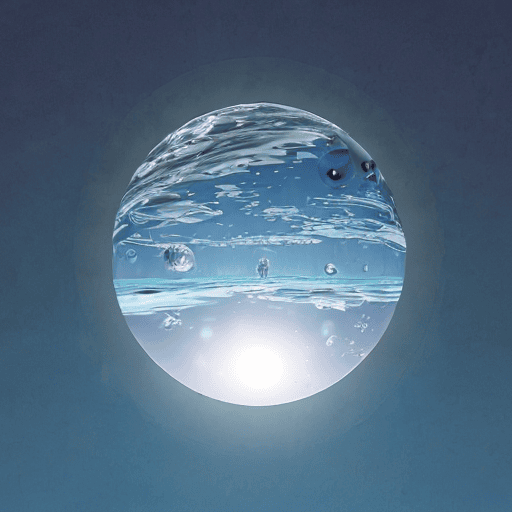}
\\

2.0
& \includegraphics[width=0.140\linewidth]{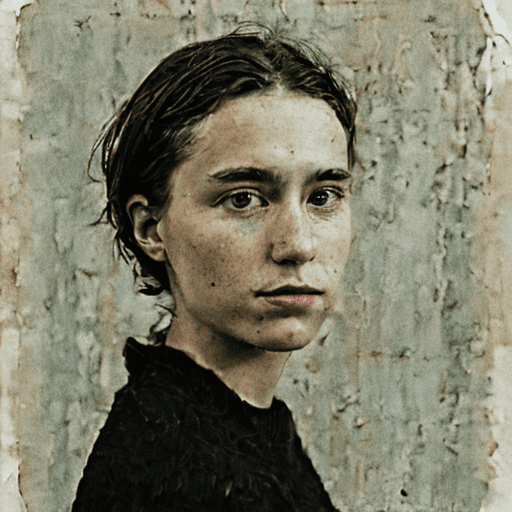}
& \includegraphics[width=0.140\linewidth]{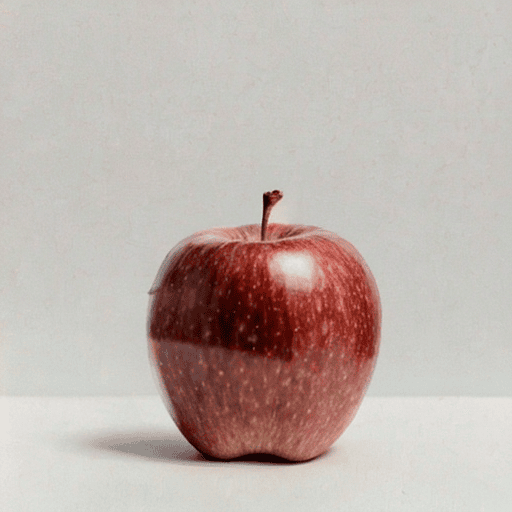}
& \includegraphics[width=0.140\linewidth]{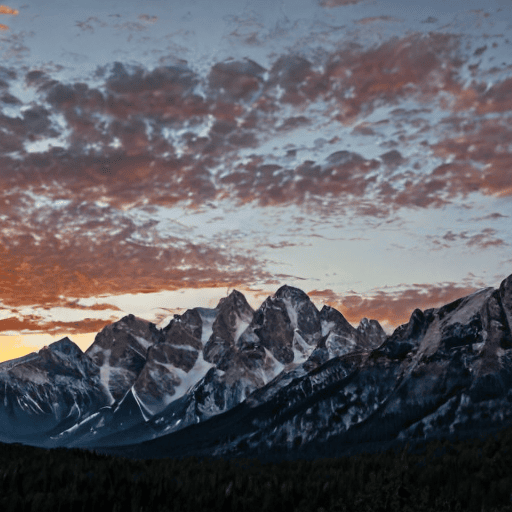}
& \includegraphics[width=0.140\linewidth]{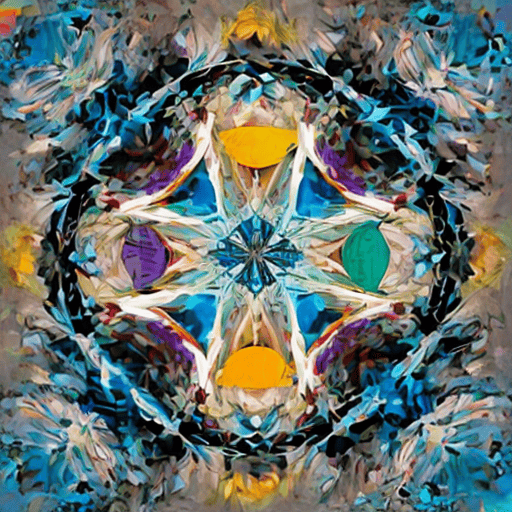}
& \includegraphics[width=0.140\linewidth]{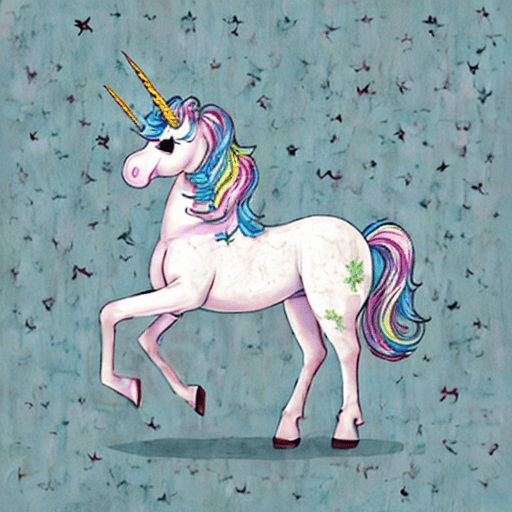}
& \includegraphics[width=0.140\linewidth]{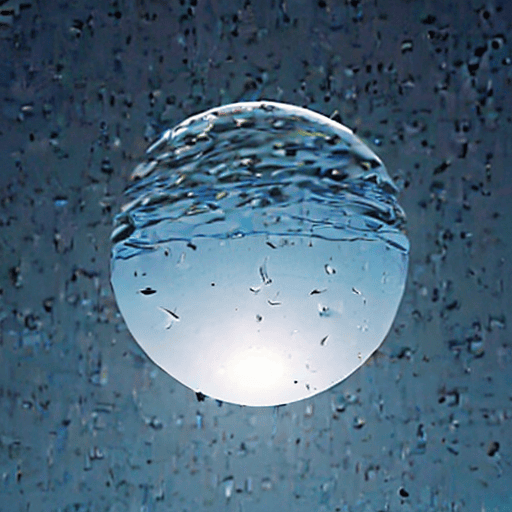}
\\
\hline
\end{tabular}

\caption{Results of bending the layer \textit{input\_blocks.1.0.out\_layers.2}; Different prompts, same seed.}
\label{fig:multiple_prompts_2}
\end{figure}

\begin{figure}[H]
\centering

\medskip

\begin{tabular}{c|cccccc}
\hline
Multiply & Prompt 1 & Prompt 2 & Prompt 3 & Prompt 4 & Prompt 5 & Prompt 6 \\
\hline

0.0
& \includegraphics[width=0.140\linewidth]{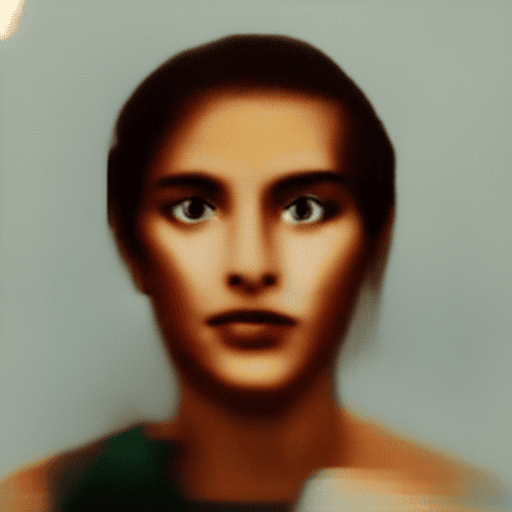}
& \includegraphics[width=0.140\linewidth]{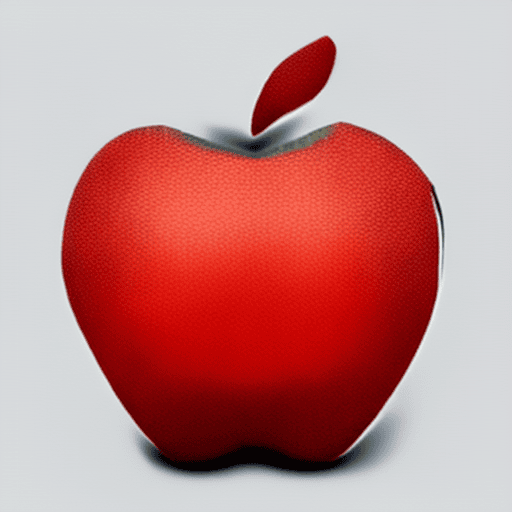}
& \includegraphics[width=0.140\linewidth]{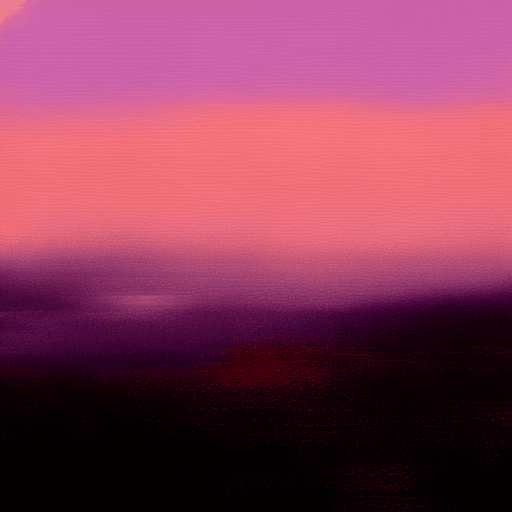}
& \includegraphics[width=0.140\linewidth]{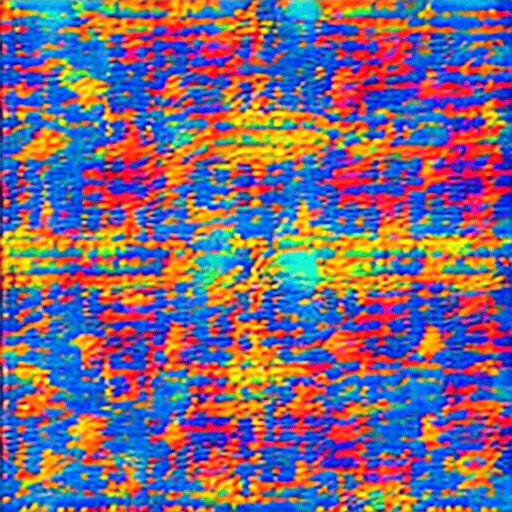}
& \includegraphics[width=0.140\linewidth]{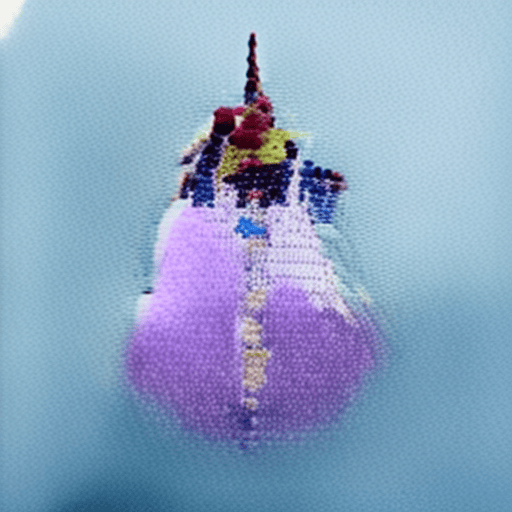}
& \includegraphics[width=0.140\linewidth]{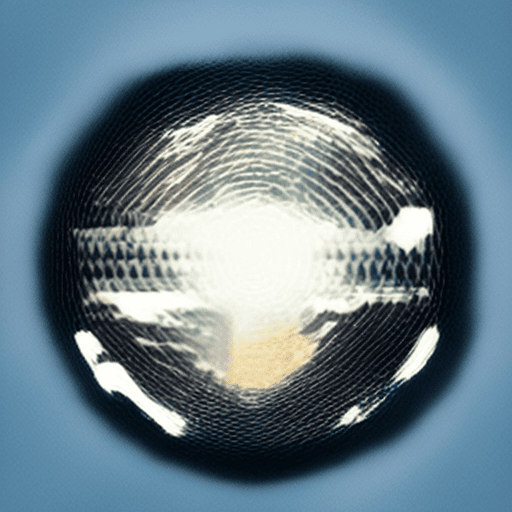}
\\

0.5
& \includegraphics[width=0.140\linewidth]{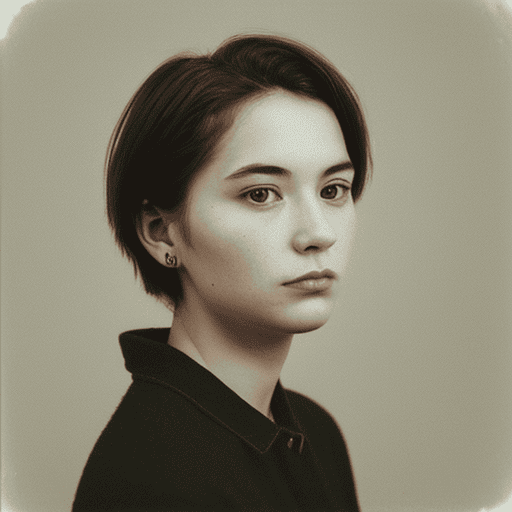}
& \includegraphics[width=0.140\linewidth]{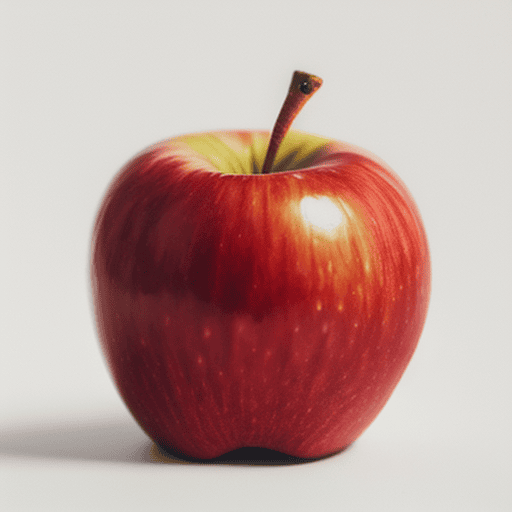}
& \includegraphics[width=0.140\linewidth]{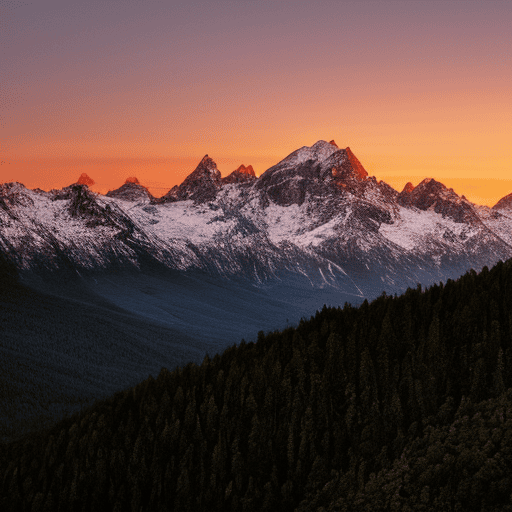}
& \includegraphics[width=0.140\linewidth]{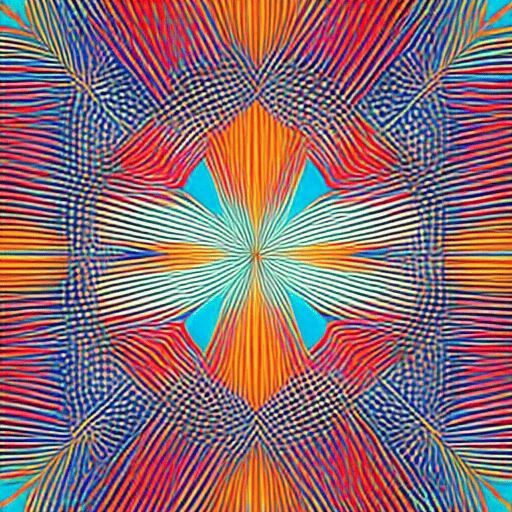}
& \includegraphics[width=0.140\linewidth]{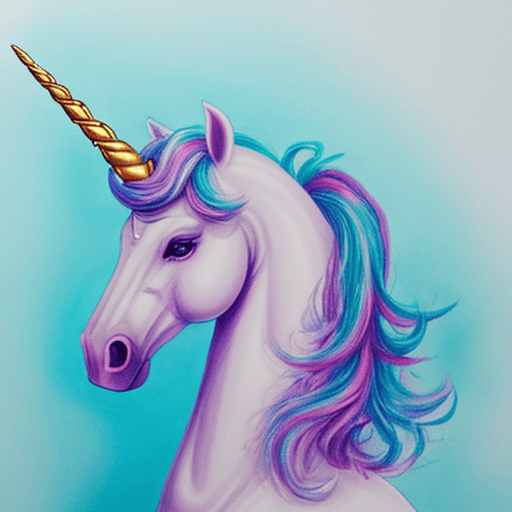}
& \includegraphics[width=0.140\linewidth]{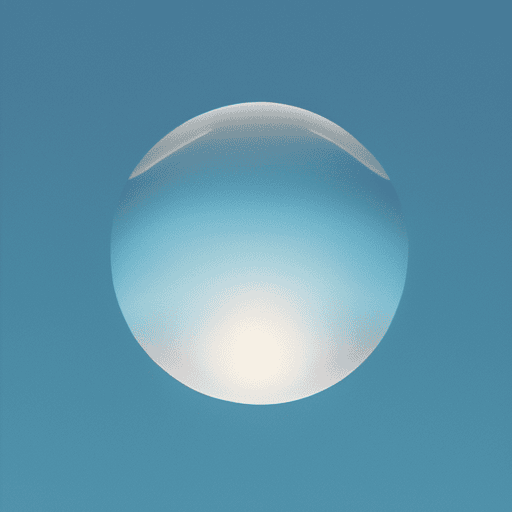}
\\

default
& \includegraphics[width=0.140\linewidth]{figs/multiple_prompts/default_0.png}
& \includegraphics[width=0.140\linewidth]{figs/multiple_prompts/default_1.png}
& \includegraphics[width=0.140\linewidth]{figs/multiple_prompts/default_3.png}
& \includegraphics[width=0.140\linewidth]{figs/multiple_prompts/default_4.png}
& \includegraphics[width=0.140\linewidth]{figs/multiple_prompts/default_5.png}
& \includegraphics[width=0.140\linewidth]{figs/multiple_prompts/default_6.png}
\\

1.5
& \includegraphics[width=0.140\linewidth]{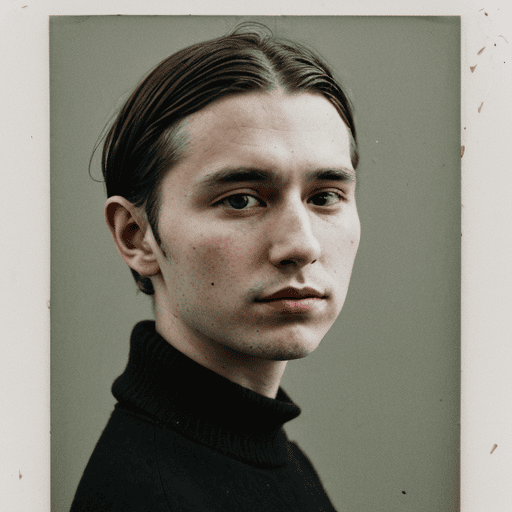}
& \includegraphics[width=0.140\linewidth]{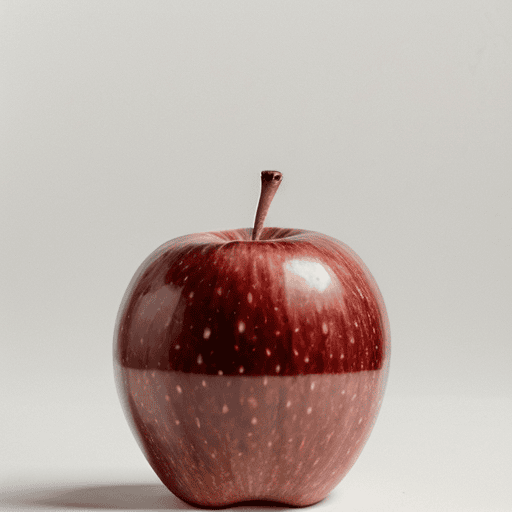}
& \includegraphics[width=0.140\linewidth]{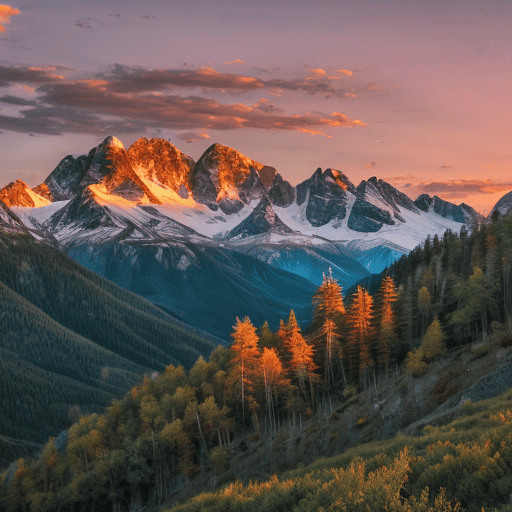}
& \includegraphics[width=0.140\linewidth]{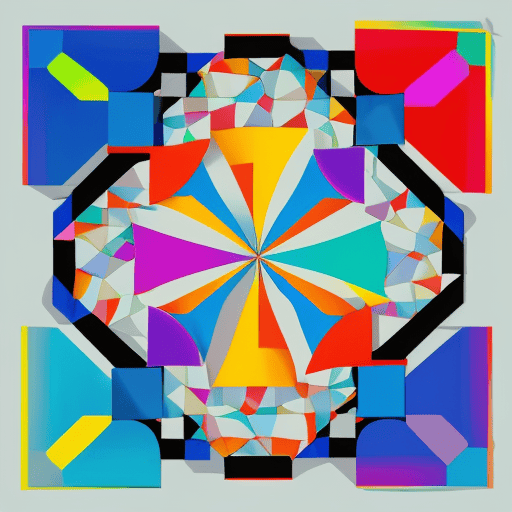}
& \includegraphics[width=0.140\linewidth]{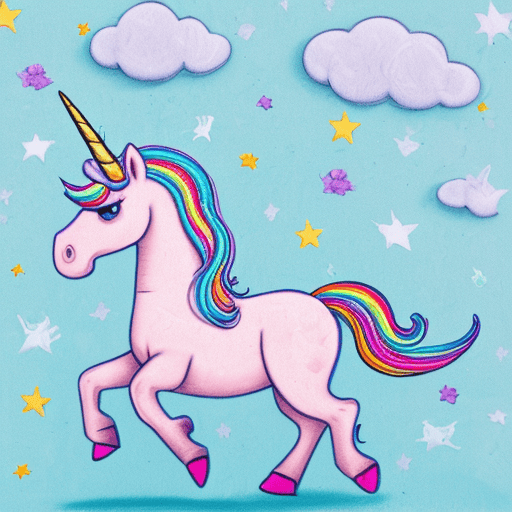}
& \includegraphics[width=0.140\linewidth]{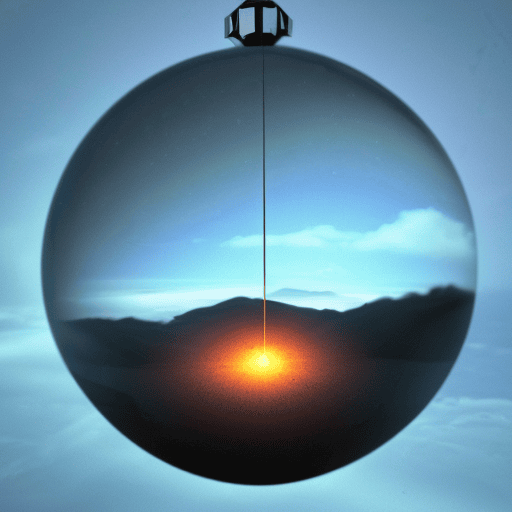}
\\

2.0
& \includegraphics[width=0.140\linewidth]{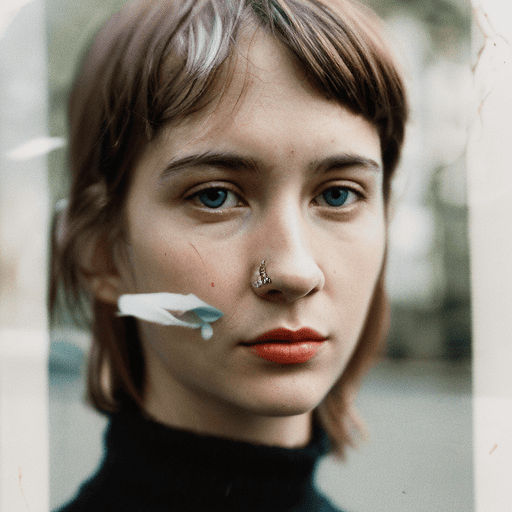}
& \includegraphics[width=0.140\linewidth]{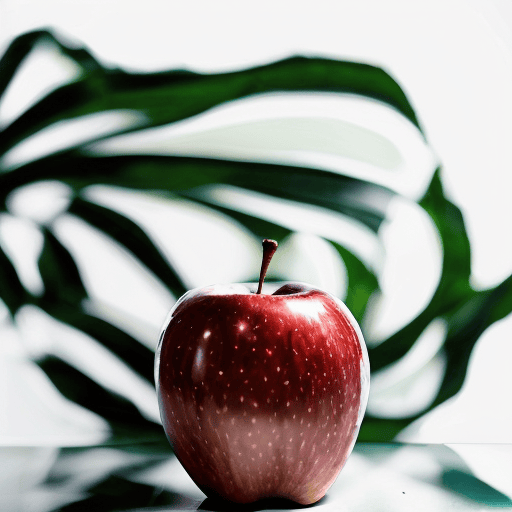}
& \includegraphics[width=0.140\linewidth]{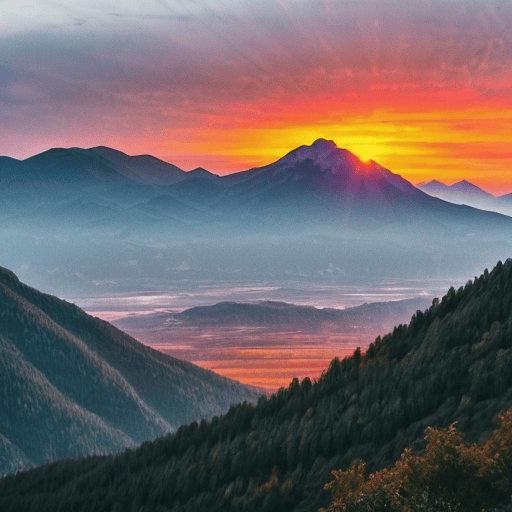}
& \includegraphics[width=0.140\linewidth]{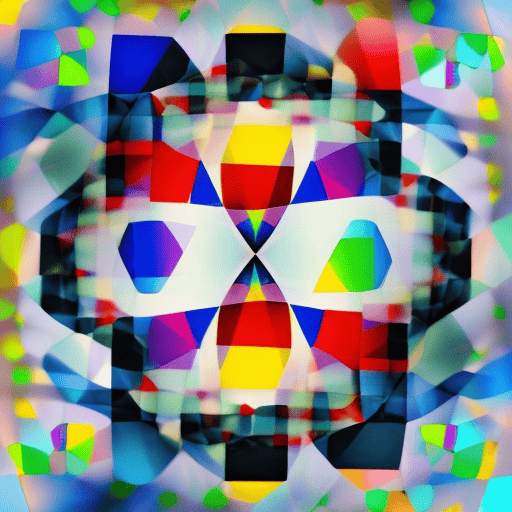}
& \includegraphics[width=0.140\linewidth]{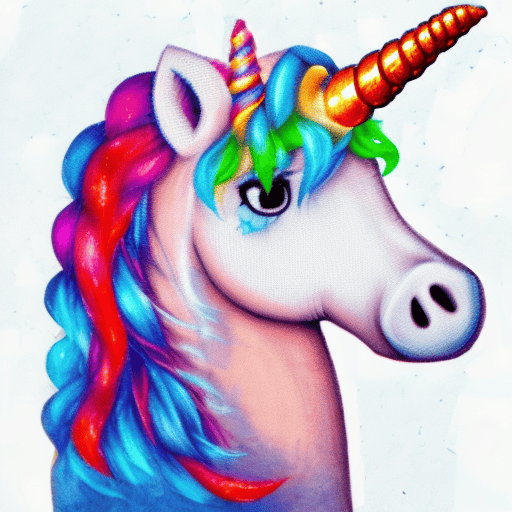}
& \includegraphics[width=0.140\linewidth]{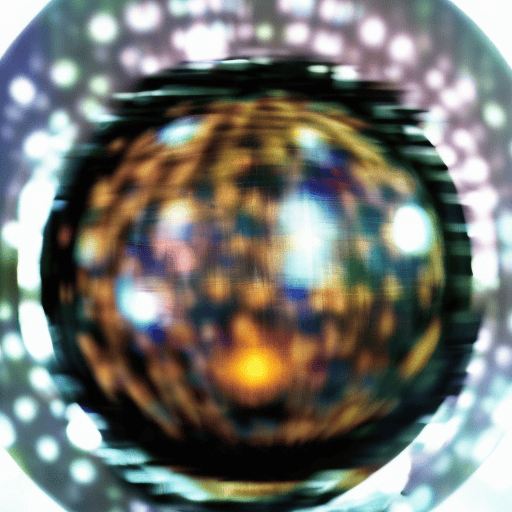}
\\
\hline
\end{tabular}

\caption{Results of bending the layer \textit{input\_blocks.4.0.skip\_connection}; Different prompts, same seed.}
\label{fig:multiple_prompts_3}
\end{figure}

\begin{figure}[H]
\centering

\medskip

\begin{tabular}{c|cccccc}
\hline
Multiply & Prompt 1 & Prompt 2 & Prompt 3 & Prompt 4 & Prompt 5 & Prompt 6 \\
\hline

0.0
& \includegraphics[width=0.140\linewidth]{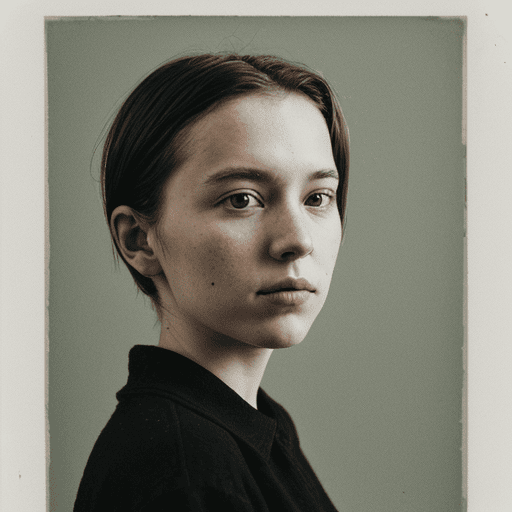}
& \includegraphics[width=0.140\linewidth]{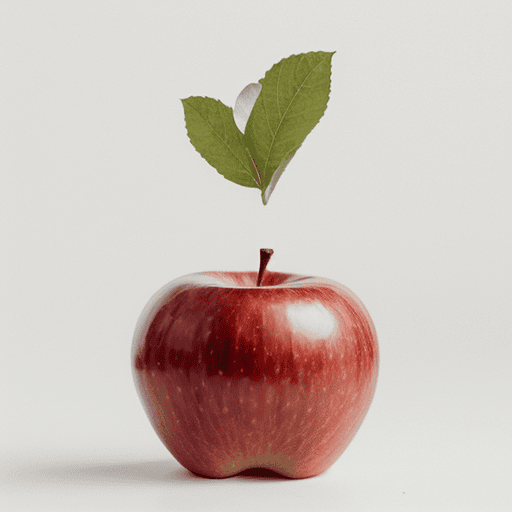}
& \includegraphics[width=0.140\linewidth]{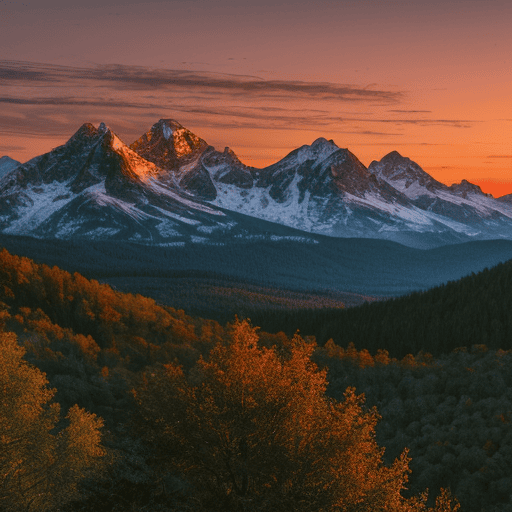}
& \includegraphics[width=0.140\linewidth]{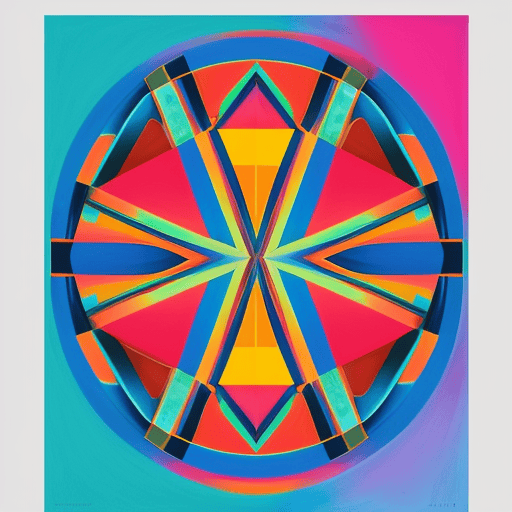}
& \includegraphics[width=0.140\linewidth]{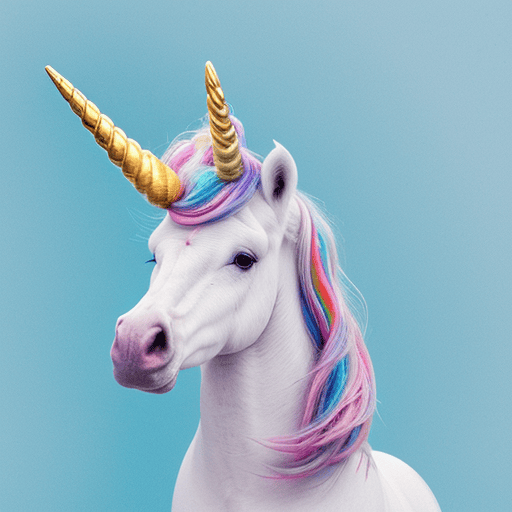}
& \includegraphics[width=0.140\linewidth]{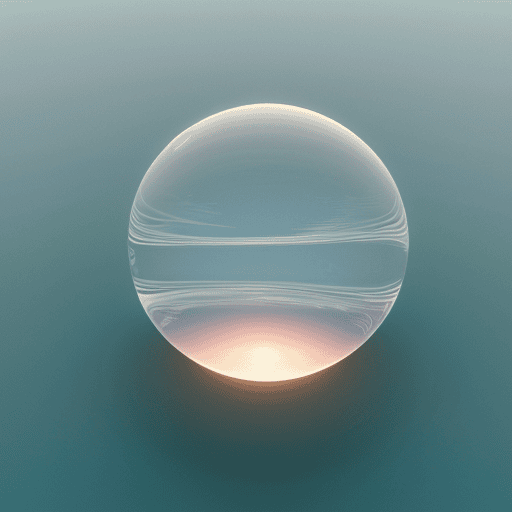}
\\

0.5
& \includegraphics[width=0.140\linewidth]{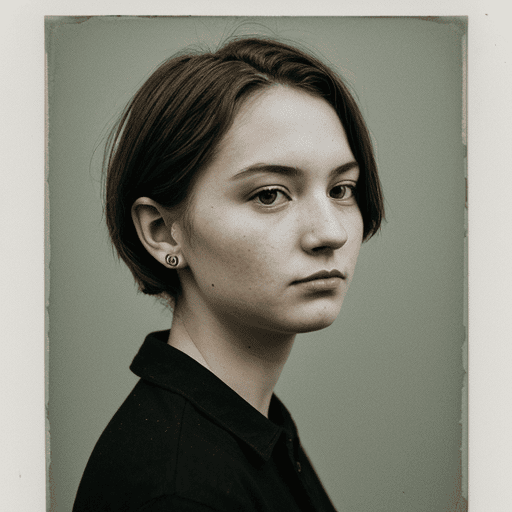}
& \includegraphics[width=0.140\linewidth]{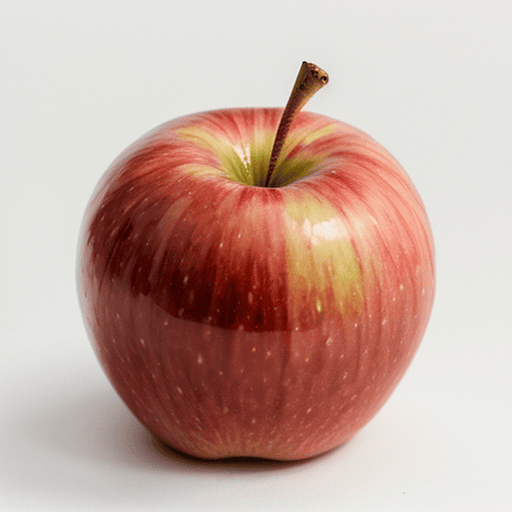}
& \includegraphics[width=0.140\linewidth]{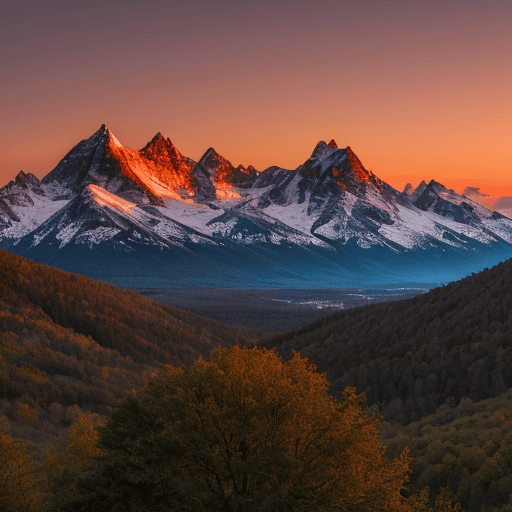}
& \includegraphics[width=0.140\linewidth]{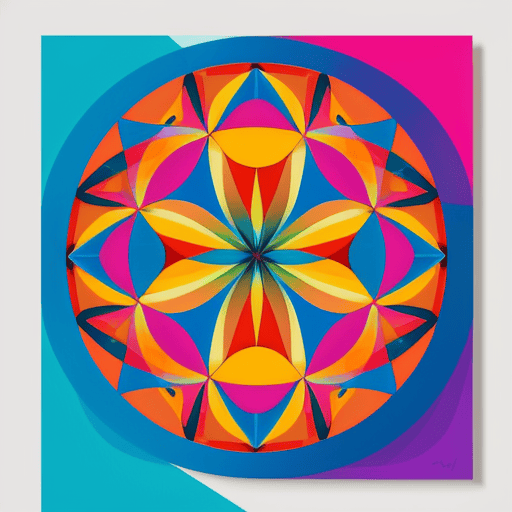}
& \includegraphics[width=0.140\linewidth]{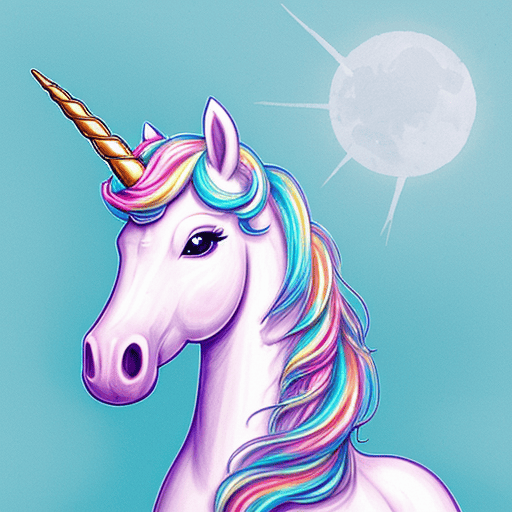}
& \includegraphics[width=0.140\linewidth]{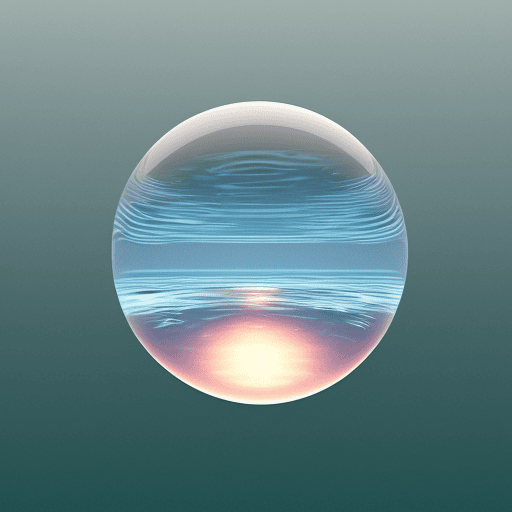}
\\

default
& \includegraphics[width=0.140\linewidth]{figs/multiple_prompts/default_0.png}
& \includegraphics[width=0.140\linewidth]{figs/multiple_prompts/default_1.png}
& \includegraphics[width=0.140\linewidth]{figs/multiple_prompts/default_3.png}
& \includegraphics[width=0.140\linewidth]{figs/multiple_prompts/default_4.png}
& \includegraphics[width=0.140\linewidth]{figs/multiple_prompts/default_5.png}
& \includegraphics[width=0.140\linewidth]{figs/multiple_prompts/default_6.png}
\\

1.5
& \includegraphics[width=0.140\linewidth]{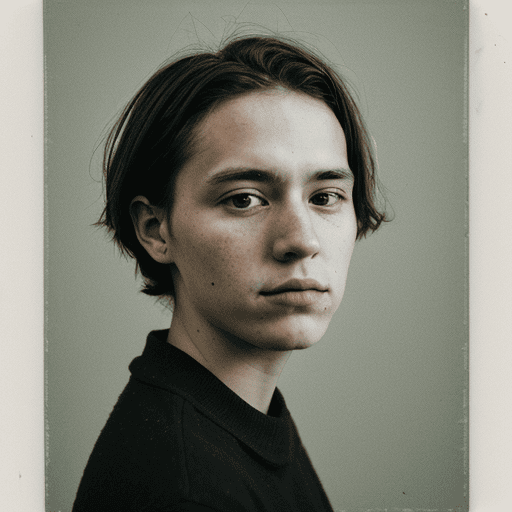}
& \includegraphics[width=0.140\linewidth]{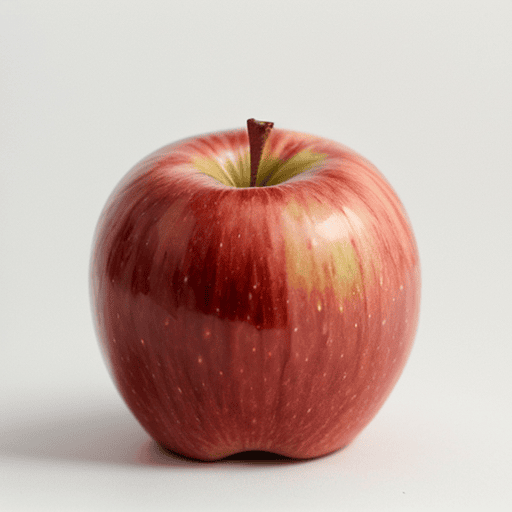}
& \includegraphics[width=0.140\linewidth]{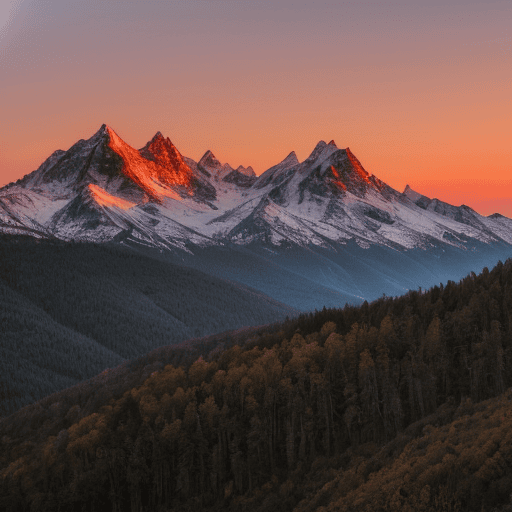}
& \includegraphics[width=0.140\linewidth]{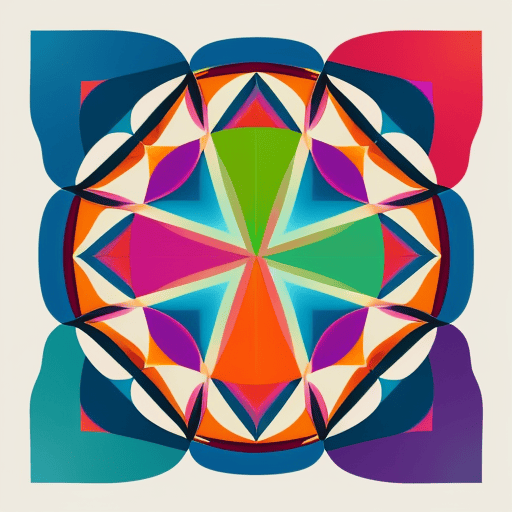}
& \includegraphics[width=0.140\linewidth]{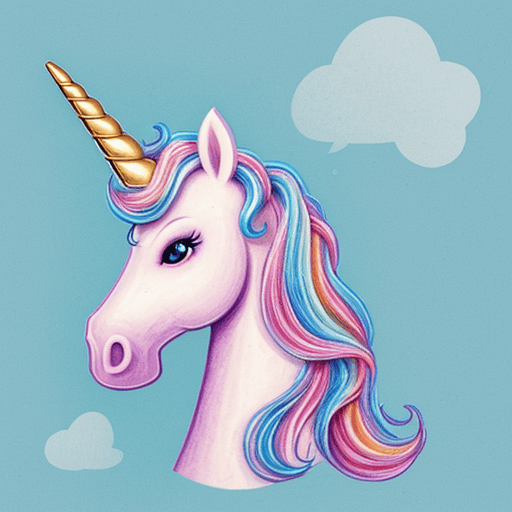}
& \includegraphics[width=0.140\linewidth]{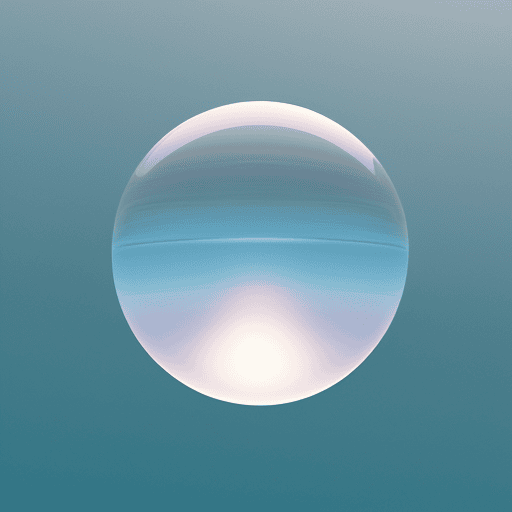}
\\

2.0
& \includegraphics[width=0.140\linewidth]{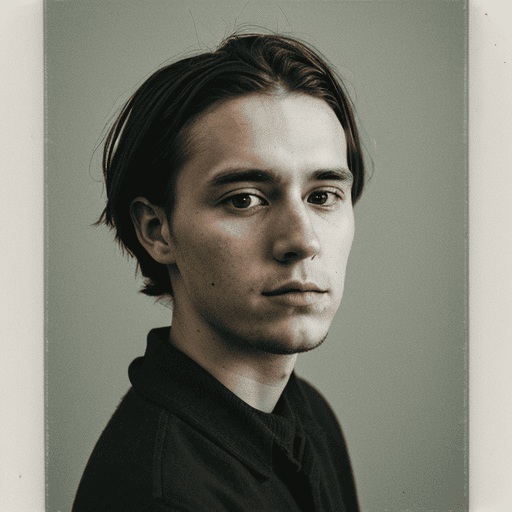}
& \includegraphics[width=0.140\linewidth]{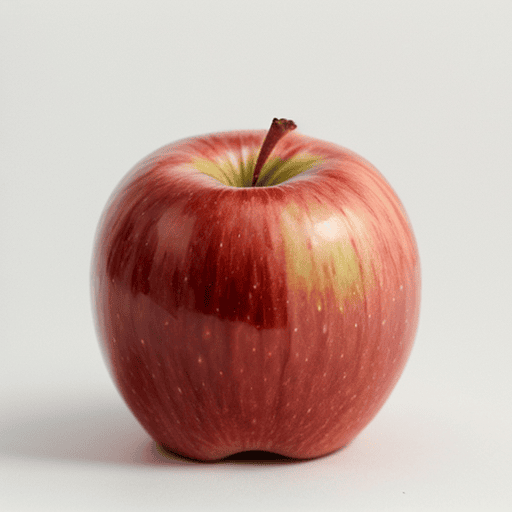}
& \includegraphics[width=0.140\linewidth]{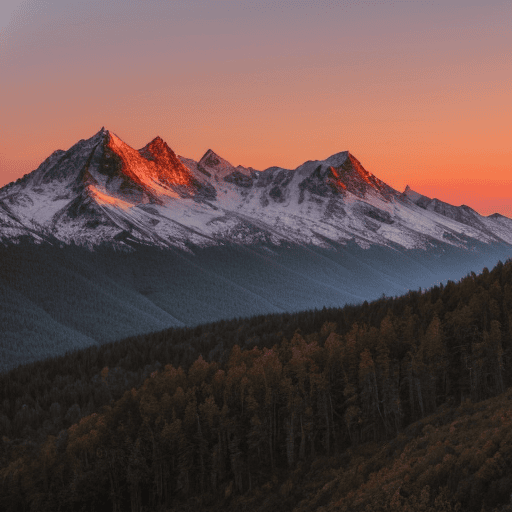}
& \includegraphics[width=0.140\linewidth]{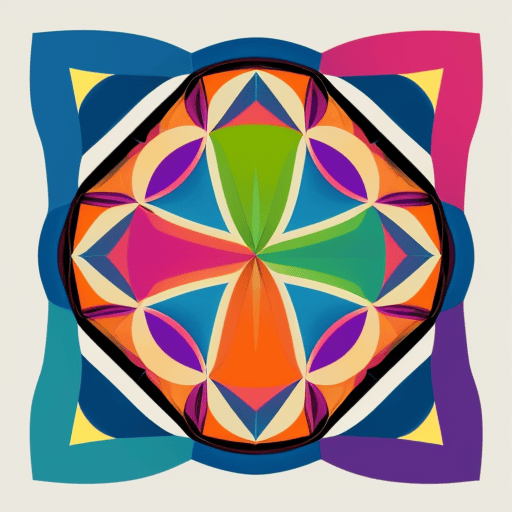}
& \includegraphics[width=0.140\linewidth]{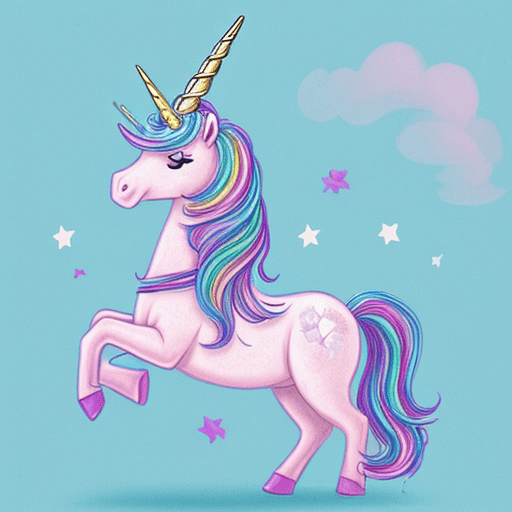}
& \includegraphics[width=0.140\linewidth]{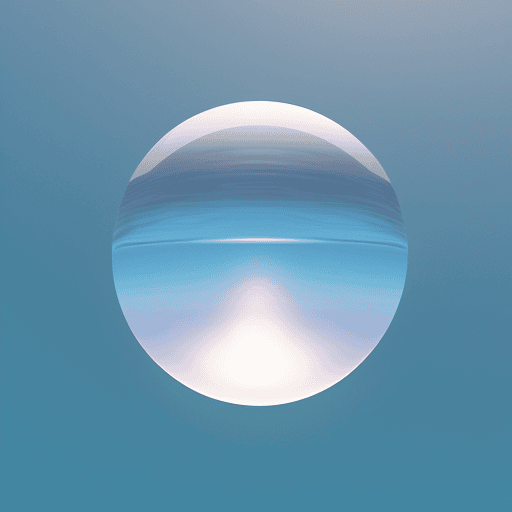}
\\
\hline
\end{tabular}

\caption{Bending \textit{middle\_block.1.transformer\_blocks.0.attn2.to\_out.0};\\ Different prompts, same seed.}
\label{fig:multiple_prompts_4}


\medskip

\begin{tabular}{c|cccccc}
\hline
Multiply & Prompt 1 & Prompt 2 & Prompt 3 & Prompt 4 & Prompt 5 & Prompt 6 \\
\hline

0.0
& \includegraphics[width=0.140\linewidth]{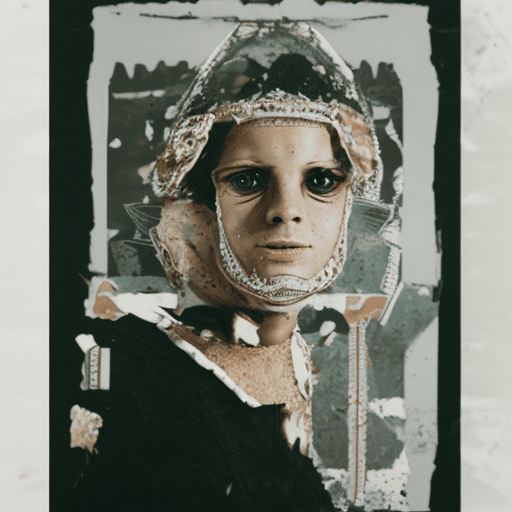}
& \includegraphics[width=0.140\linewidth]{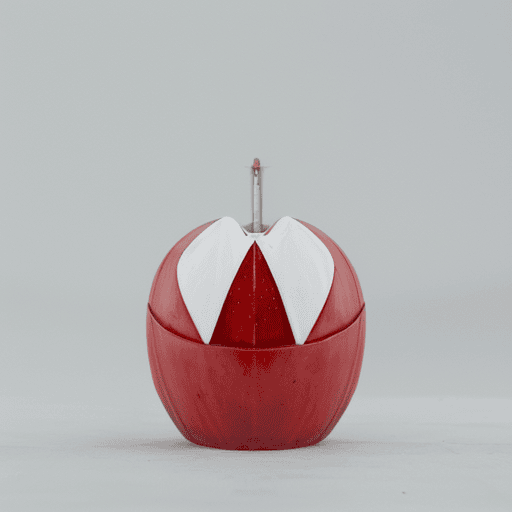}
& \includegraphics[width=0.140\linewidth]{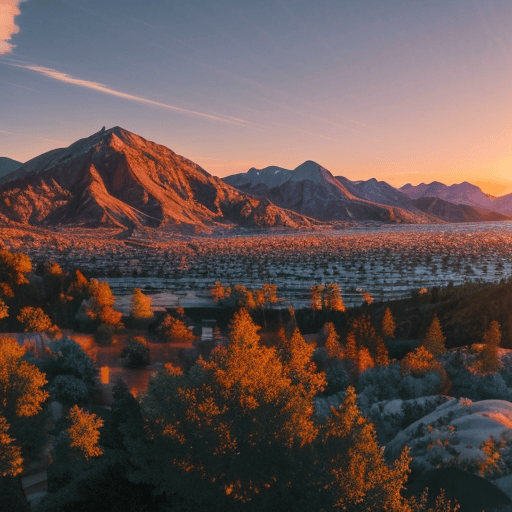}
& \includegraphics[width=0.140\linewidth]{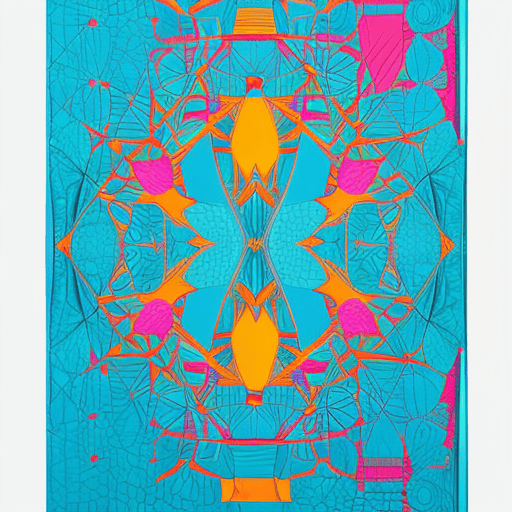}
& \includegraphics[width=0.140\linewidth]{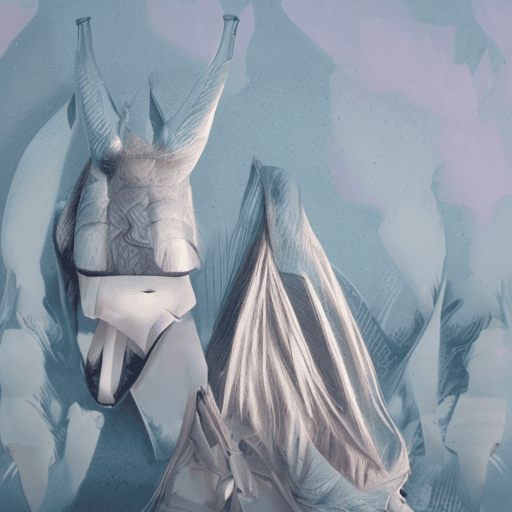}
& \includegraphics[width=0.140\linewidth]{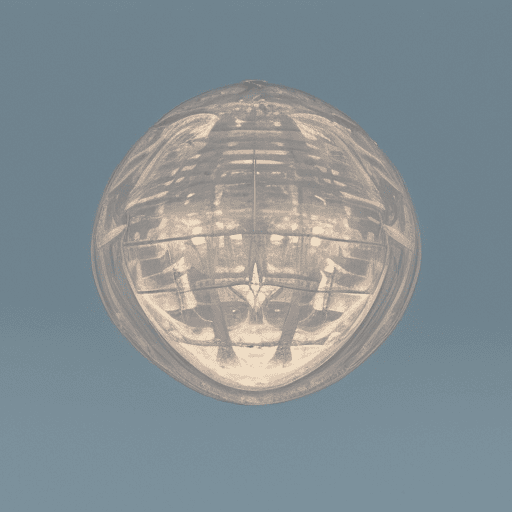}
\\

0.5
& \includegraphics[width=0.140\linewidth]{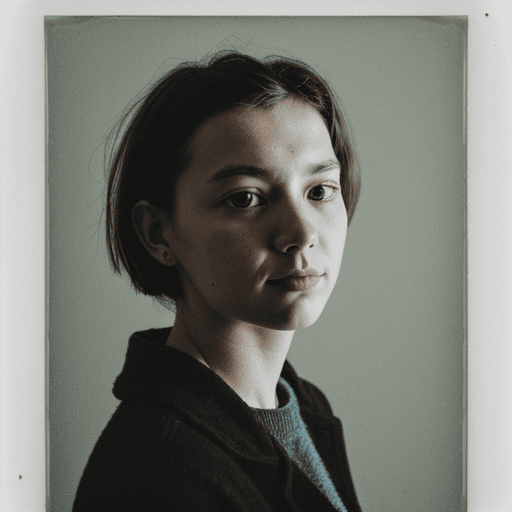}
& \includegraphics[width=0.140\linewidth]{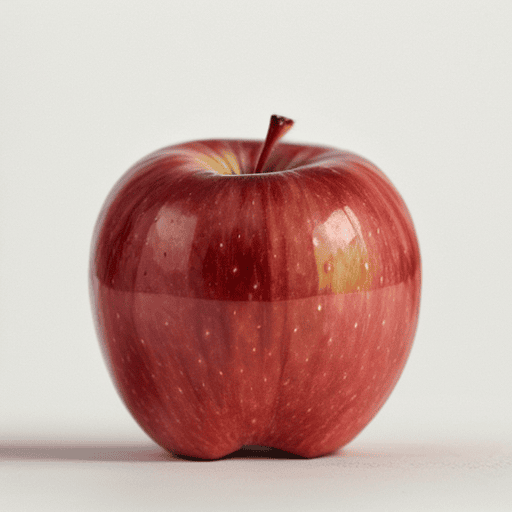}
& \includegraphics[width=0.140\linewidth]{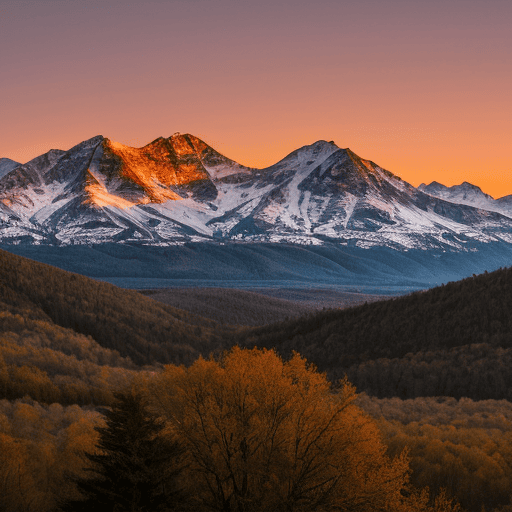}
& \includegraphics[width=0.140\linewidth]{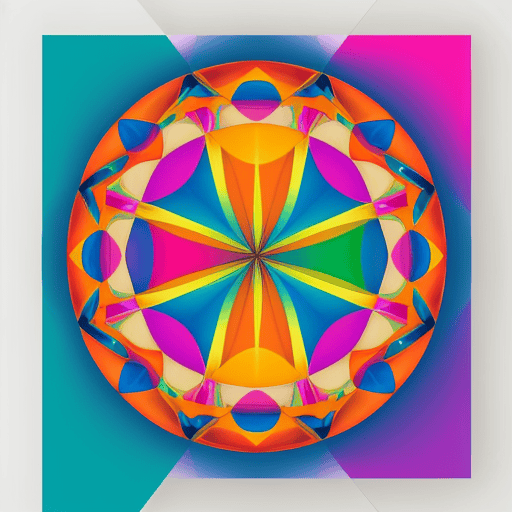}
& \includegraphics[width=0.140\linewidth]{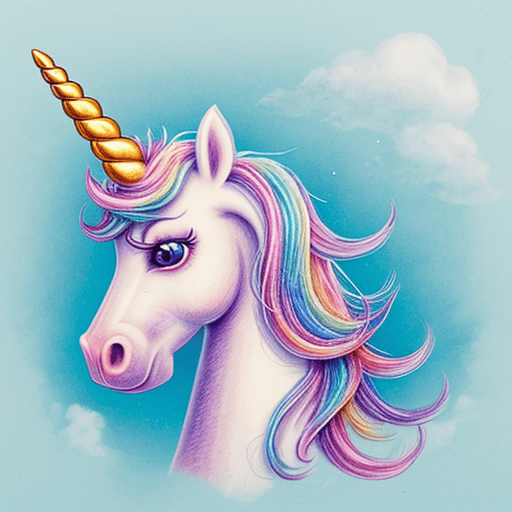}
& \includegraphics[width=0.140\linewidth]{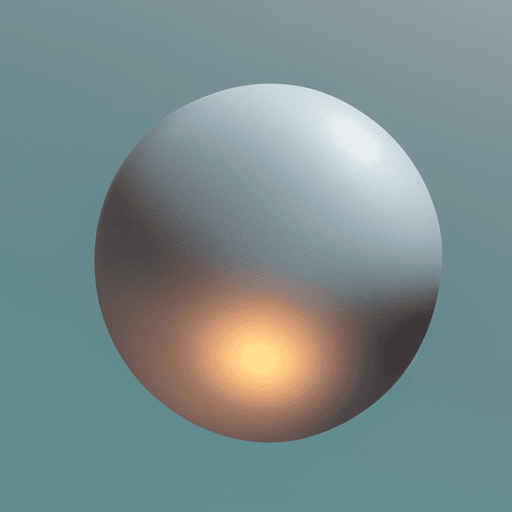}
\\

default
& \includegraphics[width=0.140\linewidth]{figs/multiple_prompts/default_0.png}
& \includegraphics[width=0.140\linewidth]{figs/multiple_prompts/default_1.png}
& \includegraphics[width=0.140\linewidth]{figs/multiple_prompts/default_3.png}
& \includegraphics[width=0.140\linewidth]{figs/multiple_prompts/default_4.png}
& \includegraphics[width=0.140\linewidth]{figs/multiple_prompts/default_5.png}
& \includegraphics[width=0.140\linewidth]{figs/multiple_prompts/default_6.png}
\\

1.5
& \includegraphics[width=0.140\linewidth]{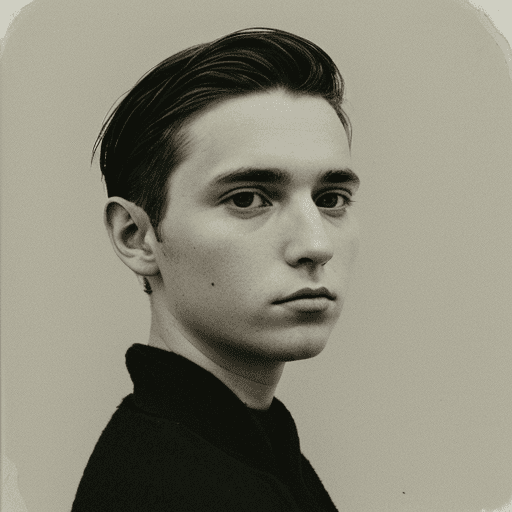}
& \includegraphics[width=0.140\linewidth]{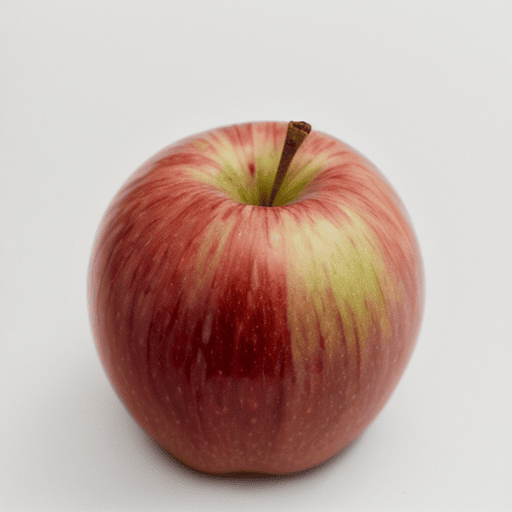}
& \includegraphics[width=0.140\linewidth]{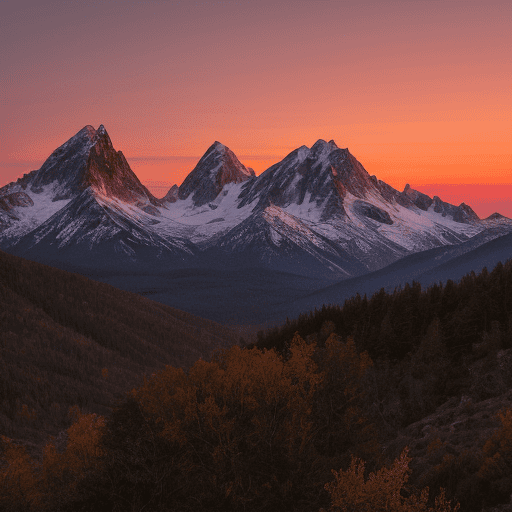}
& \includegraphics[width=0.140\linewidth]{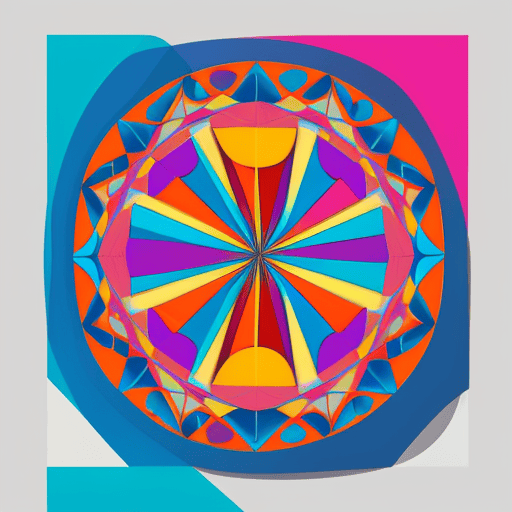}
& \includegraphics[width=0.140\linewidth]{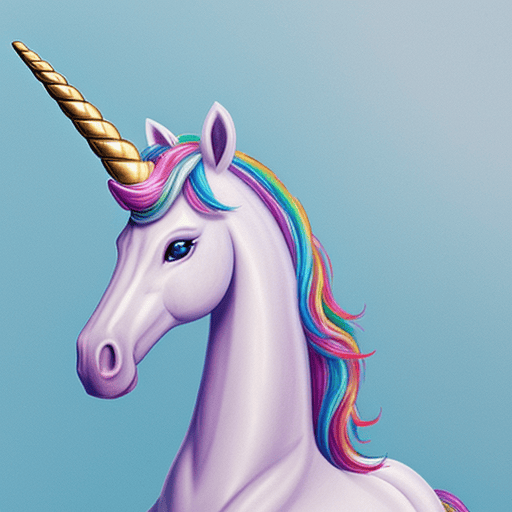}
& \includegraphics[width=0.140\linewidth]{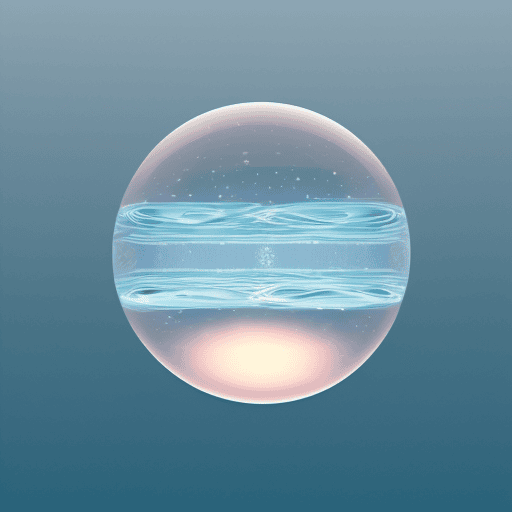}
\\

2.0
& \includegraphics[width=0.140\linewidth]{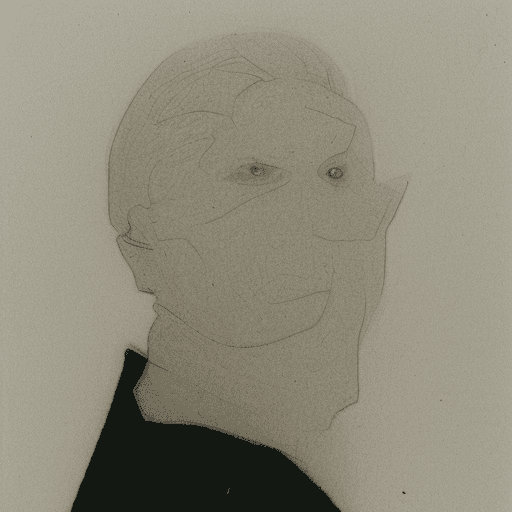}
& \includegraphics[width=0.140\linewidth]{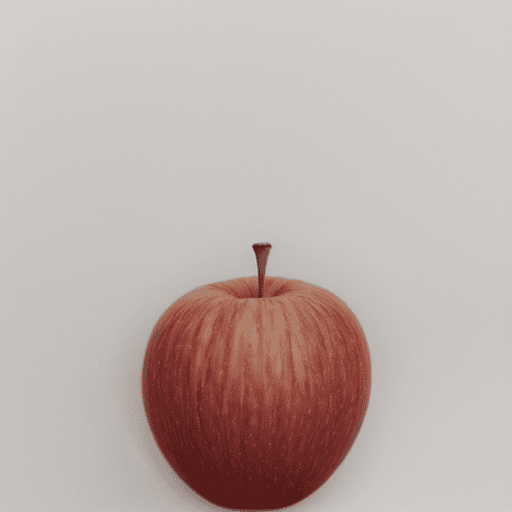}
& \includegraphics[width=0.140\linewidth]{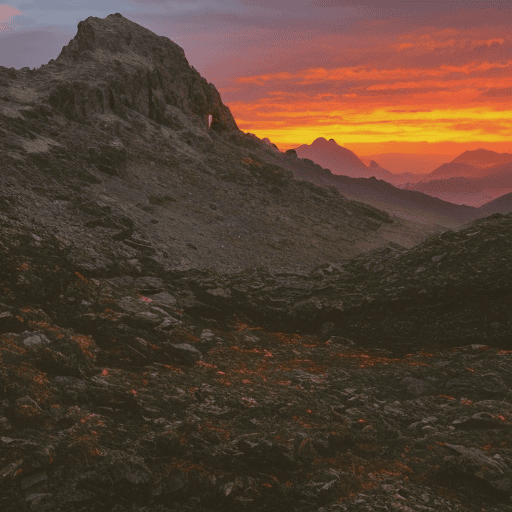}
& \includegraphics[width=0.140\linewidth]{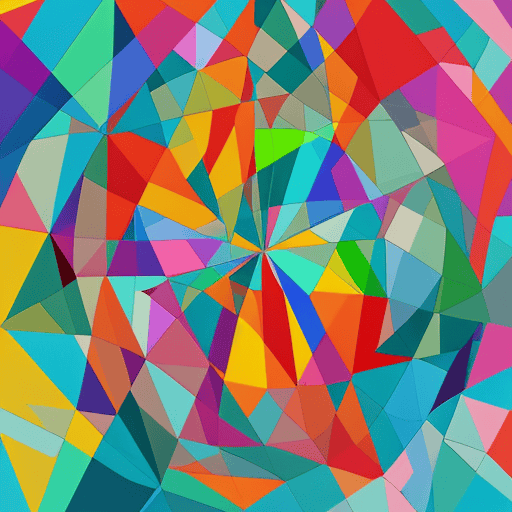}
& \includegraphics[width=0.140\linewidth]{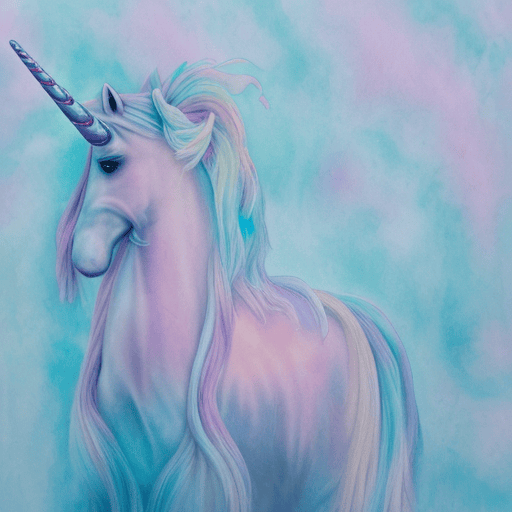}
& \includegraphics[width=0.140\linewidth]{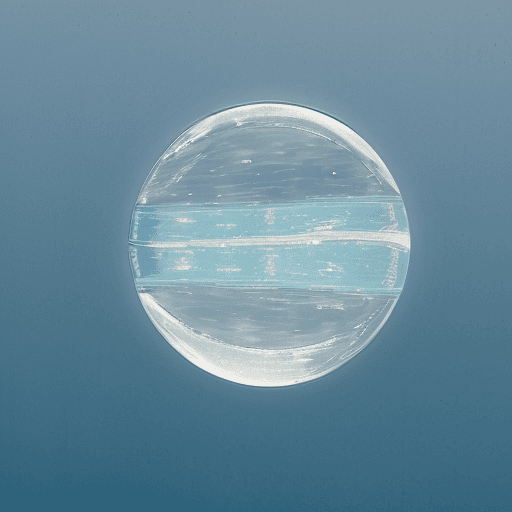}
\\
\hline
\end{tabular}

\caption{Bending the layer \textit{output\_blocks.3.1.norm}; Different prompts, same seed.}
\label{fig:multiple_prompts_5}
\end{figure}

Using the same prompt, we examine whether the same bending effects are consistent across different prompts where we change the style, composition and subject. In other words, we examine whether the bending effects relate to what a specific layer learns and its functional role during inference, or due to interactions with the prompt guidance, or both. The prompts are: (1) "Analog style portrait of a person"; (2) "A red apple on a white background"; (3) "Mountain landscape at sunset"; (4) "Abstract geometric shapes, vibrant colors"; (5) "unicorn"; (6) "a floating orb". Figures~\ref{fig:multiple_prompts_1}, ~\ref{fig:multiple_prompts_2}, ~\ref{fig:multiple_prompts_3}, ~\ref{fig:multiple_prompts_4}, ~\ref{fig:multiple_prompts_5} follow.

\subsubsection{Same Seed/Prompt/Layer, Different Timesteps}
\label{sec:results_multiple_timesteps}

\begin{figure}[H]
    \centering
    \includegraphics[width=0.50\linewidth]{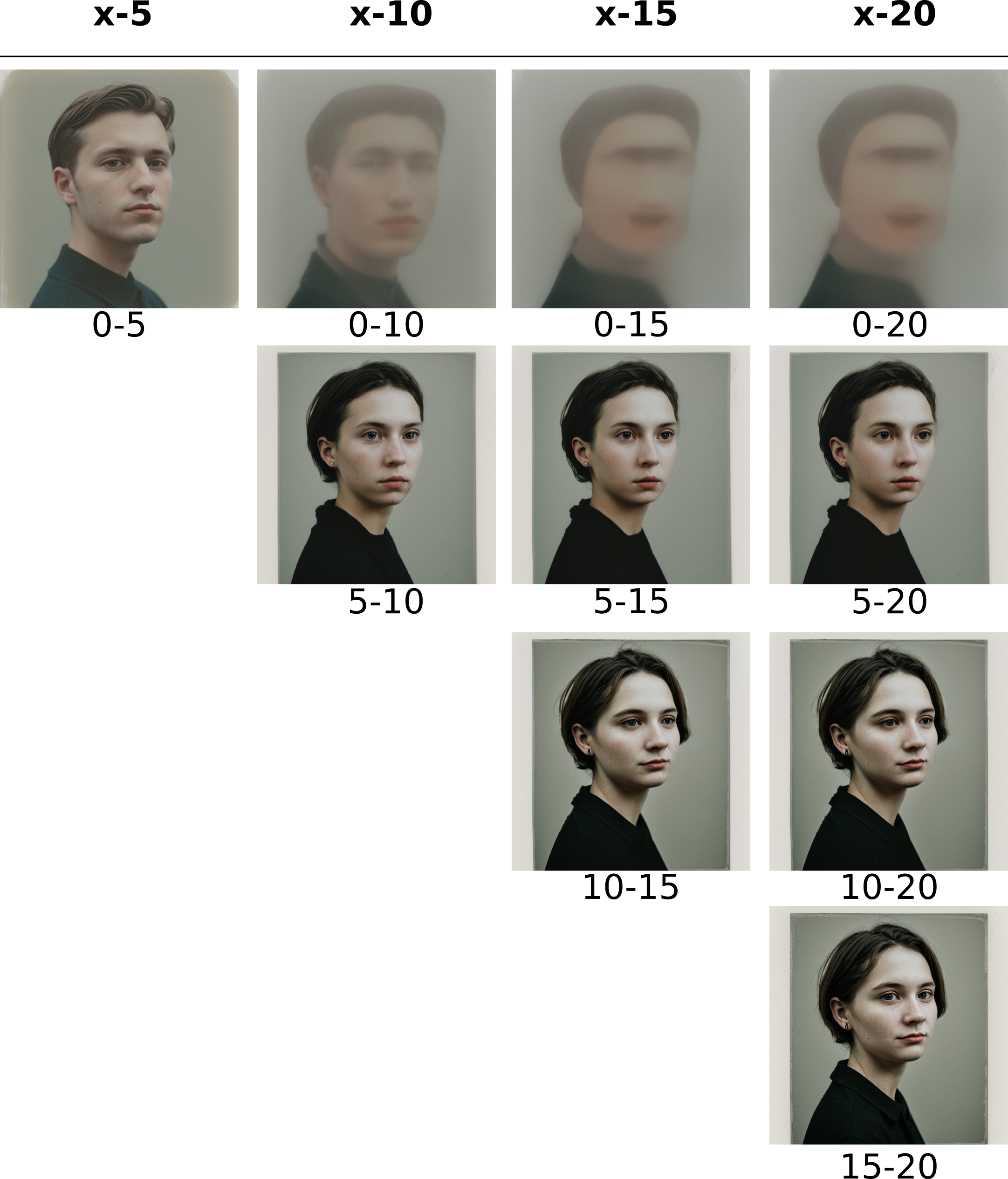}
    \caption{Varying the denoising time steps (out of 20) during which bending is applied.}
    \label{fig:placeholder}
\end{figure}

\subsection{Quantitative Results}
By analyzing the generated images and their associated latents, we can identify which parts of the model produce the most impact when bent. In particular, we can compute the cosine distance between the latents of the bent results and the latent of the default unbent result. Next, we can place bent results into groups and compare which of these groups, on average, produces a higher mean distance. The grouping can be based on whether bending was part of the input or output block, a Residual Block, a Spatial Transformer, a Convolutional layer or a Normalization layer, etc.  

Our empirical analysis reveals significant disparities in impact across different architectural components. Our findings indicate that the input region of the UNet architecture exerts a stronger influence on feature divergence than the output region, with a mean cosine distance of $0.1776 \pm 0.1674$ versus $0.1088 \pm 0.1234$. When comparing structural components, Residual Blocks (ResBlocks) demonstrated a greater impact ($0.1333 \pm 0.1541$) compared to Cross-Attention mechanisms ($0.1070 \pm 0.1110$), suggesting that the convolutional backbone is more sensitive to bending than the attention-based pathways.

At the layer level, convolutional layers (Conv2d) yielded a higher mean impact ($0.1685 \pm 0.1658$) than linear layers ($0.1185 \pm 0.1268$). Notably, normalization layers were among the most impactful atomic components, with LayerNorm ($0.1688 \pm 0.1562$) and GroupNorm ($0.1642 \pm 0.1627$) showing results comparable to convolutional layers. The most significant deviations were observed in high-level structural containers; specifically, the TimestepEmbedSequential ($0.4518 \pm 0.2453$) and Sequential ($0.4117 \pm 0.2454$) containers produced the highest mean distances. These areas, which include critical entry (e.g. timestep embedding such as \textit{time\_embed.2}) and exit points like \textit{input\_blocks.0} and \textit{out.0}, represent the most sensitive points in the model. Overall, these results suggest a clear hierarchy where Input $>$ Output and ResBlocks $>$ Cross-Attention in terms of their contribution to the total cosine distance.

\subsection{Findings}
Our qualitative results demonstrate relatively consistent visual impacts across different seeds and prompts for the same layer. In other words, the layer type/location/container and the bending magnitude ($0 <--> 2$) were more influential on the results, whereas images generated with the same bending magnitude and layer—but different seeds or prompts—tended to share greater resemblance than images where the magnitude or layer was changed. Resemblance within the same seed or prompt, keeping everything else fixed, was higher than when the seed or prompt was changed, which is expected given the role of text conditioning in guiding generation.

The type of layer appears to have a stronger impact than its location. This contrasts with StyleGAN models, where earlier layers consistently produced more drastic changes, while later layers primarily affected fine-grained features~\cite{broad_2021_NetworkbendingExpressive}. This difference is likely due to two factors. First, the UNet used in our study is composed of heterogeneous layers that serve different functions during inference. Second, diffusion models operate through multiple denoising iterations, where the input is gradually refined at each step.

In most of our study (Sections~\ref{sec:results_multiple_paths}, \ref{sec:results_multiple_seeds}, and \ref{sec:results_multiple_prompts}), we applied bending across all denoising steps, which ensured consistent results and, in effect, reinforced the bending impacts. In Section~\ref{sec:results_multiple_timesteps}, we observe that covering the initial steps (i.e., from the 0th timestep onward) produces most of the blurring visual effects seen when ablating \textit{output\_blocks.4.1.transformer\_blocks.0.norm1}. Skipping the first 5 or 10 timesteps, by contrast, primarily leads to fine-grained or decorative changes—similar to the role played by later layers in StyleGAN models.

The quantitative analysis confirms that the layer type has the strongest impact measured by the latent distance. It also confirms a big difference between input and output blocks, even if our qualitative assessment was not conclusive in that regard.

\section{Discussion}

\subsection{Model Bending Cheat Sheets}
Through conducting both qualitative and quantitative analyses, our intent is not to establish findings that generalize across different UNet architectures. Rather, we present this work as a demonstration of the kind of systematic experimentation that may support artistic practice. The observations derived from the specific UNet of SD1.5 can be translated into a model-specific “cheat sheet” that other artists using the same model may consult when directing their bending efforts, depending on the type and magnitude of effect they wish to produce.

Analyses of this kind can be carried out either through interactive interfaces (\autoref{sec:interactive_bending}) or through script-driven systematic sampling (\autoref{sec:preliminary_study}). Interactive approaches enable direct and exploratory manipulation, but diffusion models generally incur slower inference due to iterative denoising, despite emerging alternatives for near real-time generation~\cite{kodaira2025streamdiffusion}. Systematic exploration allows more exhaustive and controlled variation, yet requires computational resources and technical expertise to implement, visualize, and interpret results.

A balance between these modes may be achieved in at least two ways. First, analyses can be community-driven: if artists systematically document and share results for a given model, others can reuse and build upon this knowledge. Our preliminary findings suggest that bending effects exhibit a degree of consistency within the same model, supporting the feasibility of shared reference mappings. Second, systematic scripts can be prepared in advance and their results integrated into interactive interfaces. For example, visual indicators, such as colour-coded badges placed beside the layers/containers in \autoref{fig:interactive_bending_block}, could denote the relative impact of specific containers or layers (e.g., low, medium, or high).

Impact may be quantified through multiple complementary metrics, including perceptual difference (LPIPS)~\cite{zhang2018perceptual}, visual feature distance (e.g., DINOv2 embeddings)~\cite{oquab2023dinov2}, or semantic distance (e.g., CLIP image embeddings)~\cite{radford2021learning}. Presenting such analyses within the interface—whether generated locally or contributed by others—would allow artists to situate their experimentation within a broader, collectively developed understanding of the model’s behaviour.

\subsection{Direction Manipulation vs. Systematic Exploration}
Unlike direct manipulation, systematic exploration is relatively novel to the arts~\cite{hummel2023visual} compared to other design and engineering disciplines~\cite{sedlmair_2014_VisualParameterSpace}. For example, the study by Davis et al.~\cite{davis_2024_Fashioningcreativeexpertise} recognizes the utility of a \textit{design space exploration} framework when working with generative AI for ideation, and first-person accounts of AI-based video generation in ComfyUI~\cite{ledo_2025_GenerativeRotoscopingFirstPerson} paint a picture of current generative AI workflows as characterized by deliberation and systematic exploration rather than solely intuitive iteration.

\subsection{From Models as Commodity to Models as Material}
Model repositories like CivitAI~\cite{2025_civitai} and TensorArt~\cite{2025_tensorart} provide users with thousands of pre-trained and fine-tuned models, generated images with associated prompts and parameters, and reproducible workflows for a wide variety of use-cases and styles. The quantity and diversity of models and workflows shared online are simply unprecedented within the context of generative art, and cloud computing services are making them more accessible. A large portion of these models are adaptations and personalizations (e.g. Low-Rank Adaptations -- LoRAs~\cite{hu_2021_LoraLowrankadaptation}) of a smaller set of base models (e.g. Stable Diffusion~\cite{stabilityai_2024_stablediffusion} and Flux variants). Following Abonamah et al.~\cite{abonamah2021commoditization}, we argue that the proliferation of generative AI models is likely to lead to their commodification. Such a development may further entrench existing biases~\cite{vazquez2024taxonomy}, as machine learning models are not neutral technologies but systems shaped by the data on which they are trained. When treated as interchangeable commodities or opaque services, these embedded biases risk being obscured rather than critically examined. 

When technology becomes commodified, its design priorities shift—from serving as a raw material for creation (as with early computer terminals) to emphasizing accessibility and replaceability. This shift can bring innovation benefits, as seen with personal computers. However, in the case of generative AI, commodification is occurring at an unprecedented pace. The challenge, however, is that this rapid commodification may discourage sustained engagement with individual models. Instead of fostering deep familiarity with their inner workings, affordances, and limitations, users are incentivized to sample from an ever-expanding catalogue without developing the tacit knowledge necessary for critical or innovative use. This dynamic may also contribute to \textit{AI fatigue}—a feeling of exhaustion or overwhelm, potentially driven by the rapid pace of updates to AI models and tools, as well as the opaqueness and complexity of AI systems~\footnote{\url{https://newsletter.victordibia.com/p/you-have-ai-fatigue-thats-why-you}}. Historically, artists have cultivated a deep, material understanding of their tools—whether brushes, cameras, or software—as a prerequisite for meaningful creative expression. Generative models require the same kind of sustained engagement to be used critically and responsibly. At the same time, many artists—who are uniquely positioned to critique and subvert these systems~\cite{bryan-kinns_2025_XAIxArtsManifestoExplainablea, broad_2024_using}—are choosing not to engage with them on principle, citing concerns that these models are trained on the work of unattributed artists~\cite{kawakami2024impact}. Therefore, we argue that it's imperative to encourage and offer tools for artists to engage with AI in material~\cite{grabe_2025_HiddenLayerInteraction}, sustained~\cite{tecks_2024_ExplainabilityPathsSustained} and critical~\cite{broad_2024_using} ways, and the model understanding that is accrued through model bending could facilitate these forms of engagement.

\subsection{Manipulation for Behavioural and Theoretical Understanding}
The analyses presented here primarily contribute to what may be described as a \textit{behavioural} understanding of diffusion models: a working mental model of how specific interventions—at particular layers, timesteps, or magnitudes—affect outputs. Rather than offering a comprehensive mechanistic account of internal representations, our approach demonstrates how systematic experimentation can generate actionable knowledge about input–output relationships. In this sense, explainability becomes valuable insofar as it enables practitioners to develop reliable heuristics for achieving desired effects~\cite{bhattacharya2024towards}. The relative consistency observed across seeds and prompts within the same bending configuration further suggests that such heuristics may remain stable within a given model.

At the same time, our findings point to the importance of a degree of \textit{theoretical} understanding. Theoretical knowledge can guide intervention—for example, recognizing that text conditioning is mediated through cross-attention layers may motivate their manipulation when semantic modification is desired, such as replacing attention weights mid-inference~\cite{grabe_2025_PatchExplorerInterpreting}. Conversely, direct engagement with and manipulation of the model can reinforce \textit{theoretical} insight. Observations such as the stronger influence of layer type over spatial depth, or the disproportionate impact of early denoising steps on global structure, reflect structural properties of diffusion architectures. Attending to these characteristics moves beyond surface-level parameter exploration toward a more principled mode of manipulation—one that may enable outputs extending beyond the model’s original design constraints~\cite{broad_2021_Activedivergencegenerative}.

Taken together, these considerations suggest that art creation tools exposing generative model structure and enabling direct intervention may support the development of both \textit{behavioural} and \textit{theoretical} understanding. In addition, lightweight in-context explanations—such as tooltips describing the functional role of specific layers or modules—could further support \textit{theoretical} understanding by situating experimentation within an architectural framework. Such design elements may help users connect observed visual effects to underlying model components without requiring extensive prior technical knowledge. However, these remain design suggestions rather than empirically validated outcomes. Future work involving artist studies or longitudinal deployments would be required to assess how such interfaces shape creative workflows in practice.

\subsection{Model Bending and AI Literacy}
Creative projects have often been used to help the public engage with and understand technology—for example, game development has been widely adopted as a way of teaching programming concepts. In a similar way, artistic experimentation that involves manipulating large-scale generative models may offer an accessible path toward understanding how these systems work. By exposing and intervening in architectural components, artists and audiences can move beyond viewing generative AI as a “black box” and instead engage with it as material, similar in spirit to early circuit bending or hacking. Future work could explore installations, workshops, or participatory art experiences that contribute to AI literacy through model bending, in line with previous explorations~\cite{hemment2023ai}. Towards that end, we created a public website with pre-computed bending results: \url{https://diffusion-bending-demo.netlify.app}


\section{Limitations}
This work focuses on a single model architecture (SD1.5) and a specific UNet configuration. As such, the observations reported here should not be assumed to generalize to other diffusion models, model versions, or architectures. Differences in training data, schedulers, or architectural design may lead to different bending behaviours. Our evaluation relies primarily on qualitative inspection supported by quantitative distance metrics. While measures such as latent distance provide a useful proxy for impact, they do not fully capture perceptual, aesthetic, or semantic aspects of the generated outputs. Complementary metrics could offer a more comprehensive assessment. Ultimately, human studies would be required to evaluate how bending effects are perceived and interpreted in practice.

\section{Ethics Statement}
In this work, we explore the implications of model bending for creative exploration and AI explainability. Given the wide range of effects that bending can produce, it is possible that such interventions could bypass or disrupt guardrails placed on publicly deployed models, e.g., through Reinforcement Learning~\cite{ouyang_2022_Traininglanguagemodelsa}. At the same time, this capacity highlights bending’s potential as a tool for probing the limits, robustness, and reliability of generative systems.

\section{Future Work}
Future work could extend the interface to support additional diffusion models and architectural variants beyond SD1.5, enabling comparisons across systems. Optional visualizations of intermediate representations—such as layer activations or attention maps—could be added for users who want a closer look at how the model operates during inference. These features would remain optional, allowing users to continue experimenting based solely on outputs without requiring a technical understanding of model function.

From a research perspective, collaborations with artists and longer-term deployments would help assess how model bending fits into creative workflows. Crowd-sourced studies could also explore shared preferences regarding which model components produce compelling transformations, helping refine interface design and further develop the systematic mapping approach introduced in this work.

\section{Conclusion}
This paper reframes explainability for generative art as a form of craft knowledge built through manipulation rather than post-hoc interpretation. We argued that large diffusion models need not remain opaque commodities: when their structure can be inspected and intervened on, they can be treated as creative materials. To support this stance, we presented a ComfyUI-integrated model-bending toolkit that lets artists select internal components of a Stable Diffusion–style pipeline and apply training-free interventions, alongside an interactive interface for navigating model structure and iterating on bends. We also reported a preliminary, systematic exploration of bending in SD1.5 that combines qualitative examples with latent-space distance measures, highlighting that impact varies substantially by component type, region, and denoising-step range, and that many effects remain relatively consistent across prompts and seeds within a given configuration.

Together, these contributions suggest a practical path toward "doing-based" XAI in the arts: artists can develop actionable heuristics—what to bend, when to bend, and by how much—while also gaining architectural understanding through repeated engagement. Future work should evaluate these tools in longitudinal artist studies, expand coverage to newer diffusion architectures and additional intervention types, and support community sharing of model-specific "cheat sheets" and workflows. Ultimately, we argue that model-crafting tools like the one presented here—tools that encourage artists to treat large-scale generative models as creative materials—may contribute to ongoing conversations around authorship and agency in GenAI-assisted arts.

\printbibliography

\end{document}